\documentclass[rmp,aps,twocolumn]{revtex4}
\usepackage{graphics}
\usepackage{epsfig}
\usepackage{amsmath,amssymb}
\usepackage{natbib}

\newcommand{\bra}[1]{\langle#1|}                  
\newcommand{\ket}[1]{|#1\rangle}                  

\newcommand{\be}{\begin{equation}}
\newcommand{\ee}{\end{equation}}
\newcommand{\bea}{\begin{eqnarray}}
\newcommand{\eea}{\end{eqnarray}}
\newcommand{\beas}{\begin{eqnarray*}}
\newcommand{\eeas}{\end{eqnarray*}}
\newcommand{\spl}[1]{S^{+}_{#1}}                              
\newcommand{\smi}[1]{S^{-}_{#1}}                              

\newcommand{\ecr}[1]{c^{\dagger}_{\sigma,#1}}                 
\newcommand{\ean}[1]{c^{\phantom{\dagger}}_{\sigma,#1}}       
\newcommand{\eucr}[1]{c^{\dagger}_{\uparrow #1}}              
\newcommand{\euan}[1]{c^{\phantom{\dagger}}_{\uparrow #1}}    
\newcommand{\edcr}[1]{c^{\dagger}_{\downarrow #1}}            
\newcommand{\edan}[1]{c^{\phantom{\dagger}}_{\downarrow #1}}  
\newcommand{\fcr}[1]{c^{\dagger}_{#1}}                        
\newcommand{\fan}[1]{c^{\phantom{\dagger}}_{#1}}              
\newcommand{\bcr}[1]{b^{\dagger}_{#1}}                        
\newcommand{\ban}[1]{b^{\phantom{\dagger}}_{#1}}              
\newcommand{\fat}[1]{\mbox{\boldmath$#1$\unboldmath}}         
\newcommand{\ham}{\hat{H}}                          
\newcommand{\dm}{\hat{\rho}}                        
\newcommand{\ntot}{N_{\text{tot}}}                  
\newcommand{\imag}{\text{i}}                        
\newcommand{\impart}{\text{Im}}                     
\newcommand{\repart}{\text{Re}}                     
\newcommand{\id}{\mathbb{I}}                        
\newcommand{\BEA}{\begin{eqnarray}}
\newcommand{\EEA}{\end{eqnarray}}
\newcommand{\rar}{\rightarrow} 
 
\begin{document}
    
\title{The density-matrix renormalization group}
\thanks{Accepted by Rev.\ Mod.\ Phys.}

\author{U. Schollw\"{o}ck}

\affiliation{RWTH Aachen University, D-52056 
Aachen, Germany}

\begin{abstract}
The density-matrix renormalization group (DMRG) is a numerical 
algorithm for the efficient truncation of the Hilbert space of 
low-dimensional strongly correlated quantum systems based on a rather
general decimation prescription. This algorithm has achieved 
unprecedented precision in the description of one-dimensional quantum
systems. It has therefore quickly acquired the status of method of 
choice for numerical studies of one-dimensional quantum systems. 
Its applications 
to the calculation of static, dynamic and thermodynamic quantities
in such systems are reviewed. The potential of DMRG 
applications in the fields of two-dimensional quantum 
systems, quantum chemistry, 
three-dimensional small grains, nuclear physics, equilibrium and 
non-equilibrium statistical physics, and
time-dependent phenomena is discussed. This review also considers the 
theoretical foundations of the method, examining its relationship to
matrix-product states and the quantum information content of the density 
matrices generated by DMRG.
\end{abstract}
\maketitle
\tableofcontents

\section{INTRODUCTION}
\label{sec:introduction}
Consider a crystalline solid; it consists of some $10^{26}$ or more 
atomic nuclei and electrons for objects on a human scale. All nuclei 
and electrons are subject to the strong, long-range Coulomb interaction. 
While this system should {\em a priori} be described by 
the Schr\"{o}dinger equation, its explicit solution is 
impossible to find. Yet, it has become clear over the decades that 
in many cases the physical properties of solids can be 
understood to a very good approximation in the framework of 
some effective one-body problem. This is a 
consequence of the very efficient interplay of nuclei and electrons to 
screen the Coulomb interaction but on the very 
shortest length scales. 

This comforting picture may break down as 
various effects invalidate the fundamental tenet of weak 
effective interactions, taking us into the field of strongly 
correlated quantum systems, where the full electronic many-body problem
has to be considered. 
Of the routes to strong correlation, let me mention just two, because 
it is at their meeting point that, from the point of view of physical 
phenomena, the density-matrix renormalization group (DMRG) has its 
origin.

On the one hand, for arbitrary interactions, the Fermi liquid picture breaks down 
in one-dimensional solids and gives way to the Tomonaga-Luttinger 
liquid picture: essentially due to the small size of phase-space, 
scattering processes for particles close to the Fermi energy become 
so important that the (fermionic) quasiparticle picture of Fermi liquid theory 
is replaced by (bosonic) collective excitations.

On the other hand, for arbitrary dimensions, screening is typically so effective 
because of
strong delocalization of valence electrons over the lattice. 
In transition metals and rare earths, however, 
the valence orbitals are inner $d$- or $f$-orbitals.
Hence valence electrons are much more localized, albeit not immobile.
There is now a strong energy penalty to place two electrons 
in the same local valence orbital, and the motion of valence electrons 
becomes strongly correlated on the lattice. 

To study strongly correlated systems, simplified {\em model Hamiltonians} that try to retain just the 
core ingredients needed to describe some physical phenomenon 
and methods for their 
treatment have been designed. Localization suggests the use of
{\em tight-binding} lattice models, where local orbitals on one site can take 
$N_{\text{site}}=4$ different states of up to two electrons 
($\ket{0}$, $\ket{\!\uparrow}$, $\ket{\!\downarrow}$, $\ket{\!\uparrow\downarrow}$).

The simplest model Hamiltonian for just one valence orbital (band) 
with a kinetic energy term (electron hopping between sites
$i$ and $j$ with amplitude $t_{ij}$) and Coulomb repulsion is the 
on-site
{\em Hubbard model} \cite{Hubb63,Hubb64a,Hubb64b}, 
where just the leading on-site Coulomb repulsion 
$U$ has been retained:
\begin{equation}
    \ham_{\text{Hubbard}} = - \sum_{\langle ij \rangle,\sigma} t_{ij} 
    ( \fcr{i\sigma}\fan{j\sigma} + 
    \text{h.c.} ) + U \sum_{i} n_{i\uparrow}n_{i\downarrow} .
    \label{eq:Hubbardgeneral}
\end{equation}    
$\langle\rangle$ designates bonds. In the limit $U/t\gg 1$ 
double occupancy $\ket{\!\uparrow\downarrow}$ 
can be excluded resulting in the $N_{\text{site}}=3$ state $t$-$J$ {\em model}:
\begin{equation}
    \ham_{\text{tJ}} = - \sum_{\langle ij \rangle,\sigma} t_{ij} 
    ( \fcr{i\sigma}\fan{j\sigma} + 
    \text{h.c.} ) +  \sum_{\langle ij \rangle} J_{ij} ({\bf S}_{i}\cdot 
    {\bf S}_{j}-\frac{1}{4}n_{i}n_{j}) ,
    \label{eq:tJgeneral}
\end{equation}    
where the spin-spin interaction $J_{ij}=4t_{ij}^2/U$ is due to a 
second-order virtual hopping process possible only for
electrons of opposite spin on sites $i$ and $j$. 
At half-filling, the model simplifies 
even further to the spin-$\frac{1}{2}$ isotropic {\em Heisenberg 
model},
\begin{equation}
    \ham_{\text{Heisenberg}} = \sum_{\langle ij \rangle} J_{ij} {\bf S}_{i}\cdot 
    {\bf S}_{j} ,
    \label{eq:Heisengeneral}
\end{equation}    
placing collective (anti)ferromagnetism, which it describes, into the 
framework of strongly correlated systems. These and other model 
Hamiltonians have been modified in multiple ways.

In various fields of condensed matter physics, such as
high-temperature superconductivity, low-dimensional magnetism 
(spin chains and spin ladders), low-dimensional conductors, 
and polymer physics, theoretical research is focussing on the highly 
nontrivial properties of these seemingly simple models, which are 
believed to capture some of the essential physics. Recent
progress in experiments on ultracold atomic gases \cite{Grei02} has 
allowed the preparation of strongly correlated bosonic systems in 
optical lattices with tunable interaction parameters, attracting many solid-state
physicists to this field.
On the conceptual side, important old and new questions relating e.g.\ to
topological effects in quantum mechanics, the wide field of quantum phase 
transitions, and the search for exotic forms of order stabilized by 
quantum effects at low temperatures are at the center of the physics of 
strongly correlated quantum systems.

The typical absence of a dominant, exactly solvable contribution to the 
Hamiltonian, about which a perturbative expansion such as in 
conventional many-body physics might be attempted, goes a long way in 
explaining the inherent complexity of strongly correlated systems. This 
is why, apart from some exact solutions such as provided by the Bethe
ansatz or certain toy models, most analytical approaches are quite 
uncontrolled in their reliability. While these approaches may yield 
important insight into the nature of the physical phenomena observed, 
it is ultimately up to numerical approaches to assess the validity of 
analytical approximations. 

Standard methods in the field are the exact diagonalization of small
systems, the various quantum Monte Carlo methods, and, less widely used,
coupled cluster methods and series expansion techniques. 
Since its inception by Steve \citet{Whit92b}, the density-matrix 
renormalization group has quickly achieved the status of a 
highly reliable, precise, and versatile numerical method
in the field. Its main advantages are the ability to treat unusually
large systems at high precision at or very close to zero temperature 
and the absence of the negative sign problem that plagues the otherwise
most powerful method, quantum Monte Carlo, making the latter of limited use
for frustrated spin or fermionic systems. A first striking 
illustration of DMRG precision was given by \citet{Whit93a}. Using 
modest numerical means to study the $S=1$ isotropic antiferromagnetic 
Heisenberg chain, they calculated the ground state energy at
almost machine precision, $E_{0}=-1.401 484 038 971(4)J$, and the 
(Haldane) excitation gap as $\Delta=0.41050(2)J$. The spin-spin 
correlation length was found to be $\xi=6.03(2)$ lattice spacings. 
More examples of the versatility and precision which have made the 
reputation of the DMRG method will be found throughout the text.
The major drawback of 
DMRG is that it displays its full force mainly for one-dimensional
systems; nevertheless, interesting forays into higher dimensions have 
been made. By now, DMRG has, despite its complexity, become a standard 
tool of computational physics that many research groups in condensed matter 
physics will want to have at hand. In the simulation of model 
Hamiltonians, it ranks to me as part of a methodological trias 
consisting of exact diagonalization, quantum Monte Carlo and DMRG.

Most DMRG applications still treat 
low-dimensional strongly correlated model Hamiltonians, for which the 
method was originally developed. However, the 
core idea of DMRG, the construction of a renormalization-group flow 
in a space of suitably chosen density matrices, is quite general. In 
fact, another excellent way of 
conceptual thinking about DMRG is in the very general terms of 
quantum information theory. The versatility of the core idea has allowed the 
extension of DMRG applications to fields quite far away from its origins, such 
as the physics of (three-dimensional) small grains, equilibrium and
far-from-equilibrium problems in classical and quantum statistical mechanics 
as well as the quantum chemistry of small to medium-sized molecules. 
Profound connections to exactly solvable models in statistical 
mechanics have emerged.

The aim of this review is to provide both the algorithmic foundations 
of DMRG and an overview of its main, by now quite diverse fields of applications.

I start with a discussion of the key algorithmic 
ideas needed to deal with the most conventional DMRG problem,
the study of $T=0$ static properties of a one-dimensional quantum
Hamiltonian (Sec.\ \ref{sec:dmrg}). This includes standard improvements to the
basic algorithm 
that should always be used to obtain a competitive DMRG application, 
a discussion of the assessment of the numerical quality of DMRG 
output, as well as an overview of applications. Having read 
this first major section, 
the reader should be able to set up standard DMRG having the major algorithmic
design problems in mind. 

I move on to a section on DMRG theory (Sec.\ \ref{sec:theory}), 
discussing the properties of the quantum states DMRG generates 
(matrix product states) and the 
properties of the density matrices that are essential for its success; the 
section is closed by a reexamination of DMRG from a quantum 
information theory point of view. It might be quite useful to have 
at least some superficial grasp of the key results of this section in 
order to best appreciate the remainder of the review, while a more thorough 
reading might serve as a wrap-up in the end.
All the sections that come after these key
conceptual sections can then, to some large degree, be read independently.
        
Among the various branches of applied DMRG, the applications to dynamical 
properties of quantum systems are presented first (Sec.\ \ref{sec:dynamics}). Next, I discuss attempts to 
use DMRG for systems with potentially large numbers of local degrees of freedom
such as phononic models or Bose-Hubbard models (Sec.\ \ref{sec:many}). 

Moving beyond the world of $T=0$ physics in one dimension, DMRG as 
applied to two-dimensional quantum Hamiltonians in real space with 
short-ranged interactions is introduced and various 
algorithmic variants are presented (Sec.\ \ref{sec:twodim}).

Major progress has been made by abandoning the concept of an 
underlying real-space lattice, as various authors have developed DMRG 
variants for momentum-space DMRG (Sec.\ 
\ref{subsec:momentum}), DMRG for 
small molecule quantum chemistry (Sec.\ \ref{subsec:chemistry}),  
and DMRG for mesoscopic small grains and nuclei (Sec.\ \ref{subsec:energybasics}). All these fields are currently under very active 
development.

Early in the history of DMRG it was realized that its core 
renormalization idea might also be applied to the renormalization of 
transfer matrices that describe two-dimensional classical
(Sec.\ \ref{subsec:TMRG}) or 
one-dimensional quantum systems (Sec.\ \ref{subsec:quanttransfer}) in equilibrium 
at finite temperature.
Here, algorithmic details undergo major changes, such that this class 
of DMRG methods is often referred to as TMRG {\em (transfer matrix 
renormalization group)}.

Yet another step can be taken by moving to 
physical systems that are out of equilibrium (Sec.\ \ref{sec:nonequilibrium}). DMRG has been 
successfully applied to the steady states of nonequilibrium systems, 
leading to a non-hermitian extension of the DMRG where transition 
matrices replace Hamiltonians. Various methods for time-dependent problems are 
under development and have recently started to yield results 
unattainable by other numerical methods not only at $T=0$, but also 
for finite temperature and in the presence of dissipation. 

In this review, details of computer implementation
have been excluded. Rather, I have tried to focus on 
details of algorithmic structure and their relation to the physical 
questions to be studied using DMRG, and to give the reader some idea 
about the power and the limitations of the method.
In the more standard fields of DMRG, I have not been (able to be) exhaustive 
even in merely listing applications. 
In the less established fields of DMRG, I have tried to be 
more comprehensive and to provide as much
discussion of their details as possible, in order to make these fields better 
known and, hopefully, to stimulate thinking about further algorithmic 
progress. I have excluded, for lack of space, any extensive discussions of the 
analytical methods whose development has been stimulated by DMRG 
concepts. 
Last but not least it should be mentioned that some of the 
topics of this review have been considered by other authors: 
\citet{Whit98a} gives an 
introduction to the fundamentals of DMRG; a very 
detailed survey of DMRG as it was in late 1998 has been provided in a 
collection of lectures and articles \cite{Pesc99}; it also contains 
an account of DMRG history by White. More recently,
the application of TMRG to quantum systems and two-dimensional DMRG 
have been revieved by \citet{Shib03}. \citet{Hall03} gives a rather 
complete overview of DMRG applications. \citet{Duke04} focus on DMRG 
applications to finite Fermi systems such as small grains, small 
molecules and nuclei. 

A word on {\em notation:} All state spaces considered here can be factorized 
into local state spaces $\{ \ket{\sigma} \}$ labelled by Greek letters. 
DMRG forms blocks of lattice sites; (basis) states 
$\ket{m}$ of such blocks I denote by Latin letters. These states depend
on the size of the block; when necessary, I indicate the block length 
by subscripts $\ket{m_{\ell}}$. Correspondingly, $\ket{\sigma_{i}}$ 
is a local state on site $i$. Moreover, DMRG typically operates 
with two blocks and two sites, which are referred to as
belonging to ``system'' or ``environment''. Where this distinction matters, 
it is indicated by superscripts, $\ket{m^S}$ or $\ket{\sigma^E}$.

\section{KEY ASPECTS OF DMRG}
\label{sec:dmrg}
Historically, DMRG has its origin in the analysis by \citet{Whit92a} of
the failure of {\em real-space renormalization group} (RSRG) methods to 
yield quantitatively acceptable results for the low-energy properties 
of quantum many-body problems. Most of the DMRG algorithm can be formulated
in standard (real space) renormalization group language.  
Alternative points of view in terms of matrix product states and 
quantum information theory 
will be taken later
(Sec.\ \ref{subsec:mpa} and Sec.\ \ref{subsec:quantint}). 
Before moving on to the details, let me 
mention a debate that has been going on among DMRG practitioners on 
whether calling the DMRG algorithm a renormalization-group method is 
a misnomer or not. The answer depends on what one considers the 
essence of a RG method. If this essence is the systematic thinning out 
of degrees of freedom leading to effective Hamiltonians, DMRG is an RG 
method. However, it does not involve an ultraviolet or infrared energy 
cutoff in the degrees of freedom which is at the heart of traditional 
RG methods and hence often considered as part of the core concepts. 

\subsection{Real-space renormalization of Hamiltonians}
The set of concepts grouped under the heading of 
``renormalization'' has proven extremely powerful in providing 
succinct descriptions of the collective behavior of systems of size 
$L$
with a diverging number of degrees of freedom $N_{\text{site}}^{L}$. 
Starting from some 
microscopic Hamiltonian, degrees of freedom are iteratively 
integrated out and accounted for by  
modifying the original Hamiltonian. The new Hamiltonian 
will exhibit modified as well as new couplings, and renormalization group 
approximations typically consist in physically motivated truncations 
of the set of couplings newly generated by the {\em a priori} exact 
elimination of degrees of freedom. One obtains a simplified (``renormalized'') 
effective Hamiltonian that should catch the essential physics of the 
system under study. The success of this approach rests on 
scale separation: for continuous phase transitions, the 
diverging correlation length sets a natural long-wavelength 
low-energy scale which dominates the physical properties, and 
fluctuations on shorter length scales may be integrated out and 
summed up into quantitative modifications of the long-wavelength 
behavior. In the Kondo problem, the width of the Kondo resonance sets 
an energy scale such that the exponentially decaying 
contributions of energy levels far 
from the resonance can be integrated out. This is the essence of 
the numerical renormalization group \cite{Wils75}. 

\begin{figure}
\centering\epsfig{file=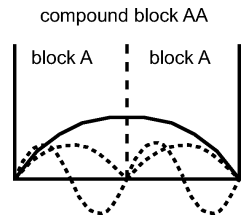,width=0.35\linewidth}
\caption{Lowest-lying eigenstates of blocks A (dashed) and AA (solid) 
for the single particle in a box problem in the continuum limit.}
\label{fig:singleparticlebox}
\end{figure}

Following up on the Kondo problem, it was hoped that thermodynamic-limit 
ground-state properties of other many-body 
problems such as the one-dimensional Hubbard or Heisenberg models 
might be treated similarly, with lattice sites replacing energy levels. 
However, results of real-space renormalization schemes turned out to 
be poor. While a precise analysis of this observation is 
hard for many-body problems, \citet{Whit92a} identified the breakdown of 
the {\em real-space renormalization group} for the toy model of a 
{\em single non-interacting 
particle} hopping on a discrete one-dimensional lattice (``particle in 
a box'' problem). For a box of size $L$, the Hilbert space spanned by $\{\ket{i}\}$
is $L$-dimensional; in state $\ket{i}$, the particle is on site $i$. The matrix elements of 
the Hamiltonian are band-diagonal, $\bra{i} \ham \ket{i} =2$; 
$\bra{i} \ham \ket{i\pm 1} =-1$ in some units.
Consider now the following {\em real-space renormalization group} procedure: 
\begin{enumerate}
    \item Interactions on an initial sublattice (``block'') A of length $\ell$ 
    are described by a block Hamiltonian $\ham_{\text A}$ acting 
    on an $M$-dimensional Hilbert space.
    \item Form a compound block AA of length $2\ell$ and the 
    Hamiltonian $\ham_{\text{AA}}$, consisting of two block Hamiltonians and 
    interblock interactions. $\ham_{\text{AA}}$ has dimension $M^2$.
    \item Diagonalize $\ham_{\text{AA}}$ to find the $M$ lowest-lying 
    eigenstates. 
    \item Project $\ham_{AA}$ onto the truncated space spanned 
    by the 
    $M$ 
    lowest-lying eigenstates, $\ham_{\text{AA}}\rightarrow 
    \ham^{\text{tr}}_{\text{AA}}$.
    \item Restart from step 2, with doubled block size: $2\ell 
    \rightarrow \ell$, AA $\rightarrow$ A, and $\ham^{\text{tr}}_{\text{AA}}
    \rightarrow\ham_{\text A}$, until the box size is reached.
\end{enumerate}
The key point is that the decimation procedure of the Hilbert space is to 
take the lowest-lying eigenstates of the compound block AA. This 
amounts to the assumption that the ground state of the entire box will 
essentially be composed of energetically low-lying states living on 
smaller blocks. The outlined
real-space renormalization procedure gives very poor results. 
The breakdown can best be understood visually (Fig.\ 
\ref{fig:singleparticlebox}): assuming an already 
rather large block size, where discretization can be neglected, the 
lowest-lying states of A all have nodes at the lattice ends, such that
all product states of AA have nodes at the compound block center. The 
true ground state of AA has its maximum amplitude right there, such that it 
cannot be properly approximated by a restricted number of block states.  

Merely considering isolated blocks imposes wrong boundary conditions, and 
\citet{Whit92a} could obtain excellent results by combining  
Hilbert spaces from low-lying states of block A obtained by assuming 
various combinations of fixed and free boundary 
conditions (i.e.\ enforcing a vanishing wave function or a vanishing 
wave function derivative respectively at the boundaries). 
They also realized that 
combining various boundary conditions for a single particle would 
translate to accounting for fluctuations between 
blocks in the case of many interacting particles. This observation in 
mind, we now return to the 
original question of a many-body problem in the thermodynamic limit 
and formulate the following strategy:
To analyze which states have to be retained for a finite-size block A, 
A has to be embedded in some environment, mimicking 
the thermodynamic limit system A is ultimately embedded in.

\subsection{Density matrices and DMRG truncation}
\label{subsec:densityflow}
Consider, instead of the
exponentially fast growth procedure outlined above, the following 
linear growth prescription \cite{Whit92b}:
Assume that for a system (a {\em block} in DMRG language) 
of length $\ell$ we have an $M^S$-dimensional Hilbert space
with states $\{ \ket{m^{S}_{\ell}}\}$. The 
Hamiltonian $\ham_{\ell}$ is given by matrix elements $\bra{m^S_{\ell}} 
\ham_{\ell} \ket{\tilde{m}^{S}_{\ell}}$. Similarly we know the matrix 
representations of local operators such as $\bra{m^S_{\ell}} 
\fan{i} \ket{\tilde{m}^{S}_{\ell}}$.

For linear growth, we now construct $\ham_{\ell+1}$ in the product basis 
$\{\ket{m^S_{\ell}\sigma}\} \equiv 
\{\ket{m^S_{\ell}}\ket{\sigma^S}\}$, where $\ket{\sigma^S}$ are 
the $N_{\text{site}}$ local states of a new site added. 

\begin{figure}
\centering\epsfig{file=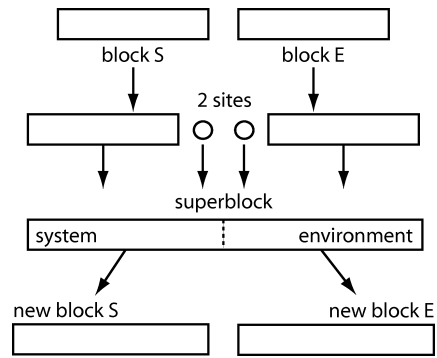,width=0.65\linewidth}
\caption{System meets environment: Fundamental DMRG construction of a superblock 
from two blocks and two single sites.}
\label{fig:systemmeetsenvironment}
\end{figure}

The thermodynamic limit is now mimicked by embedding the system in 
an {\em environment} of the same size, assumed to have been 
constructed in analogy to the system. 
We thus arrive at a {\em superblock}
of length $2\ell +2$ (Fig.\ \ref{fig:systemmeetsenvironment}), where 
the arrangement chosen is typical, but not mandatory. 

As the final goal is the ground state in the thermodynamic limit, 
one studies the best approximation to it, the ground state of 
the superblock, obtained by numerical 
diagonalization:
\begin{eqnarray}
    \ket{\psi}&=&
    \sum_{m^{S}=1}^{M^{S}} 
    \sum_{\sigma^S=1}^{N_{\text{site}}} 
    \sum_{\sigma^E=1}^{N_{\text{site}}} 
    \sum_{m^E=1}^{M^{E}} \psi_{m^S\sigma^S\sigma^E m^E}
    \ket{m^S\sigma^S}\ket{m^E\sigma^E} \nonumber \\
    &\equiv& \sum_{i}^{N^{S}} \sum_{j}^{N^{E}} \psi_{ij} 
    \ket{i}\ket{j}; \quad\quad \langle\psi|\psi\rangle = 1,
    \label{eq:be-ground}
\end{eqnarray}
where $\psi_{m^S\sigma^S\sigma^E m^E}=
\langle m^S\sigma^S ; \sigma^E m^E | \psi \rangle$.
$\{ \ket{m^S\sigma^S} \}\equiv \{ \ket{i} \}$ and $\{ \ket{m^E\sigma^E} 
\} \equiv \{ \ket{j} \}$ are the orthonormal 
product bases of system and environment (subscripts have been dropped) 
with dimensions $N^{S}=M^S N_{\text{site}}$ and $N^{E}=M^E N_{\text{site}}$
respectively (for later generalizations, we allow $N^S \neq N^E$). 
Some truncation procedure from $N^{S}$ to $M^{S}<N^{S}$ states must 
now
be implemented. Let me present three lines of argumentation on the 
optimization of some quantum mechanical quantity, all 
leading to the same truncation prescription focused on density-matrix 
properties. This is to highlight 
different aspects of the DMRG algorithm and to give confidence in 
the prescription found.

{\em Optimization of expectation values} \cite{Whit98a}: If the superblock 
is in a pure state $\ket{\psi}$ as in Eq. (\ref{eq:be-ground}), 
statistical physics describes the physical state of the system through a 
reduced density-matrix $\dm$,
\begin{equation}
    \dm = \text{Tr}_{E} \ket{\psi}\bra{\psi},
    \label{eq:dmastrace}
\end{equation}
where the states of the environment have been traced out,
\begin{equation}
    \bra{i} \dm \ket{i'} = \sum_{j} \psi_{ij} \psi^*_{i'j} .
    \label{eq:dmaselements}
\end{equation}
$\dm$ has $N^{S}$ eigenvalues $w_{\alpha}$ and orthonormal 
eigenstates
$\dm\ket{w_{\alpha}} = w_{\alpha}\ket{w_{\alpha}}$, with $\sum_{\alpha} 
w_{\alpha}=1$ and $w_{\alpha}\geq 0$. We assume the states are ordered 
such that $w_{1}\geq w_{2} \geq w_{3} \geq \ldots$.
The intuition that the ground state of 
the system is best described by retaining those $M^{S}$ states with 
largest weight $w_{\alpha}$ in the density matrix can be formalized as 
follows. Consider some bounded operator $\hat{A}$ acting on the 
system,
such as the energy per lattice bond; $\| \hat{A} \| = \max_{\phi} | \bra{\phi} 
\hat{A} \ket{\phi} / \langle \phi | \phi \rangle | \equiv c_{A}$.
The expectation value of $\hat{A}$ is found to be, 
using Eq.\ (\ref{eq:be-ground}) and 
Eq.\ (\ref{eq:dmaselements}),
\begin{equation}
    \langle \hat{A} \rangle = \bra{\psi} \hat{A} \ket{\psi}/  \langle\psi|\psi\rangle = 
    \text{Tr}_{S} \dm \hat{A} .
    \label{eq:expectationvalue}
\end{equation}    
Expressing Eq.\ (\ref{eq:expectationvalue}) in the density-matrix 
eigenbasis, one finds
\begin{equation}
    \langle \hat{A} \rangle = \sum_{\alpha=1}^{N^{S}} w_{\alpha} 
    \bra{w_{\alpha}} \hat{A} 
    \ket{w_{\alpha}} .
    \label{eq:Aindmbasis}
\end{equation}
Then, if we 
project the system state space down to the $M^{S}$ dominant 
eigenvectors $\ket{w_{\alpha}}$ with the largest eigenvalues,
\begin{equation}
    \langle \hat{A} \rangle_{\text{approx}} = \sum_{\alpha=1}^{M^{S}} w_{\alpha} 
    \bra{w_{\alpha}} \hat{A} 
    \ket{w_{\alpha}} ,
    \label{eq:Aindmbasisapprox}
\end{equation}
and the error for $\langle \hat{A} \rangle$ is bounded by
\begin{equation}
    | \langle \hat{A} \rangle_{\text{approx}} - \langle \hat{A} \rangle | \leq 
    \left( \sum_{\alpha>M^{S}}^{N^{S}} w_{\alpha} \right) c_{A} \equiv 
    \epsilon_{\rho} c_{A} .
    \label{eq:errorestimate}
\end{equation}
This estimate holds in particular for energies. Several remarks are in order. 
Technically, I have neglected to trace the fate of the denominator in 
Eq.\ (\ref{eq:expectationvalue}) upon projection; the ensuing 
correction of $(1-\epsilon_{\rho})^{-1}$ is of no relevance to the 
argument here as $\epsilon_{\rho}\rightarrow 0$. 
The estimate could be tightened for any specific operator, 
knowing $\bra{w_{\alpha}} \hat{A} \ket{w_{\alpha}}$, and a more efficient 
truncation procedure be named. For arbitrary bounded operators acting on the 
system, the 
prescription to retain the state spanned by the $M^{S}$ dominant 
eigenstates is optimal. For local quantities, such as 
energy, magnetization or density, errors are of the order of
the {\em truncated weight}
\begin{equation}
    \epsilon_{\rho}=1-\sum_{\alpha=1}^{M^{S}} w_{\alpha},
    \label{eq:truncatedweight}
\end{equation}
which emerges as the key estimate. Hence, a fast decay of density 
matrix eigenvalues $w_{\alpha}$ is essential for the performance of this truncation procedure.
The {\em truncation error} of Eq.\ (\ref{eq:errorestimate})  is the 
total error only if the system had been embedded in the final and 
exactly described
environment. Considering the iterative system and environment growth 
and their approximate representation at each step, additional sources of an {\em 
environmental error} have to be considered. In practice, therefore, 
errors for observables calculated by DMRG are often much larger than 
the truncated weight, even after additional steps to eliminate 
environmental errors have been taken. Careful extrapolation of 
results in $M^S$ (rather $\epsilon_{\rho}$) is therefore highly 
recommended, as will be discussed below.  

{\em Optimization of the wave function} \cite{Whit92b,Whit93b}: 
Quantum mechanical objects are 
completely described by their wave function. It is thus a reasonable 
demand for a truncation procedure that the approximative wave 
function $\ket{\tilde{\psi}}$ where the system space has been truncated 
to be spanned by only $M^{S}$ 
orthonormal states $\ket{\alpha}=\sum_{i} u_{\alpha i} \ket{i}$,
\begin{equation}
\ket{\tilde{\psi}} = 
\sum_{\alpha=1}^{M^{S}} \sum_{j=1}^{N^{E}} a_{\alpha 
j}\ket{\alpha}\ket{j} ,
\label{eq:psiapproximated}
\end{equation}
minimizes the distance in the quadratic norm
\begin{equation}
    \parallel \ket{\psi} - \ket{\tilde{\psi}} \parallel .
    \label{eq:distance}
\end{equation}
This problem finds a very compact solution in a singular value 
decomposition (SVD), which was the original approach of White;
SVD will be considered in a slightly different setting in the next 
section. The following is an alternative, more pedestrian approach.
Assuming real coefficients for simplicity, one has to minimize
\begin{equation}
1 - 2 \sum_{\alpha i j} \psi_{ij} a_{\alpha j} u_{\alpha i} +
\sum_{\alpha j} a^2_{\alpha j}
\label{eq:minimcond}
\end{equation}
with respect to $a_{\alpha j}$ and $u_{\alpha i}$. 
To be stationary in $a_{\alpha j}$, we must have
$\sum_{i} \psi_{ij}u_{\alpha i} = a_{\alpha j}$, finding that
\begin{equation}
1 - \sum_{\alpha i i'} u_{\alpha i}\rho_{ii'} u_{\alpha i'} 
\label{eq:psiapprox}
\end{equation}
must be stationary for the global minimum of the distance 
(\ref{eq:distance}), where we have introduced the density-matrix 
coefficients
\begin{equation}
\rho_{ii'} = \sum_{j} \psi_{ij} \psi_{i'j} .
\label{eq:densitydef}
\end{equation}
Equation (\ref{eq:psiapprox}) is stationary, according to the Rayleigh-Ritz 
principle, for $\ket{\alpha}$ being 
the eigenvectors of the density matrix. Expressing
Eq.\ (\ref{eq:psiapprox}) in the density-matrix eigenbasis, the global 
minimum is given by choosing $\ket{\alpha}$ to be the $M^{S}$ 
eigenvectors $\ket{w_{\alpha}}$ to the largest eigenvalues 
$w_{\alpha}$ of the density matrix, as they are all non-negative, and 
the minimal distance squared is, using Eq.\ (\ref{eq:truncatedweight}),
\begin{equation}
\parallel \ket{\psi} - \ket{\tilde{\psi}} \parallel^2 =
1 - \sum_{\alpha=1}^{M^{S}} w_{\alpha} = \epsilon_{\rho} .
\end{equation}
The truncation prescription now appears as a {\em 
variational principle} for the wave function.

{\em Optimization of entanglement} \cite{Gait01,Gait03,Osbo01,Lato03}: Consider the superblock 
state $\ket{\psi}$ as in Eq.\ (\ref{eq:be-ground}). The essential feature of 
a nonclassical state is its {\em entanglement}, the fact that it 
cannot be written as a simple product of one system and one 
environment state. Bipartite entanglement as relevant here can 
best be studied by representing $\ket{\psi}$ in its form after a 
{\em Schmidt decomposition} \cite{Niel00}: Assuming without loss of 
generality $N^S\geq N^E$, consider the $(N^S \times N^E)$-dimensional 
matrix $A$ with $A_{ij}=\psi_{ij}$. Singular value decomposition 
guarantees $A=UDV^T$, where $U$ is $(N^S \times N^E)$-dimensional with 
orthonormal columns, $D$ is a $(N^E \times N^E)$-dimensional diagonal 
matrix with non-negative entries $D_{\alpha\alpha}=\sqrt{w_{\alpha}}$, 
and $V^T$ is a $(N^E \times N^E)$-dimensional unitary matrix; $\ket{\psi}$
can be written as 
\begin{eqnarray}
    \ket{\psi} &=& 
    \sum_{i=1}^{N^S}\sum_{\alpha=1}^{N^E}\sum_{j=1}^{N^E} 
    U_{i\alpha} \sqrt{w_{\alpha}} V^T_{\alpha j} \ket{i}\ket{j} \\
    &=&
    \sum_{\alpha=1}^{N^E} \sqrt{w_{\alpha}} \left( \sum_{i=1}^{N^S} 
    U_{i\alpha} \ket{i} \right)
    \left( \sum_{j=1}^{N^E} V_{j\alpha} \ket{j} \right). \nonumber
\end{eqnarray}    
The orthonormality properties of $U$ and $V^T$ ensure that
$\ket{w_{\alpha}^S} = \sum_{i} U_{i\alpha}\ket{i}$ and 
$\ket{w_{\alpha}^E} = \sum_{j} V_{j\alpha}\ket{j}$ form orthonormal 
bases of system and environment respectively, in which the Schmidt 
decomposition
\begin{equation}
    \ket{\psi} = \sum_{\alpha=1}^{N_{\text{Schmidt}}} \sqrt{w_{\alpha}}
    \ket{w_{\alpha}^S}\ket{w_{\alpha}^E} 
    \label{eq:Schmidtdecomp}
\end{equation}
holds. $N^{S}N^{E}$ coefficients $\psi_{ij}$ are reduced to 
$N_{\text{Schmidt}}\leq N^E$ non-zero coefficients $\sqrt{w_{\alpha}}$,
$w_{1}\geq w_{2}\geq w_{3} \geq \ldots$. Relaxing the 
assumption $N^S\geq N^E$, one has
\begin{equation}
    N_{\text{Schmidt}} \leq \min (N^{S},N^{E}) .
    \label{eq:Schmidtmax}
\end{equation}
The suggestive labelling of states and coefficients in Eq.\ 
(\ref{eq:Schmidtdecomp}) is motivated by the observation that upon 
tracing out environment or system the reduced density matrices for 
system and environment are found to be
\begin{equation}
    \dm_{S} = \sum_{\alpha}^{N_{\text{Schmidt}}} w_{\alpha} 
    \ket{w_{\alpha}^S}\bra{w_{\alpha}^S};
    \quad
    \dm_{E} = \sum_{\alpha}^{N_{\text{Schmidt}}} w_{\alpha} 
    \ket{w_{\alpha}^E}\bra{w_{\alpha}^E}.
    \label{eq:dmbySchmidt}
\end{equation}
Even if system and environment are different (Sec.\ \ref{subsec:finite}), 
both density matrices would have the same number of non-zero eigenvalues, 
bounded by the smaller of the dimensions of system and environment, and 
an identical eigenvalue spectrum. Quantum information theory now 
provides a measure of entanglement through the von Neumann entropy,
\begin{equation}
    S_{\text{vN}}=-\text{Tr} \dm \ln_{2} \dm = -\sum_{\alpha=1}^{N_{\text{Schmidt}}} 
    w_{\alpha}\ln_{2} w_{\alpha}.
    \label{eq:vonNeumannentropy}
\end{equation}    
Hence, our truncation procedure to $M^{S}$ states preserves 
a maximum of system-environment entanglement if we retain the 
first $M^{S}$ 
states $\ket{w_{\alpha}}$ for
$\alpha=1,\ldots,M^{S}$, as $-x \ln_{2} x $ grows monotonically 
for $0<x\leq 1/e$, which is larger than typical discarded eigenvalues. 
Let me add that this optimization statement holds strictly only for 
the unnormalized truncated state. Truncation leads to a wave function 
norm $\sqrt{1-\epsilon_{\rho}}$ and enforces a new normalization 
which changes the retained $w_{\alpha}$ and $S_{vN}$. While one may 
easily construct density-matrix spectra for which upon normalization 
the truncated state produced by DMRG does no longer maximize entanglement, 
for typical DMRG density-matrix spectra the optimization statement 
still holds in practice. 
\subsection{Infinite-system DMRG}
\label{subsec:infinitesize}
Collecting the results of the last two sections, the so-called {\em 
infinite-system DMRG algorithm} can now be formulated \cite{Whit92b}. 
Details of efficient implementation will be discussed in 
subsequent sections; here, we assume that we are looking for the ground 
state:
\begin{enumerate}
    \item Consider a lattice of some small size $\ell$, forming the system block 
    S. S lives on a Hilbert space of size $M^S$ with states $\{ \ket{M^S_{\ell}} 
    \}$; the Hamiltonian $\ham^S_{\ell}$ and the operators acting on 
    the block are assumed to be known in this basis. At initialization, 
    this may still be an exact basis of the block 
    ($N_{\text{site}}^\ell \leq M^S$). Similarly, form an environment 
    block E.
    \item Form a tentative new system block S$'$ from S and one added 
    site (Fig.\ \ref{fig:systemmeetsenvironment}). S$'$ lives on a Hilbert space 
    of size $N^S=M^S N_{\text{site}}$, with a basis of
    product states $\{ \ket{M^S_{\ell}\sigma} \} \equiv \{ 
    \ket{M^S_{\ell}} \ket{\sigma} \}$.
    In principle, the Hamiltonian $\ham^S_{\ell+1}$ acting on S$'$ can 
    now be expressed in this basis (which will not be done explicitly 
    for efficiency, see Sec.\ \ref{subsec:large}). A new environment 
    E$'$ is built from E in the same way.
    \item Build 
    the superblock of length $2\ell+2$ from S$'$ and E$'$.  
    The Hilbert space is of size 
    $N^SN^E$, and the matrix elements of the Hamiltonian 
    $\ham_{2\ell+2}$ could in principle be constructed explicitly,
    but this is avoided for efficiency reasons.
    \item Find by large sparse-matrix diagonalization of 
    $\ham_{2\ell+2}$ the ground state $\ket{\psi}$. This is the most 
    time-consuming part of the algorithm (Sec.\ \ref{subsec:large}).
    \item Form the reduced density-matrix $\dm = \text{Tr}_{E} 
    \ket{\psi}\bra{\psi}$ as in Eq.\ (\ref{eq:dmaselements})
    and determine its eigenbasis 
    $\ket{w_{\alpha}}$ ordered by descending eigenvalues (weight) $w_{\alpha}$.
    Form a new (reduced) basis for S$'$ by taking the $M^S$ eigenstates 
    with the largest weights. In the product basis of S$'$, their matrix 
    elements are $\langle m^S_{\ell}\sigma | m^S_{\ell+1}\rangle$; 
    taken as column vectors, they form a $N^S \times M^S$ rectangular 
    matrix $T$. Proceed likewise for the environment.
    \item Carry out the reduced basis transformation 
    $\ham_{\ell+1}^{\text{tr}} 
    = T^{\dagger}\ham_{\ell+1} T$ onto the new $M^S$-state basis and take
    $\ham_{\ell+1}^{\text{tr}} \rightarrow \ham_{\ell+1}$ for the 
    system. Do the same for the environment 
    and restart with step (2) with block size $\ell+1$ 
    until some desired final length is 
    reached. Operator representations also have to be updated (see 
    Sec.\ \ref{subsec:correlations}).
    \item Calculate desired ground state properties (energies and 
    correlators) from $\ket{\psi}$; this step can also be carried out 
    at each intermediate length.
\end{enumerate}    
If the Hamiltonian is reflection-symmetric, one may consider system and 
environment to be identical.  One is not restricted to 
choosing the ground state for $\ket{\psi}$; any state accessible by 
large sparse-matrix diagonalization of the superblock 
is allowed. Currently available 
algorithms for this diagonalization limit us however to the 
lowest-lying excitations (see Sec.\ \ref{subsec:large}). 

\subsection{Infinite-system and finite-system DMRG}
\label{subsec:finite}
For many problems, infinite-system DMRG does not yield satisfactory 
answers: 
The idea of simulating the final system size cannot be 
implemented well by a small environment block in the early DMRG steps.
DMRG is usually canonical, working at fixed particle numbers for a
given system size (Sec.\ \ref{subsec:quantumnumbers}).
Electronic 
systems where the particle number is growing during system growth
to maintain particle density approximately constant are affected by a lack 
of ``thermalization'' of the particles 
injected during system growth; $t$-$J$ models with a relatively 
small hole density or Hubbard models far from half-filling or with 
complicated filling factors are particularly affected.
The strong physical effects of impurities or randomness in the 
Hamiltonian cannot be accounted for properly by infinite-system DMRG as 
the total Hamiltonian is not yet known at intermediate steps.
In systems with
strong magnetic fields or close to a first order transition one may be 
trapped in a metastable state favored for small system sizes e.g.\ by 
edge effects. 

\begin{figure}
\centering\epsfig{file=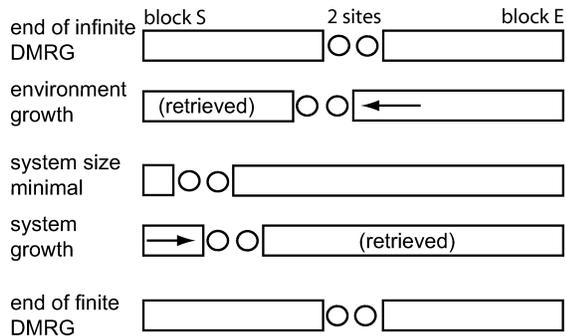,width=0.85\linewidth}
\caption{Finite-system DMRG algorithm. Block growth and shrinkage.}
\label{fig:finitesize}
\end{figure}

{\em Finite-system DMRG} manages to eliminate these concerns to a very large degree
and to reduce the error (almost) to the truncation error. 
The idea of the finite-system algorithm is to stop the infinite-system
algorithm at some preselected superblock length $L$ which is kept
fixed. In subsequent DMRG steps (Fig.\ \ref{fig:finitesize}), one applies the steps of infinite-system 
DMRG, but instead of simultaneous growth of both blocks, growth of one 
block is accompanied by shrinkage of the other block. Reduced basis 
transformations are carried out only for the growing block. Let the 
environment block grow at the expense of the system block; to describe it, 
system blocks 
of all sizes and operators acting on this block, expressed in the 
basis of that block, must have been stored previously (infinite-system 
stage or previous applications of finite-system DMRG). 
When the system block reaches some minimum size 
and becomes exact, growth direction is reversed. The system block now 
grows at the expense of the environment. All basis states are chosen 
while system and environment are embedded in the final system and in the 
knowledge of the full Hamiltonian. If the system is symmetric under 
reflection, blocks 
can be mirrored at equal size, otherwise the environment block is 
shrunk to some minimum and then regrown. A complete shrinkage and 
growth sequence for both blocks is called a {\em sweep}.

One sweep takes about two (if reflection symmetry holds) or four 
times the CPU time of the starting 
infinite-system DMRG calculation. For better performance, there is a  
simple, but powerful ``prediction algorithm'' (see Sec.\ \ref{subsec:large}), 
which cuts 
down calculation times in finite-system DMRG by more than an 
order of magnitude. In fact, it will be seen that the speed-up is the 
larger the closer the current (ground) state is to the final,
fully converged result. In practice, one therefore starts running the
infinite-system DMRG with a rather small number of states $M_{0}$, 
increasing it while running through the sweeps to some final 
$M_{\text{final}} \gg M_{0}$. The resulting slowing down
of DMRG will be offset by the increasing performance of the 
prediction algorithm. While there is no guarantee that finite-system 
DMRG is not trapped in some metastable state, it usually finds the 
best approximation to the ground state and convergence is gauged by 
comparing results from sweep to sweep until they stablize. This may 
take from a few to several dozen sweeps, with electronic problems at 
incommensurate fillings and random potential problems needing most.
In rare cases it has been observed that seemingly converged finite-system
results are again suddenly improved after some further sweeps 
without visible effect have been carried out, showing a metastable 
trapping. It is therefore advisable to carry out 
additional sweeps and to judge DMRG convergence by carrying out runs for 
various $M$. Where possible, choosing a clever sequence of finite-system DMRG steps 
may greatly improve the wave function. This has been successfully attempted in 
momentum space DMRG (Sec.\ \ref{subsec:momentum}) and quantum chemistry DMRG (Sec.\ \ref{subsec:chemistry}).

\begin{figure}
\centering\epsfig{file=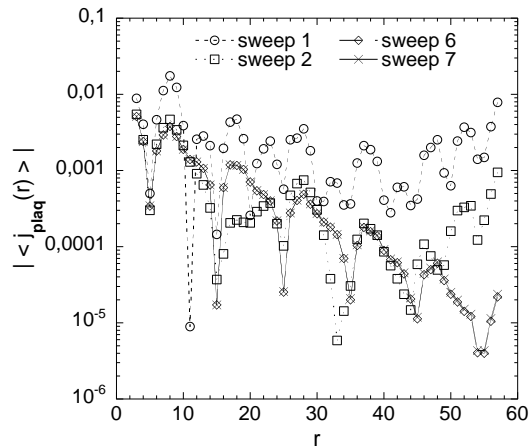,scale=0.42}
\caption{Currents on rung $r$ of a $t$-$J$ ladder induced by a source term
on the left edge, after various sweeps of finite-system DMRG. From \citet{Scho02}.}
\label{fig:currentsweeps}
\end{figure}

To show the power of finite-system DMRG, consider the calculation of the plaquette currents 
induced on a $t$-$J$ ladder by imposing a source current on one edge 
of the ladder \cite{Scho02}. Fig.\ \ref{fig:currentsweeps} shows how 
the plaquette currents along the 
ladder evolve from sweep to sweep. While they are perfectly converged 
after 6 to 8 sweeps, the final result, a modulated exponential decay, is 
far from what DMRG suggests after the first sweep, let alone in the 
infinite-system algorithm (not shown).

Despite the general reliability of the finite-system algorithm there 
might be (relatively rare) situations where its results may be 
misleading.  
It may for example be doubted whether competing or coexisting types of long-range 
order are well described by DMRG, as we will see that it produces a 
very specific kind of wave functions, so-called matrix-product 
states (Sec.\ \ref{subsec:mpa}). These states show either long-range 
order or, more typically, short-ranged correlations. In the case of 
competing forms of long-range order, the infinite-system algorithm 
might preselect one of them incorrectly e.g.\ due to edge effects, 
and the finite-system algorithm would then be quite likely to fail to 
``tunnel'' to the other, correct kind of long-range order due to the 
local nature of the improvements to the wave function. 
Perhaps such a problem is at the origin of 
the current disagreement between 
state-of-the-art DMRG \cite{Jeck02a,Jeck03a,Zhan04} and QMC 
\cite{Sand03,Sand04} on the extent of a 
bond-ordered wave phase between a spin-density-wave phase and a 
charge-density-wave phase in the half-filled extended Hubbard model; 
but no consensus has emerged yet. 

\subsection{Symmetries and good quantum numbers}
\label{subsec:quantumnumbers}
A big advantage of DMRG is that the algorithm conserves a variety of, 
but not all, symmetries and good quantum numbers of the Hamiltonians. 
They may be exploited to reduce storage and 
computation time and to thin out Hilbert space by decomposing it into 
a sum of sectors. DMRG is optimal for studying the lowest-lying states 
of such sectors, and refining any such decomposition immediately 
gives access to more low-lying states. Hence the use of symmetries is 
standard practise in DMRG implementations. Symmetries used in DMRG fall 
into three categories, continuous abelian, continuous nonabelian, and discrete.

{\em Continuous abelian symmetries.} The most frequently implemented 
symmetries in DMRG are the $U(1)$ symmetries leading 
to total magnetization $S^z_{\text{tot}}$ and total particle number 
$N_{\text{tot}}$ as good (conserved) quantum numbers. If present for 
some Hamiltonian, all operators can be expressed in matrix form as 
dense blocks of non-zero matrix elements with all other matrix 
elements zero. These blocks can be labelled by good quantum numbers. 
In DMRG, reduced basis transformations preserve these 
block structures if one fixes total magnetization and/or 
total particle number for the superblock. Assume that block and  
site states can at a given DMRG step be labeled by a good quantum number, 
say, particle number $N$. This is an essential prerequisite (cf.\ 
translational invariance leading to momentum conservation; see below). As the 
total number of particles is fixed, we have $\ntot = N_{m^S} + 
N_{\sigma^S} + N_{\sigma^E} + N_{m^E}$. Equation (\ref{eq:dmaselements})
implies that only matrix elements
$\bra{m^S\sigma^S} \dm \ket{\tilde{m}^{S}\tilde{\sigma}^{S}}$ with $ N_{m^S} + 
N_{\sigma^S} =  N_{\tilde{m}^{S}} + 
N_{\tilde{\sigma}^{S}}$ can be non-zero. The density matrix thus has block 
structure, and its eigenvectors from which the next block's eigenbasis 
is formed can again be labeled by particle number $N_{m^S} + 
N_{\sigma^S}$. Thus, for all 
operators only dense blocks of non-zero matrix elements have to 
be stored and considered. For superblock wave functions, a  
tensorial structure emerges as total particle number and/or 
magnetization dictate that only product states with compatible 
quantum numbers (i.e.\ adding up to the fixed total) may have non-zero 
coefficients.

The performance gains from implementing just the simple additive 
quantum numbers magnetization and particle number are impressive. 
Both in memory and CPU time typically significantly more than an order of 
magnitude will be 
gained.  

{\em Continuous nonabelian symmetries.} Nonabelian symmetries that have been 
considered are the quantum group symmetry $SU_{q}(2)$ \cite{Sier97b}, $SU(2)$ spin symmetry
\cite{Tats00,Xian01,McCu00,McCu01b,McCu02}, and the charge $SU(2)$ 
pseudospin symmetry \cite{McCu02}, which holds for the bipartite Hubbard 
model without field 
\cite{Yang90}: its generators are given by
$I^+ = \sum_{i} (-1)^i \eucr{i}\edcr{i}$,
$I^- = \sum_{i} (-1)^i \edan{i}\euan{i}$ and 
$I^z = \sum_{i} \frac{1}{2} ( n_{\uparrow i} + n_{\downarrow i}-1)$.

Implementation of nonabelian symmetries is much more 
complicated than that of abelian symmetries, the most performant one 
\cite{McCu02} building on Clebsch-Gordan transformations and 
elimination of quantum numbers via the Wigner-Eckart theorem. 
It might be crucial for obtaining high-quality results in 
applications with problematically large truncated weight such as in two 
dimensions, where the truncated weight is cut by several 
orders of magnitude compared to implementations using abelian 
symmetries only; the additional increase in performance is comparable to that due 
to use of $U(1)$ symmetries compared to using no symmetries at all.

{\em Discrete symmetries.} I shall formulate them 
for a {\em fermionic} Hamiltonian of spin-$\frac{1}{2}$ particles. 

{\em Spin-flip symmetry:} if the Hamiltonian $\ham$ is invariant under a general 
spin flip $\uparrow \leftrightarrow \downarrow$, one may introduce 
the spin flip operator $\hat{P} = \prod_{i} \hat{P}_{i}$, which is 
implemented locally on an electronic site as 
$\hat{P}_{i}\ket{0}=\ket{0}$,
$\hat{P}_{i}\ket{\!\uparrow}=\ket{\!\downarrow}$, 
$\hat{P}_{i}\ket{\!\downarrow}=\ket{\!\uparrow}$,
$\hat{P}_{i}\ket{\!\uparrow\downarrow}=-\ket{\!\uparrow\downarrow}$ 
(fermionic sign). As $[\ham,\hat{P}]=0$, there are common 
eigenstates of $\ham$ and $\hat{P}$ with
$\hat{P}\ket{\psi} = \pm \ket{\psi}$.

{\em Particle-hole symmetry:} if $\ham$ is invariant under 
a particle-hole transformation, $[\ham,\hat{J}]=0$ 
for the particle-hole operator $\hat{J} = \prod_{i} \hat{J}_{i}$, where the local operation is given 
by $\hat{J}_{i}\ket{0}=\ket{\!\uparrow\downarrow}$,
$\hat{J}_{i}\ket{\!\uparrow}=(-1)^i \ket{\!\downarrow}$, 
$\hat{J}_{i}\ket{\!\downarrow}= (-1)^i \ket{\!\uparrow}$,
$\hat{J}_{i}\ket{\!\uparrow\downarrow}=\ket{0}$, introducing a 
distinction between odd and even site sublattices, and eigenvalues 
$\hat{J}\ket{\psi}=\pm\ket{\psi}$.

{\em Reflection symmetry (parity):} in the case of reflection symmetric 
Hamiltonians with open boundary conditions, parity is a good 
quantum number.
The spatial reflection symmetry operator $\hat{C}_{2}$ acts globally. 
Its action on a DMRG product state is given by
\begin{equation}
    \hat{C}_{2} \ket{m^{S}\sigma^{S}\sigma^{E}m^{E}} =
    (-1)^{\eta} \ket{m^{E}\sigma^{E}\sigma^{S}m^{S}}
\end{equation}
with a fermionic phase determined by $\eta = 
(N_{m^{S}}+N_{\sigma^{S}})(N_{m^{E}}+N_{\sigma^{E}})$.
Again, eigenvalues are given by $\hat{C}_{2}\ket{\psi}=\pm\ket{\psi}$.
Parity is not a good quantum number accessible to finite-system DMRG, 
except for the DMRG step with identical system and environment.

All three symmetries commute, and an arbitrary normalized wave function 
can be decomposed into eigenstates for any desired combination of 
eigenvalues $\pm 1$ by successively calculating
\begin{equation}
    \ket{\psi_{\pm}} = \frac{1}{2} \left( \ket{\psi} \pm \hat{O} 
    \ket{\psi} \right) ,
    \label{eq:discreteproject}
\end{equation}    
where $\hat{O} = \hat{P}, \hat{J}, \hat{C}_{2}$. 

Parity may be easily implemented by starting the superblock 
diagonalization from a trial state (Sec.\ \ref{subsec:large}) that has been made (anti)symmetric 
under reflection by Eq.\ (\ref{eq:discreteproject}). As 
$[\ham,\hat{C}_{2}]=0$, the final eigenstate will have the same 
reflection symmetry. Spin-flip and particle-hole symmetries may be 
implemented in the same fashion, provided $\hat{P}$ and $\hat{J}$ are 
generated by DMRG; as they are products of local operators, this can 
be done along the lines of Sec.\ \ref{subsec:correlations}. Another way of 
implementing these two local symmetries is to realize that the argument given for magnetization and particle number 
as good density-matrix quantum numbers carries over to the spin-flip 
and particle-hole eigenvalues such that they can also be implemented 
as block labels. 

{\em Missing symmetries.} Momentum is not a good quantum number in 
real-space DMRG, even if periodic boundary conditions (Sec.\ 
\ref{subsec:boundary}) are used and translational invariance holds. 
This is because the allowed discrete momenta change during the growth 
process and, more importantly, because 
momentum does not exist as a good quantum number at the
block level. Other DMRG variants with momentum as a good quantum 
number will be considered in Sec.\ \ref{subsec:momentum}.  

\subsection{Energies: ground states and excitations}
\label{subsec:energies}
As a method working in a subspace 
of the full Hilbert space, DMRG is 
variational in energy. It provides upper bounds for energies $E(M)$ that 
improve monotonically with $M$, the number of basis states in the reduced 
Hilbert space.
Two sources of errors have been identified, the 
environmental error due to inadequate environment blocks, which can be 
amended using the finite-system DMRG algorithm, and the truncation error.
Assuming that the environmental error (which is hard to quantify 
theoretically) has been eliminated, i.e.\ finite-system DMRG has reached 
convergence after sufficient sweeping, 
the truncation error remains to be analyzed.
Rerunning the calculation of a system of size $L$ for
various $M$, one observes for sufficiently large values of $M$
that to a good approximation 
the error in energy per site scales linearly with the truncated 
weight, 
\begin{equation}
(E(M)-E_{\text{exact}})/L \propto \epsilon_{\rho} \label{eq:extrapolform},
\end{equation}
with a non-universal proportionality factor typically of order 1 to 10, sometimes 
more (this observation goes back to \citet{Whit93a}; 
\citet{Lege96} give a careful analysis). As 
$\epsilon_{\rho}$ is often of order $10^{-10}$ or less, DMRG energies can 
thus be extrapolated using Eq.\ (\ref{eq:extrapolform}) 
quite reliably to the exact $M=\infty$ result, often almost at machine precision. 
The precision desired imposes the size of $M$, which for spin 
problems is typically in the lower hundreds, for electronic problems 
in the upper hundreds, for two-dimensional and momentum-space problems 
in the lower thousands. As an example for DMRG precision, consider the
results obtained for the ground-state energy per site of the $S=1$ 
antiferromagnetic Heisenberg chain in Table \ref{tab:s1energy}.

\begin{table}
    \caption{Ground state energies per site $E_{0}$ of the $S=1$ 
    isotropic antiferromagnetic Heisenberg chain for $M$ block states 
    kept and associated truncated weight $\epsilon_{\rho}(M)$. Adapted 
    from \citet{Whit93a}.}
    \vskip 0.2truecm
    \begin{tabular}{llr}
        \hline\hline
        $M$ & $E_{0}(M)$ & $\epsilon_{\rho}(M)$ \\
        \hline
        36  & -1.40148379810  & $5.61 \times 10^{-8}$ \\
        72  & -1.40148403632  & $3.42 \times 10^{-10}$ \\
        110 & -1.40148403887  & $1.27 \times 10^{-11}$ \\
        180\quad\quad & -1.401484038970\quad\quad  & $\quad 1.4  \times 10^{-13}$ \\
        \hline\hline
    \end{tabular}       
\label{tab:s1energy}
\end{table}

Experiments relate to energy 
differences or relative ordering of levels. This raises the question 
of calculating excitations in DMRG. Excitations are easiest to 
calculate if they are the ground state in some other symmetry sector 
of the Hamiltonian and are thus algorithmically not different from 
true ground states. If, however, we are interested in some 
higher-lying state in a particular Hilbert space sector, DMRG restricts
us to the lowest-lying such states because of the restrictions of 
large sparse-matrix diagonalizations (Sec.\ \ref{subsec:large}). Excited states have to be 
``targeted'' in the same way as the ground state.  This means that they
have to be calculated for the superblock at each iteration and to be 
represented optimally, i.e.\ reduced basis states have to be chosen such that 
the error in the approximation is minimized. It can be shown quite easily that 
this amounts to considering the eigenstates of the reduced density-matrix
\begin{equation}
    \dm_{S} = \text{Tr}_{E} \sum_{i} \alpha_{i} \ket{\psi_{i}} 
    \bra{\psi_{i}} ,
    \label{eq:multipletarget}
\end{equation}
where the sum runs over all targeted states $\ket{\psi_{i}}$ (ground 
and a few excited states) and $\sum_{i} \alpha_{i}=1$. There is no 
known optimal choice for the 
$\alpha_{i}$, but it seems empirically to be most reasonable to 
weigh states roughly equally. To maintain a good overall description 
for all targeted states at a fixed $M$, typically less than 5 or so excited 
states are targeted. Best results are of course obtained by running 
DMRG for each energy level separately.
\subsection{Operators and correlations}
\label{subsec:correlations}
In general, we will also be interested in evaluating static $n$-point 
correlators $\hat{O}^{(n)}_{i_{1}\ldots i_{n}}= \hat{O}^{(1)}_{i_{1}}\ldots\hat{O}^{(1)}_{i_{n}}$ 
with respect to some 
eigenstate of the Hamiltonian. The most relevant cases are $n=1$ for density or local 
magnetization, and 
$n=2$ for two-point density-density, spin-spin or creation-annihilation
correlators, $\langle n_{i}n_{j}\rangle$, 
$\langle\spl{i}\smi{j}\rangle$ or
$\langle\fcr{i}\fan{j}\rangle$. 

Let us first consider the case $n=1$. The iterative growth strategy 
of DMRG imposes a natural three-step procedure of initializing, updating
and evaluating correlators. 

1.\ \emph{Initialization}: $\hat{O}_{i}$ acts on site $i$. When site $i$ is 
added to a block of length $\ell-1$, $\bra{\sigma} \hat{O}_{i} 
\ket{\tilde{\sigma}}$ is evaluated. With $\{ \ket{m_{\ell}}\}$ being 
the 
reduced basis of the new block incorporating site $i$ and
$\{ \ket{m_{\ell-1}} \}$ that of the old block, one has 
\be
\bra{m_{\ell}} \hat{O}_{i} \ket{\tilde{m}_{\ell}} = 
\sum_{m_{\ell-1}\sigma\tilde{\sigma}} 
\langle m_{\ell} | m_{\ell-1}\sigma \rangle \langle \sigma | \hat{O}_{i} | 
\tilde{\sigma}
\rangle \langle m_{\ell-1}\tilde{\sigma} | \tilde{m}_{\ell} \rangle .
\end{equation}
$\langle m_{\ell} | m_{\ell-1}\sigma \rangle $ is already known from the 
density-matrix eigenstates.

2.\ \emph{Update}: At each further DMRG step, an approximate
basis transformation for the block containing the site where 
$\hat{O}_{i}$ acts from $\{ \ket{m_{\ell}}\} $ to $\{ 
\ket{m_{\ell+1}}\}$ occurs. As
$\hat{O}_{i}$ does not act on the new site,
the operator transforms as
\begin{eqnarray}
& & \bra{m_{\ell+1}} \hat{O} \ket{\tilde{m}_{\ell+1}} = \nonumber \\
& & \sum_{m_{\ell}\tilde{m}_{\ell}\sigma} 
\langle m_{\ell+1} | m_{\ell}\sigma \rangle 
\langle m_{\ell} | \hat{O} | \tilde{m}_{\ell}\rangle 
\langle \tilde{m}_{\ell}\sigma | \tilde{m}_{\ell+1} \rangle .
\end{eqnarray}
This expression is evaluated efficiently by
splitting it into two $O(M^3)$ matrix-matrix multiplications.

3.\ \emph{Evaluation:} After the last DMRG step 
$\langle  m^{S} \sigma^S \sigma^E m^{E} | \psi \rangle$ 
is known and $\langle \hat{O}_{i} \rangle$ reads, assuming $\hat{O}_{i}$ to 
act on some site in the system block,
\begin{eqnarray}
& & \bra{\psi} \hat{O}_{i} \ket{\psi}  =  \sum_{m^{S}\tilde{m}^{S} 
\sigma^S \sigma^E m^{E}} \nonumber \\
& & \langle \psi| m^S \sigma^S \sigma^E m^E \rangle 
\bra{m^S} \hat{O}_{i} \ket{\tilde{m}^{S}} \times \nonumber \\
& & \langle \tilde{m}^{S} \sigma^S \sigma^E m^E |
\psi \rangle .
\end{eqnarray}

In the case of 2-point correlators two cases have to be distinguished, 
whether the locations $i$ and $j$ of the contributing 1-point operators
act on different blocks or on the same block at the last step. 
This expression is again evaluated efficiently by
splitting it into two $O(M^3)$ matrix-matrix multiplications.

If they act on different blocks, one follows through the procedure for 
1-point operators, yielding $\bra{m^S} \hat{O}_{i} 
\ket{\tilde{m}^{S}}$ and $\bra{m^E} \hat{O}_{j} \ket{\tilde{m}^{E}}$.
The evaluation is done by the following modification of the 1-point 
case,
\begin{eqnarray}
& & \bra{\psi} \hat{O}_{i}\hat{O}_{j} \ket{\psi}  =  
\sum_{m^{S}\tilde{m}^{S} 
\sigma^S \sigma^E m^{E}\tilde{m}^{E}} \nonumber \\
& & \langle \psi| m^S \sigma^S \sigma^E m^E \rangle 
\bra{m^S} \hat{O}_{i} \ket{\tilde{m}^{S}} \times \nonumber \\
& & \bra{m^E} \hat{O}_{j} \ket{\tilde{m}^{E}}
\langle \tilde{m}^{S} \sigma^S \sigma^E \tilde{m}^{E} |
\psi \rangle .
\end{eqnarray}

If they act on the same block, it is {\em wrong} to obtain 
$\bra{m} \hat{O}_{i}\hat{O}_{j} \ket{\tilde{m}}$ through
\begin{equation}
    \bra{m} \hat{O}_{i}\hat{O}_{j} \ket{\tilde{m}} =
    \sum_{m'} \bra{m} \hat{O}_{i} 
    \ket{m'}
    \bra{m'} \hat{O}_{j} \ket{\tilde{m}} \quad\text{(false)},
\end{equation}    
where an approximate partition of unity has been inserted. However, it 
is only one to a very good approximation when it is used to project 
the targeted wave function, for which it was constructed, but not in 
general.

Instead, such operators have to be built as a compound object at the moment 
when they live in a product Hilbert space, namely when one of the 
operators 
acts on a block (of length $\ell-1$), the other on a single 
site, that is being attached to the block. 
Then we know $\langle m_{\ell-1} | \hat{O}_{i} | \tilde{m}_{\ell-1}\rangle$ 
and $\bra{\sigma} \hat{O}_{j} \ket{\tilde{\sigma}}$ and within the reduced 
bases of the block of length $\ell$ 
\begin{eqnarray}
 & & \bra{m_\ell} \hat{O}_{i}\hat{O}_{j} \ket{\tilde{m}_{\ell}} =
    \sum_{m_{\ell-1}\tilde{m}_{\ell-1}\sigma\tilde{\sigma}} \nonumber \\
 & & \langle m_{\ell} | m_{\ell-1}\sigma \rangle 
    \bra{m_{\ell-1}} \hat{O}_{i} 
    \ket{\tilde{m}_{\ell-1}} \times \\
 & & \bra{\sigma} \hat{O}_{j} \ket{\tilde{\sigma}}
    \langle \tilde{m}_{\ell-1}\tilde{\sigma} | \tilde{m}_{\ell} \rangle \nonumber 
\end{eqnarray}    
is exact. Updating and final evaluation for ``compound'' operators proceed 
as for a one-point operator.

One-point operators show similar convergence behavior in $M$ as local 
energy, but at reduced precision.

While there is no exact variational principle for two-point correlations, 
derived correlation lengths are monotonically increasing in $M$, but 
always underestimated. 
The underestimation can actually be quite severe and be
of the order of several percent while the ground state energy has
already converged almost to machine precision. 

As DMRG always generates wave functions with exponentially 
decaying correlations (Sec.\ \ref{subsec:mpa}), power-law decays of correlations 
are problematic. \citet{Ande99} show that for free fermions 
the resulting correlation function mimics the correct power-law on 
short length scales increasing with $M$, but is purely exponential on larger 
scales. However, the derived correlation length diverges roughly as 
$M^{1.3}$, such that for $M\rightarrow\infty$ criticality is 
recovered.  

\subsection{Boundary Conditions}
\label{subsec:boundary}
From a physical point of view periodic boundary conditions are 
normally highly preferable to the open boundary conditions used so far
for studying bulk properties, as surface 
effects are eliminated and finite-size extrapolation works 
for much smaller system sizes. In particular, open boundaries 
introduce charge or magnetization oscillations not always easily 
distinguishable from true charge density waves or dimerization (see 
\citet{Whit02} for a thorough discussion on using bosonization to make 
the distinction). 

However, it has been observed early in the history of DMRG that ground state 
energies for a given $M$ are much less precise in the case of periodic 
boundary conditions than for open boundary conditions with differences 
in the relative errors of up to several orders of magnitude.
This is reflected in the spectrum of the reduced density-matrix, that 
decays much more slowly (see Sec.\ \ref{subsec:density}). However, it 
has been shown by \citet{Vers04} that this is an artefact of the
conventional DMRG setup and that, at some algorithmic cost, essentially the 
same precision for a given $M$ can be achieved for periodic as for 
open boundary conditions (Sec.\ \ref{subsec:mpa}).

\begin{figure}
\centering\epsfig{file=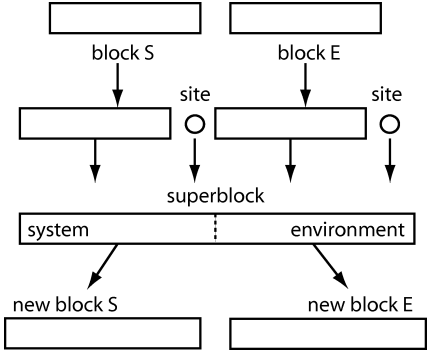,width=0.65\linewidth}
\caption{Typical system and environment growth for periodic boundary 
conditions in the infinite-system algorithm.}
\label{fig:pbcgrowth}
\end{figure}

To implement periodic boundary conditions in the infinite-system 
DMRG algorithm the block-site structure is typically changed as shown in Fig.\ 
\ref{fig:pbcgrowth}; other setups are however also feasible and used.
For finite-system DMRG, the environment block grows at the 
expense of the system block, then the system block grows back, till 
the configuration of the end of the infinite-system algorithm is 
reached. This is repeated with changed roles (unless translational 
invariance allows to identify system and environment at equal size).
A minor technical complication arises from the fact that blocks grow at both 
ends at various steps of the algorithm. 

Beyond the usual advantages of periodic boundary conditions, 
combining results for periodic and antiperiodic boundary conditions 
allows the calculation of responses to boundary conditions such as 
spin stiffness, phase sensitivity \cite{Schm96} or superfluid density \cite{Raps99}.
Periodic and antiperiodic boundary conditions $\fcr{L+1}=\pm \fcr{1}$ 
are special cases of the general complex boundary condition
$\fcr{L+1}=e^{i\phi} \fcr{1}$. Implementing the 
latter is a tedious, but straightforward generalization of real-valued DMRG;
memory doubles, computation time quadruples. Numerical stability is 
assured because the density matrix remains hermitian. This 
generalization has been used on a ring with 
interactions and impurities to determine the current $I(\phi)$ which is neither 
sawtooth-like nor sinusoidal \cite{Mede03a}, to obtain the conductance 
of interacting nanowires \cite{Mede03b}. For open boundary conditions,
complex-valued DMRG has been used to introduce 
infinitesimal current source terms for time-reversal symmetry breaking in 
electronic ladder structures \cite{Scho02}.

\subsection{Large sparse matrix diagonalization}
\label{subsec:large}
{\em Algorithms.} Key to DMRG performance is the efficient diagonalization of 
the large sparse superblock Hamiltonian. All large sparse matrix diagonalization 
algorithms iteratively calculate the desired eigenstate 
from some (random) starting state through successive costly matrix-vector multiplications.      
In DMRG, the two algorithms typically used are the Lanczos method \cite{Cull85,Golu97} and the 
Jacobi-Davidson method \cite{Slei96}. 
The pleasant feature of these algorithms is that for a $N\times N$ 
dimensional matrix it takes only a much smaller number $\tilde{N} \ll 
N$ of iterations such that iterative approximations to eigenvalues 
converge very rapidly to the maximum and minimum 
eigenvalues of $\ham$ at machine precision. With slightly more effort other 
eigenvalues at the edge of the spectrum can also be computed. 
Typical values for the number 
of iterations (matrix-vector multiplications) in DMRG calculations are of the 
order of 100.

{\em Representation of the Hamiltonian.} Naively, the superblock 
Hamiltonian is a $M^2 N_{\text{site}}^2$ dimensional matrix. As matrix-vector 
multiplications scale as (dimension)$^2$, DMRG would seem to be an 
algorithm of order $O(M^4)$. In reality, 
it is only $O(M^3)$, as typical tight-binding Hamiltonians act as sums 
over two operator terms: Assuming nearest-neighbor interactions, the superblock 
Hamiltonian decomposes as 
\begin{equation}
    \ham = \ham_{S} + \ham_{S\bullet}+ \ham_{\bullet\bullet}+ 
    \ham_{\bullet E}+ \ham_{E} .
    \label{eq:hamdecompos}
\end{equation}    
$\ham_{S}$ and $\ham_{E}$ contain all interactions within the system 
and environment blocks respectively, are hence of dimension $M$. 
Multiplying them to some state $\ket{\psi}$ is of order $M^3 
N_{\text{site}}^2$. $\ham_{S\bullet}$ and $\ham_{\bullet E}$ contain 
interactions between blocks and the neighboring sites, hence are of 
dimension $MN_{\text{site}}$. Consider a typical interaction 
$S^+_{\ell}S^-_{\ell+1}$, where $\ell$ is the last site of the block 
and $\ell+1$ a single site. Then
\begin{equation}
    \bra{ m^{S} \sigma^S} S^+_{\ell}S^-_{\ell+1} \ket{\tilde{m}^{S} 
    \tilde{\sigma}^{S}} =
    \bra{ m^{S}} S^+_{\ell} \ket{\tilde{m}^{S}} 
    \bra{ \sigma^S} S^-_{\ell+1} \ket{\tilde{\sigma}^{S}}
    \label{eq:hamfactor}
\end{equation}
and multiplying this term to $\ket{\psi}$ is best carried out in a 
two step sequence: the expression
\begin{equation}
    \langle m\sigma\tau n |\phi\rangle =
    \sum_{m'\sigma'}     \bra{ m \sigma} S^+S^-\ket{ m' \sigma'}   
    \langle m'\sigma'\tau n |\psi\rangle ,
\end{equation}
that is of order $O(M^3 N_{\text{site}}^3)$ for the determination of all state 
coefficients is decomposed as
\begin{equation}
    \langle m'\sigma\tau n |\nu\rangle =
    \sum_{\sigma'}     \bra{\sigma} S^-\ket{\sigma'}   
    \langle m'\sigma'\tau n |\psi\rangle ,
\end{equation}
of order $O(M^2 N_{\text{site}}^3)$ and
\begin{equation}
    \langle m\sigma\tau n |\phi\rangle =
    \sum_{m'}     \bra{ m } S^+\ket{ m'}   
    \langle m'\sigma\tau n |\nu\rangle ,
\end{equation}
of order $O(M^3 N_{\text{site}}^2)$, where an order of $N_{\text{site}}$ is saved, important for 
large $N_{\text{site}}$. The Hamiltonian is never explicitly constructed.
Such a decomposition is crucial when block-block 
interactions $\ham_{SE}$ appear for longer-ranged interactions. 
Considering again $S^+S^-$, with $S^-$ now acting on the environment 
block, factorization of the Hamiltonian again allows to decompose the
original term
\begin{equation}
    \langle m\sigma\tau n |\phi\rangle =
    \sum_{m'n'}     \bra{ m n} S^+S^-\ket{ m' n'}   
    \langle m'\sigma\tau n' |\psi\rangle ,
\end{equation}
that is of inconveniently high order $O(M^4 N_{\text{site}}^2)$ for the determination of all state 
coefficients, as
\begin{equation}
    \langle m'\sigma\tau n |\nu\rangle =
    \sum_{n'}     \bra{n} S^-\ket{n'}   
    \langle m'\sigma\tau n' |\psi\rangle ,
\end{equation}
of order $O(M^3 N_{\text{site}}^2)$ and
\begin{equation}
    \langle m\sigma\tau n |\phi\rangle =
    \sum_{m'}     \bra{ m } S^+\ket{ m'}   
    \langle m'\sigma\tau n |\nu\rangle ,
\end{equation}
of order $O(M^3 N_{\text{site}}^2)$, such that an order of $M$ is saved and 
the algorithm is generally of order $O(M^3)$. This is essential for 
its performance.

{\em Eigenstate prediction.} 
Providing a good guess for the final eigenstate as starting state 
of the iterative diagonalization 
allows for arbitrary cutdowns in the number of iterations in 
the iterative diagonalization procedures. 

For both the infinite-system and the finite-system DMRG it is possible to 
provide starting states which often have overlap $\approx 1$ with the 
final state, leading to a dramatic reduction of iterations often down to 
less than 10, speeding up the algorithm by about an order of magnitude.

In infinite-system DMRG, the physical system changes from step to 
step. It seems intuitive that for a very long 
system, the composition of the ground state from block and site states
may only weakly depend on its length, such that the ground state 
coefficients remain almost the same under system growth. One might therefore 
simply use the old ground state as prediction state. This fails; while
the absolute values of coefficients hardly change from step to step 
in long systems, the block basis states are fixed by the density-matrix 
diagonalization 
only up to the sign, such that the signs of the coefficients are effectively 
random. Various ways of fixing these random signs have been proposed 
\cite{Scho98b,Qin01,Sun02a}.  

In the case of the finite-system DMRG, the physical system does not 
change from DMRG step to DMRG step, just the structure of the 
effective Hilbert space changes. \citet{Whit96b} has given a prescription how to predict 
the ground state expressed in the block-site structure of the next DMRG 
step. If basis transformations were not
incomplete, one could simply transform the ground state from one basis 
to the next to obtain a prediction state. The idea is to do this even though the 
transformation is
incomplete. The state obtained turns out to be an often excellent 
approximation to the true ground state.

Let us assume that we have a system with open boundary conditions, with 
a current system block of length $\ell$ and an environment block of length 
$L-\ell-2$. The target state is known through $\langle m_{\ell} 
\sigma_{\ell+1}\sigma_{\ell+2}m_{L-\ell-2}|\psi\rangle$. Now assume 
that the system block is growing at the expense of the environment block, 
and we wish to predict the coefficients $ \langle m_{\ell+1} 
\sigma_{\ell+2}\sigma_{\ell+3}m_{L-\ell-3}|\psi\rangle$, where 
in both DMRG steps $\ket{\psi}$ is describing 
the same quantum mechanical state. As we also know $\langle 
m_{\ell}\sigma_{\ell+1}|m_{\ell+1}\rangle$ from the current DMRG 
iteration and $\langle 
m_{L-\ell-3}\sigma_{\ell+3}|m_{L-\ell-2}\rangle$ from some previous 
DMRG iteration, this 
allows to carry out two incomplete basis transformations:
First, we transform system block and site to the new system block, 
giving
\begin{eqnarray}
& & \langle m_{\ell+1}\sigma_{\ell+2}m_{L-\ell-2} |\psi\rangle = \\
& & \sum_{m_{\ell}\sigma_{\ell+1}} \langle m_{\ell+1} | 
m_{\ell}\sigma_{\ell+1} \rangle \langle m_{\ell} 
\sigma_{\ell+1}\sigma_{\ell+2}m_{L-\ell-2}|\psi\rangle ; \nonumber
\end{eqnarray}
second, we expand the current environment block into a product state 
representation of a block with one site less and a site:

\begin{eqnarray}
& & \langle m_{\ell+1}\sigma_{\ell+2}\sigma_{\ell+3}m_{L-\ell-3} 
|\psi\rangle = \\
& & \sum_{m_{L-\ell-2}} \langle 
m_{L-\ell-3}\sigma_{\ell+3}|m_{L-\ell-2}\rangle 
\langle m_{\ell+1}\sigma_{\ell+2}m_{L-\ell-2} |\psi\rangle . \nonumber
\end{eqnarray}

The calculation involves two $O(M^3)$ matrix multiplications, 
and works independently of any assumptions for the underlying model. It 
relies on the fact that by construction our incomplete basis transformations are 
almost exact when applied to the target state(s). 
\subsection{Applications}
\label{subsec:applications}
In this section, I want to give a very brief overview over 
applications of standard DMRG, which will not be covered in more 
detail in later sections. 

From the beginning, there has been a strong focus on one-dimensional 
Heisenberg models, in particular the $S=1$ case, where the Haldane 
gap $\Delta=0.41052 J$
was determined to 5 digit precision
by \citet{Whit93a}. Other authors 
have considered the equal time structure factor 
\cite{Sore94a,Sore94b,Siel00} and focused on topological order 
\cite{Lou02,Lou03,Qin03}.
There has been a particular emphasis on the study of the existence of free 
$S=\frac{1}{2}$ end spins in such chains \cite{Whit93a,Poli98,Bati98,Bati99,Jann00,Hall99},
where the open boundary conditions of DMRG are very useful; for 
the study of bulk properties authors typically attach real $S=\frac{1}{2}$ end spins
that bind to singlets with the effective ones, removing degeneracies 
due to boundary effects. Soon, studies carried over to the $S=2$ 
Heisenberg chain, where the first reliable determination of the gap 
$\Delta=0.085(5)$ and the correlation length $\xi\approx 50$ was 
provided by \citet{Scho95}, and confirmed and enhanced in other works 
\cite{Scho96a,Scho96b,Qin97a,Asch98,Wang99a,Tats00,Capo01}. The 
behavior of Haldane (integer spin) chains in (staggered) magnetic 
fields was studied by \citet{Sore93}, \citet{Lou99}, \citet{Erco00}, and
\citet{Capo01}. In general, DMRG has been very useful in the study of plateaux 
in magnetization processes of spin chains 
\cite{Yama00,Silv02,Lou00,Tand99,Hida03,Citr00,Lou00c,Kawa02}.

The isotropic half-integer spin Heisenberg chains are critical. The 
logarithmic corrections to the power-law of spin-spin correlations in 
the $S=\frac{1}{2}$ chain were 
first considered by \citet{Hall95a}, later by \citet{Hiki98}, 
\citet{Tsai00}, 
\citet{Shir01}, and \citet{Boos02}.
For the $S=\frac{3}{2}$ case, the central charge $c=1$ and again the 
logarithmic corrections to the spin-spin correlations were determined 
by \citet{Hall96}. Quasiperiodic $S=\frac{1}{2}$ chains were 
considered by 
\citet{Hida99a,Hida00a}, and the case of transverse fields by \citet{Hiei01}

Bilinear-biquadratic $S=1$ spin chains have been extensively studied 
\cite{Burs94,Scho96c,Sato98b} as well as the effect of Dzyaloshinskii-Moriya
interactions \cite{Zhao03}.

Important experimentally relevant generalizations of the Heisenberg 
model are obtained by adding frustrating interactions or dimerization,
the latter modelling either static lattice distortions or phonons in 
the adiabatic limit. \citet{Burs95,Kole96,Kole97,Kole02,Pati96,Pati97c,
Uhri99a,Uhri99b,Whit96a,Maes00,Itoi01} have extensively studied the ground state phase 
diagrams of such systems. 
DMRG has been instrumental in the discovery and description of gapped 
and gapless chiral phases in frustrated spin chains 
\cite{Kabu99,Hiki00b,Hiki01a,Hiki01b,Hiki01c,Hiki02}. Critical 
exponents for a supersymmetric spin chain were obtained by 
\citet{Sent99}.

As a first step towards two dimensions and due to the many 
experimental realizations, spin ladders have been among the first DMRG 
applications going beyond simple Heisenberg chains, starting with 
\citet{Whit94}. In the meantime, DMRG has emerged as standard tool for 
the study of spin ladders 
\cite{Whit96c,Lege97a,Trum01,Hiki01d,Zhu01,Wang00a,Wang02,Kawa03,Fath01}.
A focus of recent interest has been the effect of cyclic exchange 
interactions on spin ladders \cite{Saka99b,Laeu02,Hiki02a,Nunn02,Hond01}.

Among other spin systems, ferrimagnets have been studied by 
\citet{Pati97a,Pati97b}, \citet{Tone98}, \citet{Lang00}, and \citet{Hiki00a}.
One-dimensional toy models of the Kagome lattice 
have been investigated by \citet{Pati99a}, \citet{Whit00d}, and \citet{Wald00}, 
whereas
supersymmetric spin chains have been considered by \citet{Mars99}.

Spin-orbit chains with spin and pseudospin degrees of freedom are the 
large-$U$ limit of the two-band Hubbard model at quarter-filling and 
are hence an interesting and important generalization of the Heisenberg 
model. DMRG has allowed full clarification of the rich phase diagram  
\cite{Pati98,Ammo01,Yama98b,Pati00,Yama00a,Yama00b,Itoi00}.

DMRG in its finite-system version has also been very successful in the study of systems with
impurities or randomness, where the true ground state can be found 
with reasonable precision. Impurities have been studied by 
\citet{Wang96a},
\citet{Schm96},
\citet{Mike97}, \citet{Mart97}, \citet{Schm98}, \citet{Lauk98}, \citet{Ng00}, 
\citet{Lou98}, \citet{Zhan97c}, \citet{Zhan98b}, whereas 
other authors have focused on random interactions or fields
\cite{Hida96,Hida97a,Hida97b,Hiki99,Hida99b,Juoz99,Urba03}.
There has also been work on edges and impurities in
Luttinger liquids \cite{Qin96,Qin97b,Bedu98,Scho00}.

The study of electronic models is somewhat more complicated because 
of the larger number of degrees of freedom and because of the 
fermionic sign. However, DMRG is free of the negative sign problem of 
quantum Monte Carlo
and hence the method of choice for one-dimensional electronic models. 
Hubbard chains \cite{Saka96,Daul97,Zhan97a,Daul98,Daul00a,Aebi01,Maur00,
Nish00a,Daul00d,Alig00,Daul00f,Lepe97,Lepe00,Jeck02a} and Hubbard ladders 
\cite{Noac94,Noac95a,Noac95b,Noac97,Vojt99,Bonc00,Daul00c,Vojt01,Weih01,Hama02,Scal02,Mars02,Scho02}
have been studied in the entire range of interactions as the precision 
of DMRG is found to depend only very weakly on the interaction strength.
The three-band Hubbard model has been considered by \citet{Jeck98b} 
and \citet{Jeck02b}.
Similarly, authors have studied both the $t$-$J$ model on chains 
\cite{Chen96,Whit97b,Muto98c,Doub99b,Maur01} and on ladders
\cite{Hayw95,Whit98c,Whit97a,Romm00b,Sill01,Sill02}.

The persistent current response to magnetic fluxes through rings has 
been studied by \citet{Byrn02} and \citet{Mede03a}, also in view of possible 
quasi-exact conductance calculations \cite{Mede03b,Moli03}.

Even before the advent of the Hubbard model, a very similar model, 
the Pariser-Parr-Pople model \cite{Pari53,Popl53}, accommodating both dimerization effects 
and longer-range Coulomb interaction, had been invented in quantum 
chemistry to study conjugated polymer structures. DMRG has therefore 
been applied to polymers by many authors \cite{Rama97,Anus97,Lepe97,Shua97a,Shua97b,
Shua97c,Kuwa98,Boma98,Yaro98,Shua98a,
Rama00,Bend99,Ragh02,Barf01a,Barf02a}, moving to more and more realistic 
models of molecules, such as polydiacatylene \cite{Race01,Race03},
poly(para-phenylene) \cite{Anus97,Barf98,Burs02}, or
polyenes \cite{Burs99d,Zhan01,Barf01b}. Polymer phonons have been considered 
also in the non-adiabatic case \cite{Barf02}. There is closely 
related work on the Peierls-Hubbard model \cite{Pang95,Jeck98,Otsu98,Anus99}.

Since the early days of DMRG history \cite{Yu93} interest has also focused on 
the Kondo lattice, generic one-dimensional structures of itinerant 
electrons and localized magnetic moments, both for the one-channel
\cite{Carr96,Shib96a,Shib96b,Sikk97,Capr97a,Shib97a,Shib97c,Shib97d,Wang98,McCu99,Wata99b,Garc00,Garc02} and two-channel case \cite{More01}. 
Anderson models have been studied by \citet{Guer95}, \citet{Guer96}, and 
\citet{Guer01}.

The bosonic version of the Hubbard model has been studied by
\citet{Kuhn98}, \citet{Kuhn00}, and \citet{Koll03}, 
with emphasis on disorder effects by 
\citet{Raps99}.

Going beyond one dimension, two-dimensional electron gases in a Landau level
have been mapped to one-dimensional models suitable for DMRG 
\cite{Berg03,Shib03a,Shib01};
DMRG was applied to molecular iron rings \cite{Norm01} and 
has elucidated the lowest rotational band of the giant Keplerate 
molecule Mo$_{72}$Fe$_{30}$ \cite{Exle03}. More generic 
higher-dimensional applications will be discussed later.

Revisiting its origins, DMRG can also be used to provide 
high-accuracy solutions in one-particle quantum mechanics \cite{Mart99a};
as there is no entanglement in the one-particle wave function, 
the reduced basis transformation is formed from the $M^{S}$ 
lowest-lying states of the superblock projected onto the system (after 
reorthonormalization). It has been applied to an asymptotically free model 
in two dimensions \cite{Mart99b} and modified for up to three 
dimensions \cite{Mart01b}.
\section{DMRG THEORY}
\label{sec:theory}
DMRG practitioners usually adopt a quite pragmatic approach when 
applying DMRG to study some physical system. They consider the 
convergence of DMRG results under tuning the standard DMRG control parameters, 
system size $L$, size of the reduced block Hilbert space 
$M$, and the number of finite-system sweeps, and judge DMRG results to 
be reliable or not. Beyond empiricism, in recent years a coherent theoretical 
picture of the convergence properties and the 
algorithmic nature of DMRG has emerged, and it is fair to say that we have 
by now good foundations of a DMRG theory: DMRG generically produces a 
particular kind 
of ansatz states, known in statistical physics as matrix-product states; 
if they well approximate the true state of the system, DMRG will 
perform well. In fact, it turns out to be rewarding to reformulate 
DMRG in terms of variational optimization within classes of 
matrix-product states (Sec.\ \ref{subsec:mpa}). 
In practice, DMRG performance is best studied by 
considering the decay of the eigenvalue spectrum of the reduced 
density-matrix, which is fast for one-dimensional gapped quantum 
systems, but generically slow for critical systems in 
one dimension and all systems in higher dimensions (Sec.\ 
\ref{subsec:density}). This renders DMRG 
applications in such situations delicate at best. A very coherent 
understanding of these properties is now emerging in the framework of 
bipartite entanglement measures in quantum information theory.

\subsection{Matrix-product states}
\label{subsec:mpa} 
Like conventional RG methods, DMRG builds on Hilbert space decimation.
There is however no Hamiltonian flow to some fixed point, and no 
emergence of relevant and irrelevant operators. Instead, there is a 
flow to some fixed point in the space of the reduced density matrices. 
As has been pointed out by various authors 
\cite{Ostl95,Mart96b,Romm97,Duke98,Taka99}, this implies 
that DMRG generates {\em position-dependent 
matrix-product states} \cite{Fann89,Klum93} as block states. However, there are subtle, 
but crucial differences between DMRG states and matrix-product states 
\cite{Taka99,Vers04} that have important consequences regarding the variational 
nature of DMRG.

{\em Matrix-product states} are simple generalizations of 
product states of local states, which we take to be on a chain,
\begin{equation}
    \ket{\fat{\sigma}} = \ket{\sigma_{1}} \otimes \ket{\sigma_{2}} \otimes \ldots 
    \otimes \ket{\sigma_{L}} ,
    \label{eq:productstate}
\end{equation}
obtained by introducing linear operators $\hat{A}_{i}[\sigma_{i}]$ 
depending on the local state. These operators map from some 
$M$-dimensional auxiliary state space spanned by an orthonormal basis 
$\{\ket{\beta}\}$ to another $M$-dimensional auxiliary state space 
spanned by $\{\ket{\alpha}\}$:
\begin{equation}
\hat{A}_{i}[\sigma_{i}]=\sum_{\alpha\beta} 
(A_{i}[\sigma_{i}])_{\alpha\beta} \ket{\alpha}\bra{\beta} .
\label{eq:ansatzmatrix}
\end{equation}
One may visualize the auxiliary state spaces to be located on the 
bonds $(i,i+1)$ and $(i-1,i)$.
The operators are thus represented by
$M\times M$ matrices $(A_{i}[\sigma_{i}])_{\alpha\beta}$; $M$ will be seen 
later to be the number of block states in DMRG.
We further demand for reasons explained below that
\begin{equation}
\sum_{\sigma_{i}} \hat{A}_{i}[\sigma_{i}]\hat{A}_{i}^\dagger[\sigma_{i}] = \id .
\label{eq:mpanorm}
\end{equation}    
A position-dependent unnormalized matrix-product state for a 
one-dimensional system of size $L$ is then given by
\begin{equation}
    \ket{\psi} = \sum_{\{\fat{\sigma}\}} \left( \bra{\phi_{L}} 
    \prod_{i=1}^L 
    \hat{A}_{i} [\sigma_{i}]
    \ket{\phi_{R}} \right) \ket{\fat{\sigma}},
    \label{eq:mpsopen}
\end{equation}
where $\bra{\phi_{L}}$ and $\ket{\phi_{R}}$
are left and right boundary states in the auxiliary state spaces 
located in the above visualization to the left of the first and to the right 
of the last site. They are used to 
obtain scalar coefficients. Position-independent matrix-product states are 
obtained by making Eq.\ (\ref{eq:ansatzmatrix}) 
position-independent, $\hat{A}_{i}[\sigma_{i}]\rar \hat{A} 
[\sigma_{i}]$.
For simplicity, we shall consider only those in the following.

For periodic boundary conditions, boundary states are replaced  
by tracing the matrix-product:
\begin{equation}
    \ket{\psi} = \sum_{\{\fat{\sigma}\}} \text{Tr} \left[\prod_{i=1}^L 
    \hat{A}_{i} 
    [\sigma_{i}]
    \right] \ket{\fat{\sigma}} .
    \label{eq:mpsperiodic}
\end{equation}
The best-known matrix-product state is the valence-bond-solid ground 
state of the bilinear-biquadratic $S=1$ Affleck-Kennedy-Lieb-Tasaki 
Hamiltonian 
\cite{Affl87,Affl88}, where $M=2$.

{\em Correlations in matrix-product states:} Consider two local 
bosonic (for simplicity; see \citet{Ande99}) operators $\hat{O}_{j}$ and 
$\hat{O}_{j+l}$, acting on sites $j$ and $j+l$, applied to
the periodic boundary condition matrix-product state of Eq.\ 
(\ref{eq:mpsperiodic}). The correlator
$C(l)=\bra{\psi} \hat{O}_{j}\hat{O}_{j+l} \ket{\psi} / \langle \psi | 
\psi \rangle$
is then found, with $\text{Tr} X \text{Tr} Y = 
\text{Tr} (X\otimes Y)$ and $(ABC) \otimes (XYZ) = (A\otimes X)(B\otimes 
Y)(C\otimes Z)$, to be given by
\begin{equation}
    C(l)=
    \frac{\text{Tr} \overline{O}_{j} \overline{\id}^{l-1} \overline{O}_{j+l} 
    \overline{\id}^{L-l-1}}{\text{Tr} \overline{\id}^{L}}, 
    \label{eq:mpscorrelator}
\end{equation}    
where we have used the following mapping \cite{Romm97,Ande99} from an 
operator $\hat{O}$ acting on the local state space to a
$M^2$-dimensional operator $\overline{O}$ acting on products of
auxiliary states $\ket{\alpha\beta}=\ket{\alpha}\otimes \ket{\beta}$:
\begin{equation}
    \overline{O} = \sum_{\sigma\sigma'} \bra{\sigma'} \hat{O}
   \ket{\sigma} \hat{A}^* 
    [\sigma'] \otimes \hat{A} [\sigma] .
    \label{eq:mpsmapping}
\end{equation}
Note that $\hat{A}^*$ stands for $\hat{A}$ complex-conjugated only as opposed to 
$\hat{A}^\dagger$. Evaluating Eq.\ (\ref{eq:mpscorrelator}) in the 
eigenbasis of the mapped identity, $\overline{\id}$, we find that in 
the thermodynamic limit $L\rightarrow\infty$
\begin{equation}
    C(l)= \sum_{i=1}^{M^2} c_{i} \left( \frac{\lambda_{i}}{|\lambda_{i}|} \right)^l
    \exp (-l/\xi_{i})
    \label{eq:decaympa}
\end{equation}    
with $\xi_{i}=-1/\ln|\lambda_{i}|$. The $\lambda_{i}$ are the 
eigenvalues of $\overline{\id}$, and the $c_{i}$ depend on $\hat{O}$.
This expression holds because due to Eq.\ 
(\ref{eq:mpanorm}) $|\lambda_{i}|\leq 1$ and 
$\lambda_{1}=1$ for the eigenstate $\langle 
\alpha\beta|\lambda_{1}\rangle = \delta_{\alpha\beta}$. Equation
(\ref{eq:mpanorm}) is thus seen to ensure normalizability of matrix 
product states in the thermodynamic limit. 
Generally, all correlations in matrix-product states are either
long-ranged or purely exponentially decaying. 
They are thus 
not suited to describe critical behavior in the thermodynamic limit. 
Even for gapped one-dimensional quantum systems their utility may seem 
limited as the correlators $C(l)$ of these systems are generically of a  
two-dimensional classical Ornstein-Zernike form,
\begin{equation}
    C(l) \sim \frac{e^{-l/\xi}}{\sqrt{l}},
    \label{eq:OZ2dim}
\end{equation}
whereas exponential decay  as in Eq.\ (\ref{eq:decaympa}) is typical of one-dimensional classical 
Ornstein-Zernike forms. However, matrix-product states such as the 
AKLT ground state arise as quantum disorder points in general 
Hamiltonian spaces as the quantum remnants of classical phase 
transitions in two-dimensional classical systems; these disorder points 
are characterized by dimensional reduction of their correlations and 
typically characterize the qualitative properties of subsets of the 
Hamiltonian space, which turns them into most useful toy models, as
exemplified by the AKLT state \cite{Scho96c}. Away from the disorder points, 
choosing increasingly large $M$ as dimension of the ansatz matrices allows to 
model the true correlation form as a superposition of exponentials for 
increasingly large $l$; even for power-law correlations, this 
modelling works for not too long distances.  

{\em DMRG and matrix-product states:} To show that a DMRG calculation
retaining $M$ block states produces $M\times M$ matrix-product states, 
\citet{Ostl95} considered the reduced basis transformation to obtain the 
block of size $\ell$,
\begin{equation}
    \langle m_{\ell-1}\sigma_{\ell} | m_{\ell} \rangle \equiv 
    (A_{\ell})_{m_{\ell};m_{\ell-1}\sigma_{\ell}} \equiv 
     (A_{\ell}[\sigma_{\ell}])_{m_{\ell};m_{\ell-1}},
\label{eq:indexedtrafomatrix}
 \end{equation}    
such that 
\begin{equation}
    \ket{m_{\ell}} = \sum_{m_{\ell-1}\sigma_{\ell}} 
    (A_{\ell}[\sigma_{\ell}])_{m_{\ell};m_{\ell-1}}\ket{m_{\ell-1}}\otimes\ket{\sigma_{\ell}}.
    \label{eq:indexedtrafo}
\end{equation}    
The reduced basis transformation matrices $A_{\ell}[\sigma_{\ell}]$ 
automatically obey Eq.\ 
(\ref{eq:mpanorm}), which here ensures that $\{ \ket{m_{\ell}}\}$ is 
an orthonormal basis provided $\{ \ket{m_{\ell-1}}\}$ is one, too.
We may now use Eq.\ (\ref{eq:indexedtrafo}) for a backward recursion 
to express $\ket{m_{\ell-1}}$ via $\ket{m_{\ell-2}}$ and so forth. 
There is a (conceptually irrelevant) complication as the number of 
block states for very short blocks is less than $M$. For simplicity, 
I assume that $N_{\text{site}}^{\tilde{N}}=M$, and stop the recursion 
at the shortest block of size $\tilde{N}$ that has $M$ states, such that
\begin{eqnarray}
    \ket{m_{\ell}} &=& \sum_{m_{\tilde{N}}}
    \sum_{\sigma_{\tilde{N}+1},\ldots,\sigma_{\ell}} 
    (A_{\ell} [\sigma_{\ell}] \ldots A_{\tilde{N}+1} 
    [\sigma_{\tilde{N}+1}])_{m_{\ell}m_{\tilde{N}}}\times \nonumber 
    \\
    & & \ket{m_{\tilde{N}}} \otimes 
    \ket{\sigma_{\tilde{N}+1}\ldots\sigma_{\ell}},
\end{eqnarray}    
where we have boundary-site states 
$\ket{m_{\tilde{N}}}\equiv\ket{\sigma_{1}\ldots\sigma_{\tilde{N}}}$;
hence
\begin{equation}
    \ket{m_{\ell}} = \sum_{\sigma_{1},\ldots,\sigma_{\ell}} 
    (A_{\ell} [\sigma_{\ell}] \ldots A_{\tilde{N}+1} 
    [\sigma_{\tilde{N}+1}])_{m_{\ell},(\sigma_{1}\ldots\sigma_{\tilde{N}})} 
    \ket{\sigma_{1}\ldots\sigma_{\ell}} . 
    \label{eq:DMRGproductstate}
\end{equation}    
A comparison to Eq.\
(\ref{eq:mpsopen}) shows that DMRG generates position-dependent 
$M\times M$ matrix-product states as block states for a reduced 
Hilbert space of $M$ states; the auxiliary state space to a local state 
space is given by the Hilbert space of the block to which the local site 
is the latest attachment. 
Combining Eqs.\ (\ref{eq:be-ground}) and 
(\ref{eq:DMRGproductstate}), the superblock ground state of the full 
chain is the 
variational optimum in a space spanned by products of two local states and two
matrix-product states,
\begin{eqnarray}
    \ket{\psi} &=& \sum_{m^S m^E} \sum_{\{\fat{\sigma}\}}
    \psi_{m^S\sigma_{L/2}\sigma_{L/2+1} m^E} \times \nonumber \\
     & & (A_{L/2-1} [\sigma_{L/2-1}] \ldots A_{\tilde{N}+1} 
     [\sigma_{\tilde{N}+1}])_{m^S,(\sigma_{1}\ldots\sigma_{\tilde{N}})} \times     
     \nonumber  \\
     & & (A_{L/2+2} [\sigma_{L/2+2}] \ldots A_{L-\tilde{N}} 
     [\sigma_{L-\tilde{N}}])_{m^E,(\sigma_{L+1-\tilde{N}}\ldots\sigma_{L})} \nonumber \\
     & & \ket{\sigma_{1}\ldots\sigma_L} , \label{eq:DMRGmeetsmpa}
\end{eqnarray}    
which I have written for the case of the two single sites at the chain 
center; an analogous form holds for all stages of the finite-system 
algorithm, too.

For gapped quantum systems we may
assume that for blocks of length $\ell\gg\xi$ the reduced basis 
transformation becomes site-independent, such that the ansatz matrix 
$A$ generated by DMRG is essentially position-independent in the 
bulk. At criticality, the finite-dimensional matrix-product state 
generated introduces some effective correlation length (growing with $M$). 
In fact, this has been numerically verified for 
free fermions at criticality, where the ansatz matrices for bulk sites 
converged exponentially fast to a position-independent ansatz matrix, 
but where this convergence slowed 
down with $M$ \cite{Ande99}. 

The effect of the finite-system algorithm can be seen from Eq.\ 
(\ref{eq:DMRGmeetsmpa}) to be a sequence of {\em local} optimization 
steps of the wave function that have two effects: on the one hand, 
the variational 
coefficients $\psi_{m^S\sigma^{S}\sigma^{E} m^E}$ are optimized, 
on the other hand, a new improved 
ansatz matrix is obtained for the growing block, using the improved 
variational coefficients for a new reduced basis transformation.

In practical applications one observes that even for 
translationally invariant systems with periodic boundary conditions and 
repeated applications of finite-system 
sweeps the 
position dependency of the matrix-product state does not go away 
completely as it strictly should, indicating room for further 
improvement. \citet{Duke98} and \citet{Taka99} have pointed out and numerically 
demonstrated that finite-system 
DMRG (and TMRG; see Sec.\ \ref{sec:transfer}) results can be improved and 
better matrix 
product states for translationally invariant Hamiltonians be produced 
by switching, after convergence is reached, from the 
S$\bullet\bullet$E scheme for the finite-system algorithm to a 
S$\bullet$E scheme and to carry out 
some more sweeps. The rationale is that the variational ansatz of 
Eq.\ (\ref{eq:DMRGmeetsmpa}) generates (after the Schmidt 
decomposition and before truncation) ansatz 
matrices of dimension $MN_{\text{site}}$ at the two local sites due 
to  Eq.\ (\ref{eq:Schmidtmax}), whereas they are of dimension $M$ at all other 
sites; this introduces a notable position dependence, deteriorating 
the overall wave function. In the new 
scheme, the new ansatz matrix, without truncation, has also dimension 
$M$ (the dimension of the environment in Eq.\ (\ref{eq:Schmidtmax})), such that the local state is not favored.
The variational state 
now approaches its global optimum without further truncation, just 
improvements of the ansatz matrices.

The observation that DMRG produces variational states formed from 
products of local ansatz matrices has inspired the construction of 
variational ansatz states for the ground state of Hamiltonians 
(recurrent variational approach; see Mart\'{\i}n-Delgado and Sierra 
in \citet{Pesc99}) and the dominant eigenstate of transfer matrices 
(tensor product variational approach; 
\citet{Nish00b,Nish01b,Maes01,Gend02,Gend03}). 

Closer in spirit to the 
original DMRG concept is the {\em product wave function 
renormalization group} or PWFRG \cite{Nish95b,Hiei97}, which has been applied 
successfully to the magnetization process of spin chains in external 
field where infinite-system DMRG is highly prone to metastable trapping
\cite{Hiei97,Sato96,Okun99a,Hiei01} and to the restricted 
solid-on-solid model \cite{Akut98,Akut01a,Akut01b}. While in DMRG the 
focus is to determine for each iteration (superblock size) the wave 
function numerically as precisely as possible in order to derive the 
reduced basis transformation, PWFRG directly operates on the 
matrices $A$ itself: at each iteration, one starts with
an approximation to the wave function and local ansatz matrices 
related to it by a Schmidt decomposition.  
The wave function is then somewhat improved by carrying out 
a few Lanczos steps at moderate effort and Schmidt-decomposed again. 
The transformation matrices thus obtained are used to transform 
(improve) the local ansatz 
matrices, which in turn are combined with the decomposition weights 
to define a wave function for the next larger superblock.
The final result is reached 
when both the ansatz and transformation matrices become identical as 
they should at the DMRG fixed point for reduced basis transformations. 

{\em Variational optimization in matrix-product states.} If we compare 
Eq.\ (\ref{eq:mpsopen}) to Eq.\ (\ref{eq:DMRGmeetsmpa}), we see that 
the DMRG state is different from a true matrix-product state to describe
$\ket{\psi}$: The $A$ matrices link auxiliary state spaces of a bond 
to the right of a site to those on the left for sites in the left 
block, but vice versa in the right block. This may be mended by a 
transposition. This done, one may write the prefactors of $\ket{\sigma_{1}\ldots\sigma_L}$
as a true product of matrices by rewriting $\psi_{m^S\sigma_{L/2}\sigma_{L/2+1} m^E}$
as a $M\times M$ matrix 
$(\Psi[\sigma_{L/2}\sigma_{L/2+1}])_{m^S;m^E}$. The remaining anomaly 
is, as pointed out above, that formal ``translational invariance'' of 
this state is broken by the indexing of $\Psi$ by {\em two} sites, 
suggesting the modification of the S$\bullet\bullet$E scheme for the finite-system algorithm 
to a S$\bullet$E scheme discussed above. However, as \citet{Vers04} 
have demonstrated, it is conceptually and algorithmically worthwhile 
to rephrase DMRG consistently in terms of matrix-product states right 
from the beginning, thereby also abandoning the block concept. 

\begin{figure}
\centering\epsfig{file=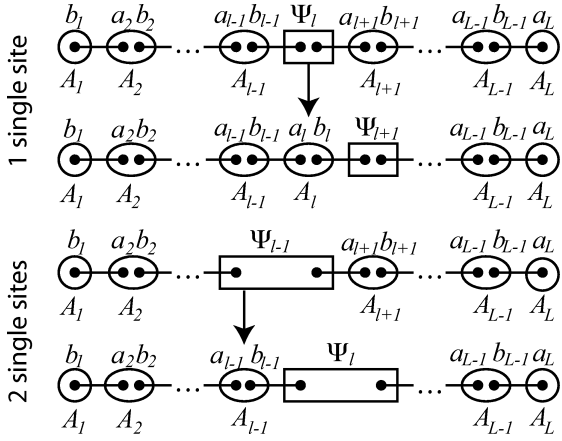,width=0.85\linewidth}
\caption{Pictorial representation of S$\bullet$E (above) and 
S$\bullet\bullet$E (below) DMRG schemes for open boundary conditions as introduced by
\citet{Vers04}. (Paired) dots represent auxiliary state spaces linked to
a local state space. Straight lines symbolize maximal entanglement, 
ellipses and rectangles maps to local state spaces as 
detailed in the text. Note the special role of the boundary sites. 
Adapted from \citet{Vers04}.}
\label{fig:mpsdmrgopen}
\end{figure}

To this end, they introduce (with the exception of the first and last 
site; see Fig.\ \ref{fig:mpsdmrgopen}) 
{\em two} auxiliary state spaces of dimension $M$, $a_{\ell}$ to 
the left and $b_{\ell}$ to the right of site $\ell$, such that on 
bond $\ell$ one has auxiliary state spaces $b_{\ell}$ and $a_{\ell+1}$. 
They now consider maps from $a_{\ell}\otimes b_{\ell} \rightarrow 
{\cal H}_{\ell}$ from the product of two auxiliary state spaces to the 
local state space which can be written using matrices 
$(A_{\ell}[\sigma])_{\alpha\beta}$: $A_{\ell} = 
\sum_{\sigma_{\ell}}\sum_{\alpha_{\ell}\beta_{\ell}}  (A_{\ell}[\sigma])_{\alpha\beta}
\ket{\sigma_{\ell}}\bra{\alpha_{\ell}\beta_{\ell}}$. The $\ket{\alpha}$ 
and $\ket{\beta}$ are states of the auxiliary state spaces. On the 
first and last site, the corresponding maps map only from one 
auxiliary state space. These maps can now be used to generate a 
matrix product state. To this purpose, \citet{Vers04} apply the string 
of maps $A_{1}\otimes A_{2} \otimes \ldots \otimes A_{L}$ to the 
product of maximally entangled states $\ket{\phi_{1}}\ket{\phi_{2}} 
\ldots \ket{\phi_{L-1}}$, where 
$\ket{\phi_{i}}=\sum_{\beta_{i}=\alpha_{i+1}} 
\ket{\beta_{i}}\ket{\alpha_{i+1}}$. The maximal entanglement, in the 
language of matrix product states, ensures that the prefactors of
$\ket{\sigma_{1}\ldots\sigma_L}$ are given by products of $A[\sigma]$ 
matrices, hence this construction is a matrix-product state. 
Comparing this state to the representation of $\ket{\psi}$ 
in Eq.\ (\ref{eq:DMRGmeetsmpa}), one finds that the maps $A$ are 
identical to the (possibly transposed) basis transformation matrices $A[\sigma]$ 
with the exception of the position of the single sites: in the 
S$\bullet\bullet$E setup (bottom half of Fig.\ \ref{fig:mpsdmrgopen}),
there are no auxiliary state spaces between the two single sites and
one map corresponds to $\Psi[\sigma_{\ell}\sigma_{\ell+1}]$.
In the S$\bullet$E setup (top half of Fig.\ \ref{fig:mpsdmrgopen}), 
this anomaly disappears and correctly normalized $A[\sigma_{\ell}]$ can be formed from 
$\Psi[\sigma_{\ell}]$.

The (finite-system) 
DMRG algorithm for the S$\bullet$E setup can now be reformulated 
in this picture as follows: 
sweeping forward and backward through the chain, one keeps for site 
$\ell$ all $A$ at other sites
fixed and seeks the $\Psi_{\ell}$ that minimizes the total energy. 
From this, one determines $A_{\ell}$ and 
moves to the next site, seeking $\psi_{\ell+1}$, and so on, until all 
matrices have converged. In the final matrix-product state one evaluates correlators 
as in Eq.\ (\ref{eq:mpscorrelator}). It is important to note that in 
this setup there is no truncation error, as explained above in the 
language of the Schmidt decomposition. 
Shifting the ``active'' site therefore does not change 
the energy, and the next minimization can only decrease the energy (or 
keep it constant). This setup is therefore truly variational in the 
space of the states generated by the maps $A$ and reaches a minimum 
of energy within that space (there is of course no guarantee to reach 
the global minimum). By comparison, the setup S$\bullet\bullet$E 
leads to a reduced basis transformation and always excludes two 
different auxiliary state spaces from the minimization procedure. It 
is hence not strictly variational.

\begin{figure}
\centering\epsfig{file=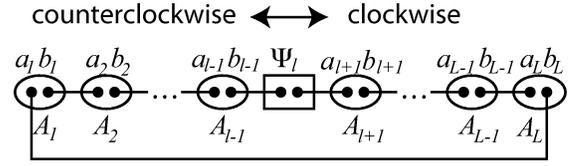,width=0.85\linewidth}
\caption{Periodic boundary condition setup used in the algorithm of
\citet{Vers04}. Labelling as in previous figure; adapted from 
\citet{Vers04}.}
\label{fig:mpsdmrgperiodic}
\end{figure}

In this setup the generalization to periodic boundary conditions is 
now easy. Additional auxiliary state spaces $a_{1}$ and $b_{L}$ are 
introduced and maximally entangled as for all other bonds. There is 
now complete formal translational invariance of the ansatz (see 
Fig.\ \ref{fig:mpsdmrgperiodic}). On this setup, one optimizes maps 
(matrices) $A$
one by one, going forward and backward.

\citet{Vers04} have shown that for a given $M$, they obtain roughly 
the same precision for periodic boundary conditions as for open 
boundary conditions. This compares extremely favorably with standard 
DMRG for periodic boundary conditions where (in the worst case) 
up to $M^2$ states are needed for 
the same precision.
\subsection{Properties of DMRG density matrices}
\label{subsec:density}
In order to gain a theoretical understanding of DMRG performance, we 
now take a look at the properties of the reduced density matrices and 
their truncation. Obviously, the ordered eigenvalue spectrum $w_{\alpha}$ of the reduced 
density-matrix $\dm$ should decay as quickly as possible to minimize 
the truncated weight $\epsilon_{\rho}=1-\sum^{M}_{\alpha=1}w_{\alpha}$ for optimal 
DMRG performance. This intuitively clear statement can be 
quantified: there are 
four major classes of density-matrix spectra, in descending order of 
DMRG performance.

(i) Density-matrix spectra for $M\times M$ {\em matrix-product states} 
as exact eigenstates of
quantum systems, with a {\em finite} fixed number of 
non-vanishing eigenvalues, leading to optimal DMRG performance.

(ii) Density-matrix spectra for non-matrix-product states of 
one-dimensional quantum 
systems with exponentially decaying correlations, with leading 
exponential decay of $w_{\alpha}$; spectra remain essentially unchanged 
for system sizes in excess of the correlation length.

(iii) Density-matrix spectra for states of one-dimensional quantum 
systems at criticality, with a decay of $w_{\alpha}$ 
that slows down with increasing system size, leading to DMRG failure 
to obtain thermodynamic limit behavior.

(iv) Density-matrix spectra for states of two-dimensional quantum 
systems both at and away from criticality, where the number of 
eigenvalues to be retained to keep a fixed truncation error grows 
exponentially with system size, restricting DMRG to very small system 
sizes.
    
All scenarios translate to classical systems of one additional dimension, 
due to the standard quantum-classical mapping from $d$- to $(d+1)$-dimensional 
systems.

{\em DMRG applied to matrix-product states.} 
A state $\ket{\tilde{\psi}}$ of the matrix-product form of Eq.\ 
(\ref{eq:mpsopen}) with dimension $\tilde{M}$, 
can be written as 
\begin{equation}
    \ket{\tilde{\psi}}= \sum_{\alpha=1}^{\tilde{M}} \ket{\tilde{\psi}^{S}_{\alpha}} 
    \ket{\tilde{\psi}^{E}_{\alpha}} \rightarrow \ket{\psi} = \sum_{\alpha=1}^{\tilde{M}} 
    \sqrt{w_{\alpha}}\ket{\psi^{S}_{\alpha}} 
    \ket{\psi^{E}_{\alpha}} ,
\end{equation}
where we have arbitrarily cut the chain into (left) system 
and (right) environment with
\begin{equation}
\ket{\tilde{\psi}^{S}_{\alpha}} = \sum_{\{\fat{\sigma}^S\}}  \bra{\phi_{S}} 
 \prod_{i\in S} A [\sigma_{i}] \ket{\alpha} 
\ket{\fat{\sigma}^S} ,
    \label{eq:mpshalf}
\end{equation}
and similarly $\ket{\tilde{\psi}^{E}_{\alpha}}$; $\ket{\psi}$, 
$\ket{\psi^{S}_{\alpha}}$, and $\ket{\psi^{E}_{\alpha}}$
are the corresponding normalized states. An appropriate treatment of 
boundary sites is tacitly implied [cf.\ the discussion before Eq.\ 
(\ref{eq:DMRGproductstate})].
Then the density matrix $\dm_{S}=\sum_{\alpha=1}^{\tilde{M}} w_{\alpha} \ket{\psi^{S}_{\alpha}} 
\bra{\psi^{S}_{\alpha}}$ and has a finite spectrum of $\tilde{M}$ 
non-vanishing eigenvalues. The truncated weight will thus be zero 
if we choose $M>\tilde{M}$ for DMRG, as DMRG generates these states 
[see Eq.\ (\ref{eq:DMRGproductstate})].

\begin{figure}
\centering\epsfig{file=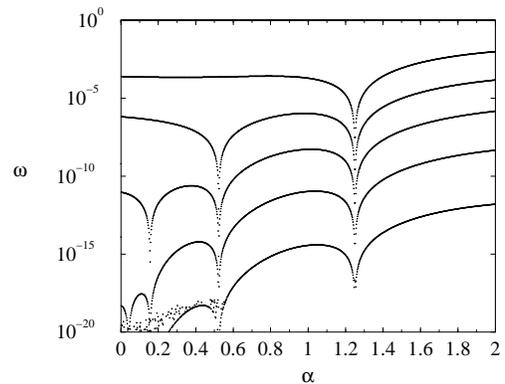,scale=0.35}
\caption{Largest density-matrix eigenvalues vs.\ a tuning 
parameter $\alpha$ in a $q=1/4$-symmetric Heisenberg chain. Largest 
eigenvalue $\approx 1$ invisible on logarithmic scale; eigenvalue collapse 
indicates pure matrix-product states of small finite dimension. From 
Kaulke and Peschel in \citet{Pesc99}. Reprinted with permission.}
\label{fig:eigenvaluecollapse}
\end{figure}

In such cases, DMRG may be expected to become an exact method up to 
small numerical inaccuracies. This has been observed recurrently; 
Kaulke and Peschel (in \citet{Pesc99}) provide an excellent example, 
the non-hermitean $q$-symmetric Heisenberg model with an additional boundary 
term.
This Hamiltonian is known to have matrix-product ground states of 
varying complexity (i.e.\ matrix sizes $\tilde{M}$) for particular 
choices of parameters (see also \citet{Alca94}).  Monitoring the eigenvalue spectrum 
of the DMRG density matrix for a sufficiently long system, they 
found that indeed it collapses at these particular values: 
all eigenvalues but the $\tilde{M}$ largest vanish at these points (Fig.\ 
\ref{fig:eigenvaluecollapse}). Similarly, a class of two-dimensional quantum 
Hamiltonians with exact matrix-product ground states has been studied 
using DMRG by \citet{Hiei99}.

{\em DMRG applied to generic gapped one-dimensional systems.}
It is quite easy to observe numerically that for gapped 
one-dimensional quantum systems the eigenvalue spectrum of the density 
matrices decays essentially exponentially such that the truncated
weight can be reduced exponentially fast by increasing $M$, which is 
the hallmark of DMRG success. Moreover, the eigenvalue spectrum 
converges to some thermodynamic-limit form.  

\citet{Pesc99b} have confirmed these numerical observations by studying 
exact density-matrix spectra that may be calculated for 
the one-dimensional Ising model in a transverse field and the XXZ 
Heisenberg chain in its antiferromagnetic gapped regime, using a
corner transfer matrix method \cite{Baxt82}. 

\begin{figure}
\centering\epsfig{file=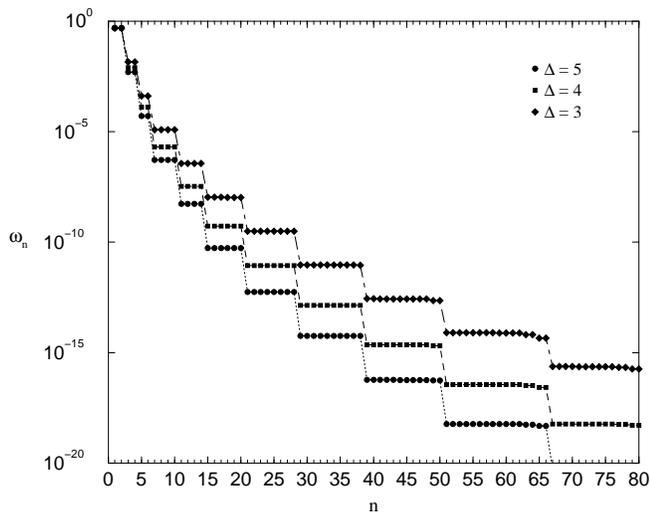,scale=0.5}
\caption{Density-matrix eigenvalues $w_{n}$ vs.\ eigenvalue number $n$ for 
a gapped XXY Heisenberg chain of $L=98$ and three values of anisotropy 
$\Delta>\Delta_{c}=1$. Degeneracies are as predicted analytically. From 
\citet{Pesc99b}. Reprinted with permission.}
\label{fig:heisenbergeigenvalues}
\end{figure}

In those cases, the eigenvalues of $\dm$ are given, up to a global 
normalization, as 
\begin{equation}
    w\propto\exp\left(-\sum_{j=0}^\infty \epsilon_{j}n_{j}\right)
\label{eq:dmspectrumfromctm}
\end{equation}    
with ``fermionic'' occupation numbers $n_{j}=0,1$ and an essentially 
{\em linear} ``energy spectrum'' $\epsilon_{j}$: these are typically 
some integer multiple of a fundamental scale $\epsilon$. 
The density-matrix eigenvalue spectrum shows 
clear exponential decay only for large eigenvalues due to the increasing degree 
of degeneracy of the eigenvalue spectrum, as the number of possible partitions of 
$\epsilon n$ into $\epsilon_{j}n_{j}$ grows. DMRG density matrices 
perfectly reproduce this behavior (Fig.\ 
\ref{fig:heisenbergeigenvalues}). Combining such exactly known corner 
transfer matrix spectra with results from the theory of partitions, 
\citet{Okun99c} have derived the asymptotic form
\begin{equation}
   w_{\alpha} \sim \exp \left( - \text{const.} \times \ln^2 \alpha \right)
   \label{eq:asymptoticeigenvalue}
\end{equation}   
for the $\alpha$th eigenvalue (see also \citet{Chan02a}).

These observations imply that for a desired truncated weight 
the number of states to be kept remains finite 
even in the thermodynamic limit and that the truncated weight decays 
exponentially fast in $M$, with some price to be paid due to the 
increasing degree of degeneracy occuring for large $M$.

{\em DMRG applied to one-dimensional systems at criticality.} 
Numerically, one observes at 
criticality that the eigenvalue spectrum decays dramatically slower and 
that for increasing system size this phenomenon tends to aggravate 
\cite{Chun01},
with errors in e.g.\ ground-state energies increasing by several 
orders of magnitude compared to gapped systems \cite{Lege96}.

Hence, the double question of the decay of the eigenvalue spectrum for 
the density matrix of a given system size and the size dependency 
of this result arises. \citet{Chun01} have shown for generic 
Hamiltonians quadratic in fermionic operators that the density 
matrix spectrum is once again of the form of Eq.\ (\ref{eq:dmspectrumfromctm}),
but the ``energies'' $\epsilon_{j}$ are now no longer given by a 
simple relationship linear in $j$. Instead, they show much slower, 
curved growth, that slows down with system size. Translating this into 
actual eigenvalues of the density matrix, they show less-than 
exponential decay slowing down with system size. This implies that 
for one-dimensional quantum systems at criticality numerical 
convergence for a fixed system size will no longer be exponentially 
fast in $M$. Maintaining a desired truncated weight in the 
thermodynamic limit implies a diverging number $M$ of states to be kept. 

\begin{figure}
\centering\epsfig{file=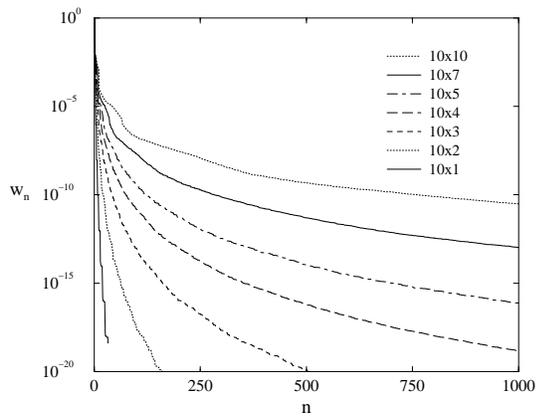,scale=0.4}
\caption{Density-matrix eigenvalues $w_{n}$ for rectangular two-dimensional 
gapped systems of interacting harmonic oscillators. The left-most curve 
corresponds to the one-dimensional case. From \citet{Chun00b}. 
Reprinted with permission.}
\label{fig:2dhoeigenvalues}
\end{figure}

{\em DMRG in two-dimensional quantum systems.} 
Due to the large interest in two-dimensional quantum systems, we now 
turn to the question of DMRG convergence in gapped and critical 
systems. In the early days of DMRG, \citet{Lian94} had observed 
numerically that to maintain a given precision, an exponentially growing 
number of states $M \sim \alpha^{L}$, $\alpha>1$, had to be kept for system sizes 
$L\times L$. However, reliable information from numerics is very 
difficult to obtain here, due to the very small system sizes in 
actual calculations. \citet{Chun00b} 
have studied a (gapped) system of interacting harmonic oscillators, where the 
density matrix can be written as the bosonic equivalent of Eq.\ (\ref{eq:dmspectrumfromctm}),
with eigenvalues, again up to a normalization, 
\begin{equation}
    w\propto\exp\left( -\sum_{j=0}^\infty 
    \epsilon_{j}\bcr{j}\ban{j}\right) .
\label{eq:dmspectrumbosons}
\end{equation}    
Considering strip systems of size $L\times N$ with $N\leq L$, 
numerical evaluations have shown that
\begin{equation}
    \epsilon_{j} \sim \text{const.}\times j/N,
    \label{eq:stripedependency}
\end{equation}
hence the eigenvalue decay slows down exponentially with inverse system 
size. This $1/N$ behavior can be understood by considering the 
spectrum of $N$ chains without interchain interaction, which is given by the 
spectrum of the single chain with $N$-fold degeneracy introduced. 
Interaction will lift this degeneracy, but not 
fundamentally change the slowdown this imposes on the decay of the 
density-matrix spectrum [see also \citet{Croo98} for the generic 
argument].
Taking bosonic combinatorics into account, one finds
\begin{equation}
    w_{\alpha} \sim \exp \left( - (\text{const.}/N) \ln^2 \alpha \right),
\label{eq:2ddecay}
\end{equation}
which is a consistent extension of Eq.\ (\ref{eq:asymptoticeigenvalue}).
For a system of size $10\times 10$, typical truncation errors of 
$10^{-5}$ for $M=100$, $10^{-7}$ for $M=500$ and $10^{-8}$ for $M=1000$ 
have been reported (cf.\ Fig.\ \ref{fig:2dhoeigenvalues}), reflecting the very slow convergence of DMRG in 
this case.

For a critical system, the non-interacting fermion model may be 
used once again as a model system \cite{Chun01}. For 
$M=2000$ states, for this simple model, the resulting truncation error 
is $5 \times 10^{-2}$ for systems of size $12\times 12$, $5 \times 
10^{-1}$ for $16\times 16$ and $10^{-1}$ for size $20\times 20$. 
Here, DMRG is clearly at its limits. 

\begin{figure}
\centering\epsfig{file=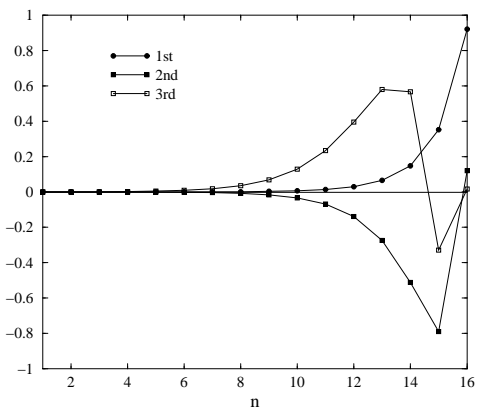,scale=0.4}
\caption{Amplitudes of highest-weight single-particle eigenstates of the left-block 
density-matrix
for a one-dimensional open chain of $L=32$ interacting harmonic 
oscillators. From \citet{Chun00b}. Reprinted with permission.}
\label{fig:localizedeigenstates}
\end{figure}

{\em DMRG precision for periodic boundary conditions.} 
While for periodic boundary conditions the overall properties of 
DMRG density matrices 
are the same as those of their open boundary condition counterparts, 
their spectra have been observed numerically to decay much more 
slowly. Away from criticality, this is due to some (usually only approximate) additional factor two 
degeneracy of the eigenvalues. This can be explained by studying the 
amplitudes of density-matrix eigenstates. \citet{Chun00b} have demonstrated 
in their non-critical harmonic oscillator model that the 
density-matrix eigenstates associated to high 
eigenvalue weight are strongly located close to the boundary between 
system and environment (Fig.\ \ref{fig:localizedeigenstates}). 
Hence, in a periodic system, 
where there are 
two boundary points, there are for the high-weight eigenvalues 
two essentially identical sets of eigenstates localized at the left 
and right boundary respectively, leading to the approximate double degeneracy of 
high-weight eigenvalues. At criticality, no such simple argument 
holds, but DMRG is similarly affected by a slower decay of spectra.    

In a recent study, \citet{Vers04} have shown that this strong 
deterioration of DMRG is essentially due to its particular setup for simulating
periodic boundary conditions, and provided a new formulation of the 
algorithm which produces results of the same quality as for open 
boundary conditions for the same number of states kept, at the cost 
of losing matrix sparseness (see Sec.\ \ref{subsec:mpa}).

\subsection{DMRG and quantum information theory}
\label{subsec:quantint}

As understood in the early phases of DMRG 
development\cite{Whit92a,Whit92b}, the reason for the success of the 
method is that no system is considered in isolation, but embedded in a 
larger entity. In fact, as discussed in Sec.\ \ref{subsec:densityflow}, 
DMRG truncation can be understood in the language of quantum 
information theory as preserving the maximum 
entanglement between system and environment as measured by the von 
Neumann entropy of entanglement,
\begin{equation}
    S =-\text{Tr} \hat{\rho} \ln_{2} \hat{\rho} =
    - \sum_{\alpha} w_{\alpha} \ln_{2} w_{\alpha} 
\end{equation}
in an $M$-dimensional block state space.

\begin{figure}
\centering\epsfig{file=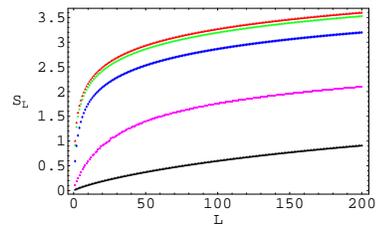,scale=0.6}
\caption{Diverging von Neumann entropy $S_{L}$ vs.\ block length $L$ for a 
critical isotropic XX chain in external fields $H<H_{\text{crit}}$. 
Increasing field strengths suppress entropy. From \citet{Lato03}. 
Reprinted with permission.}
\label{fig:entanglementxxchaininfield}
\end{figure}

\begin{figure}
\centering\epsfig{file=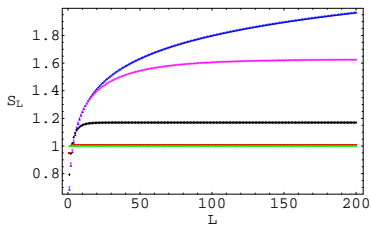,scale=0.6}
\caption{Saturating density-matrix entropy $S_{L}$ vs.\ block length 
$L$ for an Ising chain in a transverse field $H\leq H_{\text{crit}}$. Saturation 
entropy grows with $H\rightarrow H_{\text{crit}}$; divergence is recovered at
criticality (top curve). From \citet{Lato03}. Reprinted with 
permission.}
\label{fig:entanglementisingchaininfield}
\end{figure}

\citet{Lato03} have calculated the entropy of 
entanglement $S_{L}$ for systems of length $L$ embedded in infinite 
(an)isotropic XY chains 
and for systems embedded in finite, periodic Heisenberg chains of 
length $N$, both with and without external field. 
Both models show critical and noncritical behavior 
depending on anisotropy and field strength. In the limit
$N\rightarrow\infty$, $S_{L}\leq L$, which
is obtained by observing that entropy is maximal if all $2^L$ states 
are equally occupied with amplitude $2^{-L}$.
They find that $S_{L}\rightarrow\infty$ for 
$L\rightarrow\infty$ at criticality, but saturates as 
$S_{L}\rightarrow S^*_{L}$ for $L\approx\xi$ in the non-critical regime. 
At 
criticality, however, $S_{L}$ grows far less than permitted by 
$S_{L}\leq L$, but obeys logarithmic behavior
\begin{equation}
    S_{L} = k \log_{2}L + \text{const.}
\label{eq:entropyL}
\end{equation}    
The constant $k$ is found to be given by $k=1/6$ for the anisotropic 
XY model at the critical field $H_{c}=1$, and by $k=1/3$ for the 
isotropic XY model at $H_{c}=1$ as well as the isotropic Heisenberg 
model for $H\leq H_{c}=1$ (for an isotropic XX model in field 
$H<H_{crit}$, see Fig.\ \ref{fig:entanglementxxchaininfield}). Away from criticality, the saturation 
value $S^{*}_{L}$ decreases with decreasing correlation length $\xi$ 
(Fig.\ \ref{fig:entanglementisingchaininfield}).

One-dimensional quantum spin chains at criticality are described by 
conformally invariant (1+1)-dimensional field theories.
In fact, Eq.\ (\ref{eq:entropyL}) can be linked \cite{Gait03,Lato03} to the 
geometric entropy \cite{Call94} of such field theories,
\begin{equation}
    S^{geo}_{L} = \frac{c+\overline{c}}{6} \log_{2}L ,
\label{eq:entropy_geo}
\end{equation}    
where $c$ ($\overline{c}$) are the central charges, if one observes 
that for the anisotropic XY model $c=\overline{c}=1/2$ and 
for the Heisenberg model and isotropic XY model $c=\overline{c}=1$.

Geometric entropy arguments for $(d+1)$-dimensional field theories 
use a  
bipartition of $d$-dimensional space by a $(d-1)$-dimensional hypersurface, which 
is shared by system S and environment E. By the Schmidt decomposition, 
S and E share the same reduced density-matrix spectrum, hence 
entanglement entropy, which is now argued 
to reside essentially 
on the shared hypersurface (cf.\ the locus of highest weight 
density-matrix eigenstates in Fig.\ \ref{fig:localizedeigenstates}; 
see also \citet{Gait01}). 
Taking the thermodynamic (infrared) limit,
entropy scales as the hypersurface area,
\begin{equation}
    S_{L} \propto \left( \frac{L}{\lambda} \right)^{d-1} ,
\label{eq:entropy_higher}
\end{equation}    
where $\lambda$ is some ultraviolet cutoff which in condensed matter 
physics we may fix at some lattice spacing. Introducing a gap (mass), 
an essentially infrared property, into this field theory does not modify this 
behavior generated on ultraviolet scales on the hypersurface.
In $d=1$, a more careful 
argument shows that there is a (subleading) logarithmic correction as 
given above at criticality, saturation otherwise. 

$S_{L}$ is the number of qubits corresponding to the entanglement 
information. To code this information in DMRG, one needs a system Hilbert 
space of size $M\geq 2^{S_{L}}$; in fact, numerical results indicate 
that $M$ and $2^{S_{L}}$ are -- to a good approximation -- proportional. Taking the 
linear dimensions of total space and embedded system 
to $N,L\rightarrow\infty$, quantum information theory hence 
provides us with a unified picture of DMRG performance, which is in 
perfect agreement with the observations obtained by studying density 
matrix spectra:

(i) In one-dimensional quantum systems away from criticality, DMRG yields 
very precise results for the thermodynamic limit for some finite 
number of states kept, $M \sim 2^{S^*_{L}}$, which grows with the 
correlation length.

(ii) In one-dimensional quantum systems at criticality, the number of states 
that has to be kept, will diverge as
\begin{equation}
    M(L) \sim L^k ,
\end{equation}
with $k$ from Eq. (\ref{eq:entropyL}). This explains the failure of 
DMRG for critical one-dimensional systems as $L\rightarrow\infty$. As 
$k$ is small, this statement has to be 
qualified; DMRG still works for rather large finite systems.

(iii) In higher-dimensional quantum systems, the number of states to be kept 
will diverge as
\begin{equation}
    M(L) \sim 2^{L^{d-1}} ,
\end{equation}
rendering the understanding of thermodynamic limit behavior by 
conventional DMRG 
quite impossible; information is beyond retrieval just as in a
black hole -- whose entropy scales with its surface, like the 
entanglement entropy would in a three-dimensional DMRG application!
In any case, even for higher-dimensional systems, DMRG 
may be a very useful method as long as system size is kept resolutely 
finite, such as in nuclear physics or quantum chemistry applications. 
Recent proposals \cite{Vers04b} also show that it is possible to formulate 
generalized DMRG
ansatz states in such a way that entropy shows correct size dependency 
in two-dimensional systems (see Sec. \ref{sec:twodim}).

\citet{Lege03a} have carried the analysis of DMRG 
state selection using entanglement entropy 
even further, arguing that the acceptable truncated weight -- 
which is not identical to, but closely related to the entropy of 
entanglement, which emerges as the key quantity -- should be kept fixed 
during DMRG runs, determining how many states $M$ have to be retained 
at each iteration. This {\em dynamical block state selection} has 
already been applied in various contexts \cite{Lege03b,Lege03c}. More 
recently, \citet{Lege04} have tightened the relationship between 
quantum information theory and DMRG state selection by proposing a 
further refinement of state selection. $M$ is now chosen variably to keep loss 
of quantum information below some acceptable threshold. They argue 
that this loss is given as
\begin{equation}
    \chi({\cal E}) \equiv S(\dm) - p_{typ}S(\dm_{typ}) - 
    (1-p_{typ})S(\dm_{atyp}).
\end{equation}    
Here, $\dm = p_{typ}\dm_{typ} + (1-p_{typ})\dm_{atyp}$. For a given 
$M$, $\dm_{typ}$ is formed from the $M$ dominant eigenstates of 
$\dm$, $\dm_{atyp}$ from the remaining ones, with $1-p_{typ}$ being the 
truncated weight and $\dm_{typ,atyp}$ scaled to have trace one.  
\citet{Lege04} report a very clear linear relationship between DMRG 
errors and $\chi({\cal E})$.

\section{ZERO-TEMPERATURE DYNAMICS}
\label{sec:dynamics}
As we have seen, DMRG is an excellent method 
to calculate ground states and selected excited eigenstates at almost 
machine precision. On the other hand, the targeting of specific states 
suggests that DMRG is not suitable to calculate dynamical 
properties of strongly correlated systems even at $T=0$ as the time evolution of 
general excited states will explore large parts of the Hilbert space. 
Closer inspection 
has revealed, however, that the relevant parts of Hilbert space can 
be properly addressed by DMRG. For some 
operator $\hat{A}$, one may define a (time-dependent) Green's function at $T=0$ 
in the Heisenberg picture by
\begin{equation}
    \imag G_{A}(t'-t) = \bra{0} \hat{A}^\dagger(t') \hat{A}(t) \ket{0}  
    \label{eq:deftimecorrel}
\end{equation}
with $t' \geq t$ for a time-independent Hamiltonian $\ham$. Going to
frequency-range, the Green's function reads
\begin{equation}
    G_{A}(\omega+\imag\eta)= \bra{0} \hat{A}^\dagger \frac{1}{E_{0}+\omega+\imag\eta-\ham} 
    \hat{A} \ket{0} ,
    \label{eq:Greensfunction}
\end{equation}
where $\eta$ is some positive number to be taken to 
zero at the end. We may also use the spectral or Lehmann 
representation of correlations in the eigenbasis of $\ham$,
\begin{equation}
    C_{A}(\omega) = \sum_{n} |\bra{n} \hat{A} \ket{0}|^2 
    \delta(\omega+E_{0}-E_{n}).
    \label{eq:Lehmann}
\end{equation}
This frequency-dependent correlation function is related to $G_{A}(\omega+\imag\eta)$ as
\begin{equation}
    C_{A}(\omega) = \lim_{\eta\rightarrow 0^+} - \frac{1}{\pi} \impart G_{A}(\omega+\imag\eta) .
    \label{eq:etalimit}
\end{equation}    
In the following, I shall also use $G_{A}(\omega)$ and 
$C_{A}(\omega+\imag\eta)$ where the limit $\eta\rightarrow 0^+$ 
will be assumed to have been taken in the former and omitted in the 
latter case. 
The role of $\eta$ in DMRG calculations is threefold:
First, it ensures causality in Eq.\ (\ref{eq:Greensfunction}). 
Second, it introduces a 
finite lifetime $\tau\propto 1/\eta$ to excitations, such that on 
finite systems they can be prevented from travelling to the open 
boundaries where their reflection would induce spurious effects. 
Third, $\eta$ provides a Lorentzian broadening of $C_{A}(\omega)$,
\begin{equation}
    C_{A}(\omega+\imag\eta) = \frac{1}{\pi} \int d\omega' C_{A}(\omega') 
    \frac{\eta}{(\omega-\omega')^2 + \eta^2},
    \label{eq:lorentzian}
\end{equation}
which serves either to broaden the numerically obtained discrete spectrum of 
finite systems into some ``thermodynamic limit'' behavior or to 
broaden analytical results for $C_{A}$ for comparison to numerical 
spectra where $\eta>0$.

Most DMRG approaches to dynamical correlations center 
on the evaluation of Eq.\ (\ref{eq:Greensfunction}). The first, which I 
refer to as Lanczos vector dynamics, has been 
pioneered by \citet{Hall95}, and calculates highly time-efficient, but comparatively 
rough approximations to dynamical quantities adopting the 
Balseiro-Gagliano method to DMRG. The second approach, which is 
referred to as correction vector method \cite{Rama97,Kuhn99}, is yet another older scheme 
adopted to DMRG, much more precise, but numerically by far more 
expensive. A third approach, called DDMRG (dynamical DMRG), has been 
proposed by \citet{Jeck02}; while on the surface it is only a minor 
reformulation of the correction vector method, it will be seen to be 
much more precise. 

Very recently, dynamical correlations as in Eq.\ 
(\ref{eq:deftimecorrel}) have also been calculated directly in real 
time using time-dependent DMRG with adaptive Hilbert spaces 
\cite{Dale04,Whit04,Vida03} as will be discussed in Sec.\ 
\ref{subsec:time}. These may then be Fourier transformed to a frequency
representation. As time-dependent DMRG is best suited for short 
times (i.e.\ high frequencies) and the methods discussed in this 
Section are typically best for low frequencies, this alternative 
approach may be very attractive to cover wider frequency ranges. 

All approaches share the need for precision control and the 
elimination of finite-system and/or boundary effects. Beyond DMRG specific 
checks of convergence, precision may 
be checked by comparisons to independently obtained equal-time 
correlations using the following sum rules:
\begin{eqnarray}
    \int_{-\infty}^{\infty} d\omega\, C_{A}(\omega) = \bra{0} 
    \hat{A}^\dagger \hat{A} \ket{0} \\
    \int_{-\infty}^{\infty} d\omega\, \omega C_{A}(\omega) = \bra{0} 
    \hat{A}^\dagger [\hat{H},\hat{A}] \ket{0} \\
    \int_{-\infty}^{\infty} d\omega\, \omega^2 C_{A}(\omega) = \bra{0} 
    [\hat{A}^\dagger,\hat{H}] [\hat{H},\hat{A}] \ket{0} ,
    \label{eq:sumrules}    
\end{eqnarray}  
where the first equation holds for $\eta\geq 0$ and the latter two 
only as $\eta\rightarrow 0$. As DMRG is much more precise 
for equal-time than dynamical correlations, comparisons are
made to DMRG results which for the purpose may be 
considered exact. Finite-size effects due to boundary conditions can 
be treated in various ways. They can be excluded completely by 
the use of periodic boundary conditions at the price of much lower 
DMRG precision \cite{Hall95}. In cases where open boundary conditions 
are preferred, two situations should be distinguished. If $\hat{A}$ acts 
locally, such as in the calculation of an optical conductivity, one 
may exploit  that finite $\eta$ exponentially suppresses 
excitations \cite{Jeck02}. As they travel at some speed $c$ through the system, a 
thermodynamic limit $L\rightarrow\infty, \eta\rightarrow 0$ with
$\eta=c/L$ may be taken consistently. For the calculation of dynamical structure 
functions such as obtained in elastic neutron scattering, $\hat{A}$ is 
a spatially delocalized Fourier transform, and another approach must be 
taken. The open boundaries introduce both
genuine edge effects and a hard cut to the wave functions of excited 
states in real space, leading to a large spread in momentum space. To 
limit bandwidth in momentum space, filtering is necessary. The 
filtering function should be narrow in 
momentum space and broad in real space, while simultaneously strictly
excluding edge sites. \citet{Kuhn99}
have introduced the so-called Parzen filter 
$F_{\text{P}}$,
\begin{equation}
F_{\text{P}}(x) = \left\{ 
\begin{array}{ll}
1-6|x|^2+6|x|^3 & \quad 0\leq|x|\leq 1/2 \\
2(1-|x|)^3      & \quad 1/2 \leq |x| \leq 1
\end{array}
\right. ,
\end{equation}
where $x=i/L_{P} \in [-1,1]$, the relative site position in the filter 
for a total filter width $2L_{P}$. 
In momentum space $F_{\text{P}}$ has a wave vector uncertainty 
$\Delta q=2\sqrt{3}/L_{\text{P}}$, which scales as $L^{-1}$
if one scales $L_{P}$ with system size $L$. For finite-size 
extrapolations it has to be ensured that filter size does not 
introduce a new size dependency. This can be ensured by introducing
a Parzen filter prefactor given by $\sqrt{140\pi/151L_{P}}$.
\subsection{Continued fraction dynamics}
\label{subsec:lanczosdynamics}
The technique of {\em continued fraction dynamics} has first been 
exploited by \citet{Gagl87} in the framework of exact ground state 
diagonalization. Obviously, the calculation of Green's functions as in Eq.\ 
(\ref{eq:Greensfunction}) involves the inversion of $\ham$ (or more 
precisely, $E_{0}+\omega+\imag\eta-\ham$), a typically 
very large sparse hermitian matrix. This inversion is carried out in 
two, at least formally, exact steps. First, an iterative basis 
transformation taking $\ham$ to a tridiagonal form is carried out.
Second, this tridiagonal matrix is then inverted, allowing the 
evaluation of Eq.\ (\ref{eq:Greensfunction}).

Let us call the diagonal elements of $\ham$ in the tridiagonal form $a_{n}$ and 
the subdiagonal elements $b_{n}^2$. 
The coefficients $a_{n}$, $b_{n}^2$ are obtained as the 
Schmidt-Gram coeffcients in the generation of a Krylov subspace of
unnormalized states starting from some arbitrary state, which we take 
to be the excited state $\hat{A}\ket{0}$:
\begin{equation}
    \ket{f_{n+1}} = \ham\ket{f_{n}} - a_{n} \ket{f_{n}} - b_{n}^2 
    \ket{f_{n-1}} ,
    \label{eq:Krylovsequence}
\end{equation}
with 
\begin{eqnarray}
    \ket{f_{0}} &=& \hat{A}\ket{0}, \\
    a_{n} &=& \frac{\bra{f_{n}} \ham\ket{f_{n}}}{\langle f_{n} | 
    f_{n} \rangle}, \\
    b_{n}^2 &=&\frac{\bra{f_{n-1}} \ham\ket{f_{n}}}{\langle f_{n-1} | 
    f_{n-1} \rangle} = \frac{\langle f_{n} |f_{n}\rangle}{\langle f_{n-1} | 
    f_{n-1} \rangle}.
\end{eqnarray}    
The global orthogonality of the states $\ket{f_{n}}$ (at least in 
formal mathematics) and the tridiagonality of the new representation (i.e.\
$\bra{f_{i}} \ham \ket{f_{j}}=0$ for $|i-j|>1$) follow 
by induction. It can then be shown quite 
easily by an expansion of determinants that the inversion of 
$E_{0}+\omega+\imag\eta-\ham$ leads to a continued fraction such that 
the Green's function $G_{A}$ reads
\begin{equation}
    G_{A}(z) = \frac{\bra{0} \hat{A}^\dagger \hat{A} \ket{0}}{
    z-a_{0}-\frac{b_{1}^2}{z-a_{1}-\frac{b_{2}^2}{z-\ldots}}} ,
    \label{eq:continuedfraction}
\end{equation}
where $z=E_{0}+\omega+\imag\eta$. This expression can now be 
evaluated numerically, giving access to dynamical correlations. 

In practice, several limitations occur. The iterative generation of the 
coefficients $a_{n}$, $b_{n}^2$ is equivalent to a Lanczos diagonalization of 
$\ham$ with starting vector $\hat{A}\ket{0}$. Typically, the convergence of the 
lowest eigenvalue of the transformed tridiagonal Hamiltonian to the 
ground state eigenvalue of $\ham$ will have happened after 
$n \sim O(10^2)$ iteration steps for standard model Hamiltonians. 
Lanczos convergence is however accompanied by numerical loss of global 
orthogonality which computationally is ensured only locally,
invalidating the inversion procedure. The generation of coefficients 
has to be stopped before that. \citet{Kuhn99} have proposed to 
monitor, for normalized vectors, $\langle f_{0} | f_{n} \rangle > \epsilon$
as termination criterion. The precision of this approach therefore 
depends on whether the continued fraction has sufficiently converged 
at termination. With $\hat{A}\ket{0}$ as starting vector, convergence 
will be fast if $\hat{A}\ket{0}$ is a long-lived excitation (close to an eigenstate) 
such as would be the 
case if the excitation is part of an excitation band; this will 
typically not be the case if it is part of an excitation continuuum.

It should also be mentioned that beyond the above complications also 
arising for exact diagonalization with an exact $\ham$, additional 
approximations are introduced in DMRG as the Hamiltonian itself is of course 
not exact.

Instead of evaluating the continued fraction of Eq.\ (\ref{eq:continuedfraction}), 
one may also exploit 
that upon normalization of the Lanczos vectors $\ket{f_{n}}$ and accompanying 
rescaling of the $a_{n}$ and $b_{n}^2$, the Hamiltonian is 
iteratively transformed into a tridiagonal form in a new approximate 
orthonormal basis. Transforming 
the basis $\{ \ket{f_{n}} \}$  by a diagonalization of the 
tridiagonal Hamiltonian matrix to the 
approximate energy eigenbasis of $\ham$, $\{ \ket{n} \}$ with 
eigenenergies $E_{n}$, the Green's function can be written within this 
approximation as
\begin{eqnarray}
    & & G_{A}(\omega+\imag\eta) = \\
    & & \sum_{n} \bra{0} \hat{A}^\dagger \ket{n}\bra{n}  
    \frac{1}{E_{0}+\omega+\imag\eta-E_{n}}  \ket{n}\bra{n} 
    \hat{A} \ket{0} \nonumber ,
\end{eqnarray}    
where the sum runs over all approximate eigenstates. The 
dynamical correlation function is then given by
\begin{equation}
    C_{A}(\omega+\imag\eta) = \frac{\eta}{\pi} \sum_{n} \frac{|\bra{n} 
    \hat{A} \ket{0}|^2}{(E_{0}+\omega-E_{n})^2 +\eta^2} ,
\end{equation}
where the matrix elements in the numerator are simply the 
$\ket{f_{0}}$ expansion coefficients of the approximate eigenstates 
$\ket{n}$.

For a given effective Hilbert space dimension $M$, optimal precision 
within these constraints can be obtained by targeting not only the 
ground state, but also a selected number of states of the Krylov sequence. 
While a first approach is to take arbitrarily the first $n$ states 
generated (say, 5 to 10) at equal weight, the approximate eigenbasis 
formulation gives direct access to the relative importance of the 
vectors. The importance of a Lanczos vector $\ket{f_{n}}$ is given by,
writing $\ket{A} = \hat{A}\ket{0}$,
\begin{equation}
    \lambda_{n} = \sum_{m} |\langle A | m \rangle|^2 |\langle 
    m | f_{n}\rangle|^2 ,
    \label{eq:weightoftarget}
\end{equation}    
which assesses the contribution the vector makes to an approximative 
eigenstate $\ket{m}$, weighted by that eigenstate's contribution 
to the Green's function. The density matrix may then be constructed 
from the ground state and a number of Lanczos vectors weighted 
according to Eq.\ (\ref{eq:weightoftarget}).
The weight distribution indicates the 
applicability of the Lanczos vector approach: for the spin-1 
Heisenberg chain at $q=\pi$, where there is a one-magnon band at 
$\omega=\Delta=0.41J$, three target states carry 98.9 \% of total 
weight, for the spin-1/2 Heisenberg chain at $q=\pi$ with a two-spinon 
continuum for $0\leq\omega\leq \pi J$, the first three target states 
carry only 28.0 \%, indicating severe problems with precision \cite{Kuhn99}. 

There is currently no precise rule to assess the dilemma of wanting 
to target as many states as possible while retaining sufficient precision
in the description of each single one.
        
\begin{figure}
\centering\epsfig{file=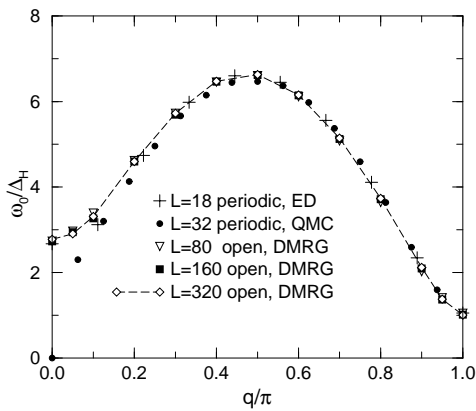,angle=270,scale=0.32}
\caption{Single magnon line of the $S=1$ Heisenberg AFM from exact 
diagonalization, quantum Monte Carlo and DMRG for various system 
sizes and boundary conditions. From \citet{Kuhn99}. Reprinted with 
permission.}
\label{fig:singlemagnon_spin1}
\end{figure}

As an example for the excellent performance of this method, one may 
consider the isotropic spin-1 Heisenberg chain, where the single 
magnon line is shown in Fig.\ \ref{fig:singlemagnon_spin1}. Exact 
diagonalization, quantum Monte Carlo and DMRG are in excellent 
agreement, with the exception of the region $q\rightarrow 0$, where 
the single-magnon band forms only the bottom of a magnon continuum. 
Here Lanczos vector dynamics does not correctly reproduce the 
$2\Delta$ gap at $q=0$, which is much better resolved by quantum Monte 
Carlo. 
        
\begin{figure}
\centering\epsfig{file=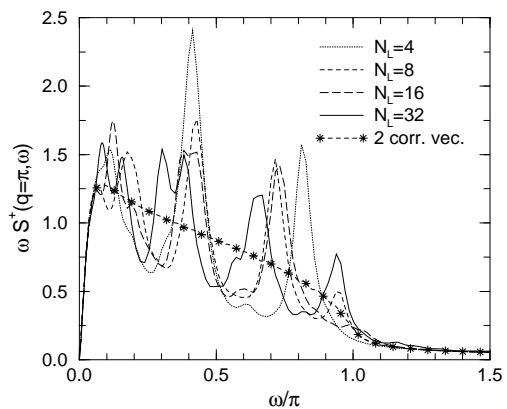,angle=270,scale=0.32}
\caption{Spectral weight $S^+(q=\pi,\omega)$ 
of the $S=1/2$ Heisenberg AFM from Lanczos 
vector and correction vector DMRG. $N_{L}$ indicates the number of 
target states; $M=256$. Note that spectral weight times $\omega$ is 
shown. From \citet{Kuhn99}. Reprinted with permission.}
\label{fig:singlemagnon_spin12}
\end{figure}

The intuition that excitation continua are badly approximated by 
a sum over some $O(10^2)$ effective excited states is further 
corroborated by considering the spectral weight function $S^+(q=\pi,\omega)$ 
[use $A=S^+$ in Eq.\ (\ref{eq:etalimit})] for a 
spin-1/2 Heisenberg antiferromagnet. As shown in Fig.\ 
\ref{fig:singlemagnon_spin12}, Lanczos vector dynamics roughly 
catches the right spectral weight, including the $1/\omega$ 
divergence, as can be seen from the essentially exact correction 
vector curve, but no convergent behavior can be observed 
upon an increase of the number of targeted vectors. The very fast 
Lanczos vector method is thus certainly useful to get a quick 
overview of spectra, but not suited to detailed quantitative 
calculations of excitation continua, only excitation bands. 
Nevertheless, this method has been applied successfully to the 
$S=\frac{1}{2}$ antiferromagnetic Heisenberg chain \cite{Hall95}, the 
spin-boson model \cite{Nish99a}, the Holstein model \cite{Zhan99a}, and 
spin-orbital chains in external fields \cite{Yu01}. \citet{Okun01} 
have used it to extract spin chain dispersion relations. 
\citet{Garc04} have used continued-fraction techniques to provide a 
self-consistent impurity solver for dynamical mean-field calculations 
\cite{Metz89,Geor96}.

\subsection{Correction vector dynamics}
\label{subsec:correctiondynamics}
Even before the advent of DMRG, another way to obtaining more precise 
spectral functions had been proposed by \citet{Soos89}; it was 
first applied using DMRG by \citet{Rama97} and \citet{Kuhn99}. After 
preselection of a fixed frequency $\omega$ one may introduce a 
{\em correction vector} 
\begin{equation}
    \ket{c(\omega+\imag\eta)} = \frac{1}{E_{0}+\omega+\imag\eta-\ham}\hat{A} \ket{0} ,
    \label{eq:correctionvector}
\end{equation}
which, if known, allows for trivial calculation of the Green's 
function and hence the spectral function at this particular frequency:
\begin{equation}
    G_{A}(\omega+\imag\eta)=\langle A \ket{c(\omega+\imag\eta)} .
    \label{eq:Greensfunctionfromcorrection}
\end{equation}    
The correction vector itself is obtained by solving the large sparse linear 
equation system given by
\begin{equation}
(E_{0}+\omega+\imag\eta-\ham)\ket{c(\omega+\imag\eta)} =   \hat{A} \ket{0} .
\label{eq:sparseequation}
\end{equation}
To actually solve this nonhermitean equation system, the current 
procedure is to split the correction vector into real and imaginary 
part, to solve the hermitean equation for the imaginary part and 
exploit the relationship to the real part: 
\begin{eqnarray}
    & & [(E_{0}+\omega-\ham)^2+\eta^2] \impart \ket{c(\omega+\imag\eta)} =  -\eta 
    \hat{A} \ket{0} \\
    & & \repart \ket{c(\omega+\imag\eta)} = \frac{\ham-E_{0}-\omega}{\eta} 
    \impart \ket{c(\omega+\imag\eta)}
    \label{eq:correctionvectorsystem}
\end{eqnarray}    
The standard method to solve a large sparse linear equation system
is the conjugate-gradient method \cite{Golu97}, which effectively generates a Krylov
space as does the Lanczos algorithm. The main implementation work in this method 
is to provide $\ham^2\impart \ket{c}$. Two remarks 
are in order. The reduced basis representation of $\ham^2$ is obtained by squaring the effective Hamiltonian 
generated by DMRG. This approximation is found to work extremely 
well as long as both real and imaginary part of the correction vector 
are included as target vectors: While the real part is not needed for 
the evaluation of spectral functions, $(E_{0}+\omega-\ham)\impart \ket{c} 
\sim \repart \ket{c}$ due to Eq.\ (\ref{eq:correctionvectorsystem}); and
targeting $\repart \ket{c}$ ensures minimal truncation errors in  
$\ham\impart \ket{c}$.
The fundamental drawback of using a 
squared Hamiltonian is that for all iterative eigenvalue or equation solvers
the speed of convergence is determined by the matrix condition number
which drastically deteriorates by the squaring of a matrix.
Many schemes 
are available to improve the convergence of conjugate-gradient methods,
usually based on providing the solution to some related, but trivial
equation system, such as that formed from the diagonal elements of 
the large sparse matrix \cite{Golu97}. 

In the simplest form of the correction vector method, the density matrix is 
formed from targeting four states, $\ket{0}$, 
$\hat{A}\ket{0}$, $\impart \ket{c(\omega+\imag\eta)}$ and
$\repart \ket{c(\omega+\imag\eta)}$.

\begin{figure}
\centering\epsfig{file=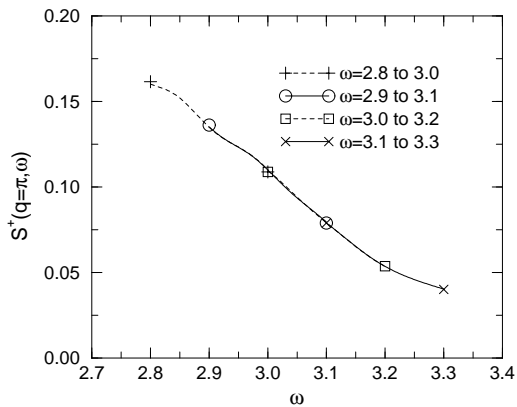,angle=270,scale=0.32}
\caption{Spectral weight of the $S=1/2$ Heisenberg 
AFM from correction vector DMRG. $M=256$ states kept. 
Spectral weights have been calculated for 
$\omega$-intervals starting from various anchoring frequencies for 
the correction vector. From \citet{Kuhn99}. Reprinted with permission.}
\label{fig:correctionvector}
\end{figure}

As has been shown by \citet{Kuhn99}, it is not necessary to 
calculate a very dense set of correction vectors in $\omega$-space to 
obtain the spectral function for an entire frequency interval.
Assuming that the finite 
convergence factor $\eta$ ensures that an entire range of energies of
width $\approx \eta$ is described quite well by the correction 
vector, they have applied the Lanczos vector method as detailed in 
the last section to $\hat{A}\ket{0}$, using the 
effective Hamiltonian obtained by also targeting the correction vector, 
to obtain the spectral function in some
interval around their anchor value for $\omega$. Comparing the results 
for some $\omega$ obtained by starting from neighboring anchoring 
frequencies allows for excellent convergence checks (Fig.\ 
\ref{fig:correctionvector}). In fact, they 
found that best results are obtained for a two-correction vector 
approach where two correction vectors are calculated and targeted for 
two frequencies $\omega_{1}$, $\omega_{2}=\omega_{1}+\Delta \omega$ 
and the spectral function is obtained for the interval 
$[\omega_{1},\omega_{2}]$. This method is, for example, able to provide 
a high precision result for the spinon continuum in the $S=1/2$ 
Heisenberg chain where standard Lanczos dynamics fails (Fig. 
\ref{fig:singlemagnon_spin12}). 

Reducing the broadening factor $\eta$, it is also possible to resolve 
finite-system peaks in the spectral function, obtaining to some 
approximation both location and weight of the Green's function's poles. 

The correction vector method has been applied to determine the 
nonlinear optical coefficients of Hubbard chains and derived models 
by \citet{Pati99b}; \citet{Kuhn00} have extracted the ac conductivity 
of the Bose-Hubbard model with nearest neighbor interactions. 
\citet{Raas03} have used it to study the dynamic correlations of 
single-impurity Anderson models and were able to resolve sharp 
dominant resonances at high energies, using optimized algorithms for 
the matrix inversion needed to obtain the correction vector.
\subsection{Dynamical DMRG}
\label{subsec:dynamicalDMRG}
A further important refinement of DMRG dynamics is obtained by a 
reformulation of the correction vector method in terms of a 
minimization principle, which has been called
``dynamical DMRG'' \cite{Jeck02}. While the fundamental approach remains 
unchanged, the large sparse equation system is replaced by a 
minimization of the functional 
\begin{eqnarray}
    & & W_{A,\eta}(\omega,\psi) = \label{eq:minfunctional} \\
    & & \bra{\psi} (E_{0}+\omega-\ham)^2 + \eta^2 \ket{\psi} +
    \eta\langle A | \psi\rangle +
    \eta\langle \psi | A\rangle . \nonumber
\end{eqnarray}
At the minimum, the minimizing state is
\begin{equation}
    \ket{\psi_{\text{min}}} = \impart \ket{c(\omega+\imag\eta)} .
\end{equation}
Even more importantly, the value of the functional itself is 
\begin{equation}
     W_{A,\eta}(\omega,\psi) = -\pi\eta C_{A} (\omega+\imag\eta) ,
\end{equation}
such that for the calculation of the spectral function it is not 
necessary to explicitly use the correction vector. The huge advantage 
is that if the correction vector is known to some precision 
$\epsilon$ (which will be essentially identical for the equation 
solver and the minimizer), the value of the functional itself is by 
general properties of variational methods known to the much higher precision 
$\epsilon^2$. Hence the DMRG algorithm is essentially implemented as 
in the correction vector method, with the same target vectors, until 
they converge under sweeping, but with the minimization of 
$W_{A,\eta}(\omega,\psi)$ replacing the sparse equation solver.
Results that are obtained for a sequence of $\omega_{i}$ may then be 
extended to other $\omega$ by suitable interpolation \cite{Jeck02}, 
also exploiting first derivatives of spectral functions numerically directly accessible at 
the anchor points.

For dynamical quantities, there are no strict statements on convergence. 
However, convergence for large $M$ seems to be monotonic, with $C_{A}$ 
typically underestimated. 

The high-quality numerical data obtained from dynamical DMRG in 
fact allow for an extrapolation to the thermodynamic limit. As pointed 
out by \citet{Jeck02}, the double limit 
\begin{equation}
    C_{A}(\omega) = \lim_{\eta\rightarrow 0} \lim_{L\rightarrow\infty} 
    C_{A}(L;\omega+\imag\eta) ,
    \label{eq:doubleextrapolation}
\end{equation}
where limits may not be interchanged and which is very hard to take 
numerically, may be taken as a single limit
\begin{equation}
    C_{A}(\omega) = \lim_{\eta(L)\rightarrow 0}  
    C_{A}(L;\omega+\imag\eta(L)) ,
    \label{eq:singleextrapolation}
\end{equation}
provided that $\eta(L)\rightarrow 0$ as $L\rightarrow\infty$ is chosen such
that the finiteness of the system is not visible for the chosen 
$\eta(L)$ and it thus seems to be in the thermodynamic limit from a 
level spacing perspective. This implies that $\eta(L) > \delta\omega(L)$, 
the maximum level spacing of the finite system around energy $\omega$.
For a typical tight-binding Hamiltonian such as the Hubbard model one finds 
\begin{equation}
    \eta(L) \geq \frac{c}{L} ,
\end{equation}
where $c$ is the bandwidth (and in such Hamiltonians also a measure 
of propagation velocity). The key argument is now that if $\eta(L)$ is 
chosen according to that prescription the scaling of numerical results 
broadened by $\eta(L)$ is the same as for some thermodynamic limit 
form known or conjectured analytically subject to Lorentzian 
broadening using the same $\eta$. From Lorentzian broadening of model 
spectra one can then show that a local $\delta$-peak in an 
otherwide continuous spectrum with weight $C^0$ scales as 
$C^0/\pi\eta(L)$, and that a power-law divergence 
$(\omega-\omega_{0})^{-\alpha}$ at a band edge is signalled 
by a scaling as $\eta(L)^{-\alpha}$. More model spectra are discussed 
by \citet{Jeck02}.

Dynamical DMRG has been extensively used to study the spectrum and 
the optical conductivity of the extended Hubbard 
model \cite{Jeck00,Essl01,Jeck03,Kim03}. The optical conductivity may be 
calculated as
\begin{equation}
    \sigma_{1}(\omega) = - \frac{1}{L\omega} \lim_{\eta\rightarrow 0}
    \impart G_{J}(\omega+\imag\eta),
    \label{eq:opticalcondfromcurrent}
\end{equation}
with the current operator
$ \hat{J}= -\imag e \sum_{i\sigma} t_{i} (\ecr{i}\ean{i+1} - \ecr{i+1}\ean{i}) $.
Another possibility is to use
\begin{equation}
    \sigma_{1}(\omega) = - \frac{1}{L\omega} \lim_{\eta\rightarrow 0}
    \impart \left[(\omega+\imag\eta) G_{D}(\omega+\imag\eta) \right],
    \label{eq:opticalcondfromdipole}
\end{equation}
with the dipole operator $\hat{D}= -e \sum_{i} i (n_{i}-1)$.

\begin{figure}
\centering\epsfig{file=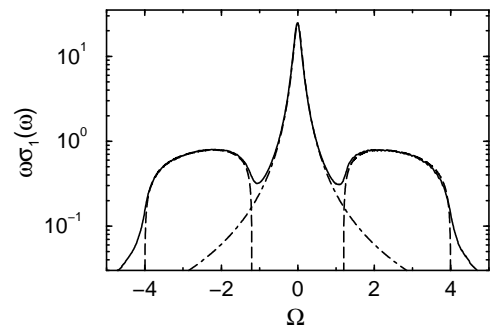,scale=0.4}
\caption{Optical conductivity for a Peierls-Hubbard model in the $U\gg 
t$ limit, $L=128$, $\eta=0.1$, $\Omega=\omega -U$: dynamical DMRG (solid) vs. 
broadened exact $\delta$-contribution (dot-dashed) and unbroadened 
thermodynamic limit bands (dashed). Note log-linear scale. From \citet{Jeck02}.
Reprinted with permission.}
\label{fig:strongcoup}
\end{figure}

In the noninteracting limit $U=0$ the optical conductivity is 
nonvanishing in a band $2|\Delta| < \omega < 4t$ with square-root 
edge divergences. Dynamical DMRG reproduces all features of the exact 
solution including the divergence quantitatively. It is even possible 
to extract the degree of the singularity. Moving to the strongly 
interacting case $U\gg t$, one expects two continuous bands due to the 
dimerization at $2|\Delta|\leq |\omega-U| \leq 4t$ and the Hubbard 
resonance at $\omega=U$. Fig.\ \ref{fig:strongcoup} which compares 
to analytical solutions shows both that 
(up to broadening at the edges) the bands are reproduced with 
excellent quality and that the $\delta$-singularity is measured with 
extremely high precision. Similarly, very precise spectral functions 
for the Hubbard model away from half-filling have been obtained 
\cite{Bent04}. \citet{Nish03a} have shown 
that dynamical DMRG may also be applied to impurity problems such as 
the flat-band single-impurity Anderson model upon suitable 
discretization of the band. Recently, \citet{Nish04} have extended 
this to a precise calculation 
of the Green's function of a single-impurity Anderson model with 
arbitrary band, thereby providing a high-quality self-consistent impurity solver 
needed for dynamical mean-field calculations \cite{Metz89,Geor96}.


\section{BOSONS AND DMRG}
\label{sec:many}

So far, our discussion of DMRG applications has been largely 
restricted to quantum spin and electronic systems, which are 
characterized by a fixed, usually small number of degrees of freedom 
per site. As algorithmic performance relies heavily on $N_{\text{site}}$ 
small (formally $O(N_{\text{site}}^3)$, in practice rather 
$O(N_{\text{site}}^2)$), one may wonder whether DMRG is applicable to
bosonic degrees of freedom with $N_{\text{site}}=\infty$. Such bosonic 
degrees of freedom occur for example in the Bose-Hubbard model,
\begin{equation}
    \ham_{BH}=-t \sum_{i} \bcr{i+1}\ban{i} + \bcr{i}\ban{i+1} + 
    \frac{U}{2} \sum_{i} n_{i}(n_{i}-1),
    \label{eq:bosehubbard}
\end{equation}
that has come to the forefront of reseach due to the realization of a 
tunable quantum phase transition from a Mott-insulating to a superfluid
phase in ultracold bosonic atomic gases in optical lattices 
\cite{Grei02}. Another model of interest is the
Holstein model, where electrons (also spinless fermions 
or XY spins) couple to local (quantum) phonons that react 
dynamically and are not a priori in some adiabatic limit,
\begin{eqnarray}
    \ham_{\text{Hol}}& =& -t \sum_{i} \ecr{i+1}\ean{i} + 
    \ecr{i}\ean{i+1} \label{eq:Holstein} \\
    & & -
    \gamma \sum_{i} (\bcr{i}+\ban{i})(n_{i}-1) + \omega \sum_{i} 
    (\bcr{i}\ban{i}+ 1/2). \nonumber
\end{eqnarray}    
It models electrons in a vibrating lattice and opens the way to polaron 
physics. Yet another model is the spin-boson model, 
which models the dissipative coupling of a two-state system to a thermal 
reservoir given by bosonic oscillators:
\begin{equation}
    \ham_{SB}=-\frac{\Delta}{2}\sigma^x + \sum_{i} \omega_{i} 
    \bcr{i}\ban{i} +
    \frac{\sigma^z}{2} \sum_{i} f_{i} (\bcr{i}+\ban{i}) .
    \label{eq:spinboson}
\end{equation}

\subsection{Moderate number of degrees of freedom}
\label{sec:moderate}
The simplest conceptual approach is to arbitrarily truncate the local 
state space to some $N_{\text{max}}$ for the bosonic degrees of freedom, and 
to check for DMRG convergence both in $M$ and in $N_{\text{max}}$. This 
approach has been very successful in the context of the Bose-Hubbard 
model where the onsite Coulomb repulsion $U$ suppresses large 
occupation numbers. It has been used for the standard Bose-Hubbard 
model \cite{Kuhn98,Kuhn00}, with a random potential \cite{Raps99}, in a 
parabolic potential due to a magnetic trap for cold atoms 
\cite{Koll03} and with non-hermitian hopping and a pinning impurity 
to model superconductor flux lines \cite{Hofs03}. Typically, allowing 
for about three to five times the average occupation is sufficient as 
$N_{\text{max}}$. Similarly, the physics of the fluctuation
of confined membranes \cite{Nish03,Nish02a,Nish02b} and quantum strings
\cite{Nish01a} necessitated the introduction of a larger number of 
local vibrational states.
Other applications that are more problematic have been 
to phonons, both with \cite{Caro96,Maur00,Maur01} and without 
\cite{Caro97} coupling to magnetic or fermionic 
degrees of freedom. While they are believed to be reliable to give a generic 
impression of physical phenomena, for more precise studies more 
advanced techniques are necessary (compare results of \citet{Caro96} and 
of \citet{Burs99a}).

\subsection{Large number of degrees of freedom}
\label{sec:large}
Essentially three approaches have been taken to reduce the large, 
possibly divergent number of states per site to a small number 
manageable by DMRG.

\citet{Burs99a} has proposed a so-called {\em four block approach} that is 
particularly suited to periodic boundary conditions and is a mixture 
of Wilson numerical renormalization group and DMRG ideas. 
Starting from 4 initial blocks of size 1 with $M$ states (this may be 
a relatively large number of electronic and phononic degrees of 
freedom) forming a ring, one solves for the ground state of that $M^4$ state 
problem; density matrices are then formed to project out two blocks, 
and form a new block of double size with $M^2$ states, which are 
truncated down to $M$ using the density-matrix information. From 4 of 
these blocks, a new 4 block ring is built, leading to a doubling of 
system size at every step. Calculations may be simplified beyond the 
usual time-saving techniques by reducing 
the number of $M^4$ product states to some smaller number of states
for which the product of their weights in the density matrices is in excess 
of some very small $\epsilon$, of the order of $10^{-25}$ to 
$10^{-10}$. Translation operators applied to the resulting ground 
state of the 4 blocks on the ring allow the explicit construction of 
specific momentum eigenstates, in particular for $k=0$ and $k=\pi$. 
For both an XY-spin \cite{Burs99a} and isotropic spin \cite{Burs99b} variety of 
the Holstein model, this method allows to trace out very precisely a 
Kosterlitz-Thouless phase transition from a quasi-long-ranged 
antiferromagnet for small spin-phonon coupling to a dimerized, 
gapped phase. As Kosterlitz-Thouless 
phase transitions exhibit an exponentially slow opening of the gap
\cite{Okam92}, the exact localization of the transition by analyzing the gap 
is problematic. It is rather convenient to apply the level 
spectroscopy technique \cite{Nomu94} which locates the phase transition 
by a small-system finite-size extrapolation of the crossing points $g^*$ of the lowest 
lying excitation of different symmetries, including $k=0$ and $k=\pi$ 
states that are easily accessible in the 4 block approach. It was 
found that in the neighborhood of the phase transition, roughly 30 
phonon states were sufficient to obtain 
converged results. A similar scenario of a KT transition was obtained 
for spinless fermions in the Holstein model \cite{Burs98a}.  

\begin{figure}
\centering\epsfig{file=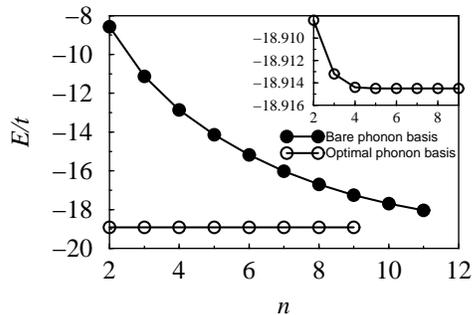,scale=0.75}
\caption{Optimal vs.\ bare phonon states: ground state energy 
convergence for a 4-site Holstein model at half-filling vs.\ number of 
(optimal, bare) phonon states kept. Taken from \citet{Zhan98c}.}
\label{fig:optimalbare}
\end{figure}

An approach more in the spirit of DMRG, the so-called {\em local state 
reduction}, was introduced by \citet{Zhan98c}. While it can
also be combined with exact diagonalization, I want 
to formulate it in a DMRG setup. Assuming a chain with 
fermionic and a small number of bosonic degrees of freedom on each 
site, one of the sites is chosen as ``big site'' to which a 
further number of bare bosonic degrees of freedom is added. Within a DMRG calculation, a density 
matrix is formed for the big site to truncate the number of degrees 
of freedom down to some fixed number of ``optimal'' degrees of freedom. 
This procedure is repeated throughout the lattice in the finite-system algorithm,
sweeping for convergence and for adding further bosonic degrees 
of freedom. The standard prediction algorithm makes the calculation 
quite fast. Physical quantities are then extracted within the optimal   
phononic state space. 
As can be seen from Fig.\ \ref{fig:optimalbare}, merely keeping 
two or three optimal states, in which high-lying bare states have 
non-negligible weight, may be as efficient as keeping of the 
order of hundred bare states. 
This approach has 
allowed to demonstrate the strong effect 
of quantum lattice fluctuations in trans-polyacetylene 
\cite{Barf02}. Combined with Lanczos vector dynamics, very precise 
dynamical susceptibilities have been extracted for spin-boson models 
\cite{Nish99a}. Extensions of the method are found in \citet{Frie00} and 
\citet{Fehs00}.

\citet{Jeck98a} have devised a further approach where $2^n$ bosonic 
degrees of freedom are embodied by $n$ {\em fermionic pseudosites}: writing 
the number of the bosonic degree of freedom as a binary number, the 
degree of freedom is encoded by empty and full fermionic pseudosites 
as dictated by the binary number. All operators on the bosonic 
degrees of freedom can now be translated into (rather complicated) 
operators on the fermionic pseudosites. Finite-system DMRG is then applied 
to the resulting Hamiltonian. They have been able to study polaronic 
self-trapping of electrons in the Holstein model for up to 128 phonon 
states and have located very precisely the metal-insulator transition 
in this model \cite{Jeck99}.  

\section{TWO-DIMENSIONAL QUANTUM SYSTEMS}
\label{sec:twodim}

Not least since the spectacular discovery of high-$T_{c}$ 
superconductivity related to CuO$_{2}$ planes, there has been a 
strong focus on two-dimensional quantum systems. Early on, 
it was suggested that the Hubbard model or 
the $t$-$J$ model away from half-filling might be simple yet powerful 
enough to capture essential features of high-$T_{c}$ 
superconductivity. The analytical study of these quantum systems is 
plagued by similar problems as the one-dimensional case, as in many 
effective field theories $d=2$ is the lower critical dimension and 
few exact results are available. Numerically, exact 
diagonalization methods are even more restricted in two dimensions 
and quantum Monte Carlo is difficult to use for fermionic models away 
from half-filling due to the negative sign problem. Can DMRG help?

The first step in any two-dimensional application of DMRG is to identify blocks and 
sites in order to apply the strategies devised in the one-dimensional 
case. Assuming nearest-neighbor interactions on a square 
lattice one might organize columns of sites into one supersite, such 
that blocks are built from supersites, sweeping through the system 
horizontally. While this approach maintains short-ranged interactions, 
it must fail for any two-dimensional strip of appreciable width $L$ as 
the number of states per supersite grows exponentially as 
$N^L_{\text{site}}$ and is 
only useful for narrow ladder systems.

\begin{figure}
\centering\epsfig{file=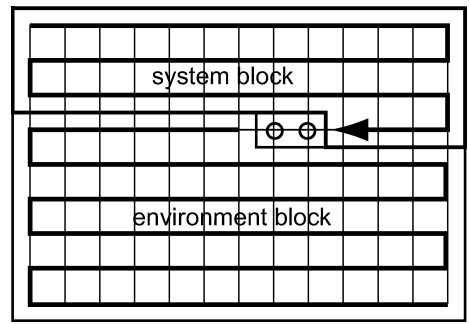,width=0.7\linewidth}
\caption{Standard reorganization of a two-dimensional lattice as a 
zig-zag one-dimensional chain with long-ranged interactions for DMRG 
treatment. Typical blocks during a finite-system DMRG application are 
shown.}
\label{fig:zigzag}
\end{figure}

The standard approach (Noack, White, and Scalapino in \citet{Land94}, 
\citet{Lian94}, \citet{Whit96b}), known as {\em multichain 
approach} is to define a suitable 
one-dimensional path over all sites of the two-dimensional lattice, 
such as the configuration shown in Fig.\ \ref{fig:zigzag}. 

One may now apply standard one-dimensional finite-system DMRG at the price of 
long-ranged interactions of range $2L$ both within and between blocks,
as indicated in Fig.\ \ref{fig:zigzag}.
An inherent difficulty is the preparation of 
blocks and operators for application of the finite-system algorithm, as 
there is no precursor infinite-system DMRG run in some sequence of 
smaller systems possible in this setup. Few compositions of smaller 
blocks in this scheme resemble the final system at all. While 
\citet{Lian94} have simply switched off all non-nearest neighbor 
interactions in the mapped one-dimensional system and applied 
standard infinite-system DMRG in order to switch on all interactions in 
finite-system runs, one can also grow blocks
in a standard infinite-system DMRG run, where some very short exactly 
solvable system are used as environment such that the entire superblock 
remains treatable. This and 
similar warmup procedures will generate starting states for 
finite-system DMRG far from anything physically realistic. As finite-system DMRG 
provides only sequential local updates to that state, there is no 
guarantee that the system does not get trapped in some local minimum 
state. That such trappings do exist is seen from frequent observations that 
seemingly converged systems may, after many further sweeps,
suddenly experience major ``ground state'' energy drops into 
some new (possibly) ground state. 

White and coworkers have followed the approach to exploit local 
trappings by theorizing ahead of using DMRG on possible ground state 
types using other analytical or numerical techniques and to force 
these types of states onto the system by the application of suitable local 
magnetic fields and chemical potentials. These external fields are 
then switched off and the convergence behavior of the competing 
proposed states under finite-system DMRG observed. This procedure 
generates a set of states each of which corresponds to a local energy 
minimum. The associated  
physical properties may be measured and compared (see \citet{Whit98b} 
and \citet{Whit00c} for a discussion). In the multichain approach, the 
two-dimensional Heisenberg model \cite{Whit96b}, the two-dimensional
$t$-$J$-model \cite{Whit98b,Whit98d,Whit00c,Kamp01,Cher03} with 
particular emphasis on stripe formation, and the 6-leg ladder Hubbard model 
\cite{Whit03} have been studied. 
 
Among competing setups \cite{Croo98,Hene99,Farn03,Mouk03a}, 
\citet{Xian01}  have set up a two-dimensional algorithm that 
uses a true DMRG calculation all along and builds 
$L\times L$ systems from previously generated $(L-1)\times(L-1)$ 
systems while keeping the lattice structure intact. It can be applied 
to all lattices that can be arranged to be square with suitable 
interactions: e.g., a triangular lattice is a square lattice 
with additional next-nearest neighbor interactions along one diagonal 
in each square. 

\begin{figure}
\centering\epsfig{file=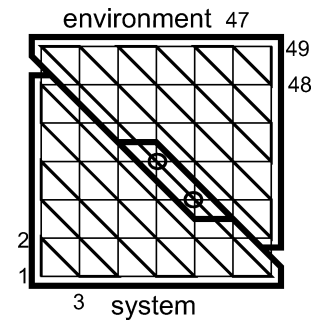,width=0.45\linewidth}    
\caption{Diagonal reorganization of a two-dimensional lattice as used 
by \citet{Xian01}. Typical blocks during a finite-system DMRG application are 
shown.}
\label{fig:diagonal}
\end{figure}

\begin{figure}
\centering\epsfig{file=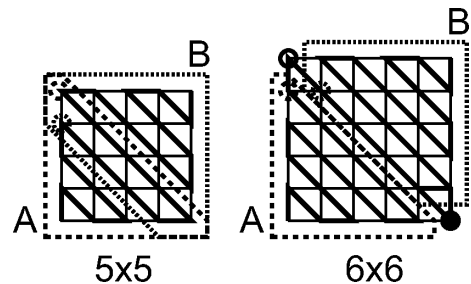,width=0.7\linewidth}
\caption{Composition of blocks from a $(L-1)\times(L-1)$ system and 2 free sites 
into a $L\times L$ system as used 
by \citet{Xian01}. The fat solid line indicates the one-dimensional 
path through the lattice. The lattice subsets surrounded by the broken 
lines are blocks A and B; the last added sites are indicated by broken circles.
Note that blocks overlap for the smaller lattice and are ``pulled 
apart'' for the big lattice. The open and full circles stand for the 
new sites added.}
\label{fig:combine}
\end{figure}

A one-dimensional path is organized as shown in Fig.\ 
\ref{fig:diagonal}, where the pair of free site may be zipped through the
square along the path using the standard finite-system algorithm, 
yielding arbitrary block sizes. In 
particular, one may obtain triangular upper or lower blocks as shown 
in Fig.\ \ref{fig:combine}. Combining these blocks from a $(L-1)\times(L-1)$ 
system and adding two free sites at the corners, one arrives at a 
$L\times L$ system. Here, the upper left free site can be zipped to be a 
neighbor of the lower right free site, as it sits next to active 
block ends (i.e.\ the ends where new sites are added). The pair of free sites can now be zipped through the 
system to yield the desired triangular blocks for the step 
$L\rightarrow L+1$. 

Both for two-dimensional square and triangular $S=1/2$ Heisenberg models
this approach systematically yields lower energies than the multichain 
approach with the exception 
of very small systems. Even for a relatively modest number of states kept ($M=300$) 
\citet{Xian01} report thermodynamic limit extrapolations in good 
agreement with large-scale quantum Monte Carlo simulations: For the square 
lattice, they find $E_{\infty}=-0.3346$ versus a QMC result 
$E_{\infty}=-0.334719(3)$ \cite{Sand97}. The potential of this 
approach seems far from exploited at the moment.

So far, I have been tacitly assuming that interactions are of the 
same order along both spatial directions. Experimentally, a relatively 
frequent situation is that one-dimensional quantum systems 
are weakly coupled along a second axis. 
At high enough temperatures this interaction will washed out, 
but at sufficiently low temperatures there will be a crossover 
from one- to two-dimensional behavior. Such systems may be studied 
by attempting a precise one-dimensional description and introducing 
the weak interchain interaction on the mean field or some other 
perturbative level. \citet{Mouk03a} and \citet{Mouk03b} have used DMRG  
in a similar spirit by solving a one-dimensional system using 
standard DMRG with $M$ states, and determining the $M'$ lowest-lying states 
for the 
superblock. Chain lengths are chosen such that the lowest lying 
excitation of the finite chain is down to the order of the interchain 
coupling. These states are taken to be the states of a ``site'', 
which are combined to a new effective Hamiltonian, that 
then is once again treated using DMRG. Results for weakly 
coupled spin chains are in good agreement with quantum Monte 
Carlo results, however $M'$ is severely limited to currently several tens.

A severe limitation on two-dimensional DMRG is provided by 
the exponential growth of $M$ with $L$ for a pre\-se\-lec\-ted truncated 
weight or ground state precision (Sec.\ \ref{subsec:density}). For 
frustrated and fermionic systems beyond the very small exact 
diagonalization range ($6\times 6$ for $S=1/2$ spins) DMRG may yet 
be the method of choice as quantum Monte 
Carlo suffers from the negative sign problem and even medium-sized 
fermionic systems of size, say, $12\times 12$ sites would be
most useful; in models with non-abelian symmetries, the 
implementation of the associated good quantum numbers has been 
shown to reduce drastically the truncation error for a given number 
of states \cite{McCu00,McCu01a,McCu01b,McCu02}. 

Very recently, \citet{Vers04b} have proposed a new approach to 
two-dimensional quantum systems combining a generalization of their matrix 
product formulation of DMRG \cite{Vers04} and imaginary-time evolution 
\cite{Vers04a}, discussed in Sec.\ \ref{subsec:mpa} and Sec.\ 
\ref{subsec:time}. One-dimensional 
matrix-product state are formed from matrix products of $M\times M$ matrices 
$(A[\sigma])_{\alpha\beta}$, with $M$-dimensional auxiliary state 
spaces on the bond to the left and right of each site. 
The two-dimensional generalization for a square lattice is given by the contraction of tensors
$(A[\sigma])_{\alpha\beta\gamma\delta}$ with $M$-dimensional 
auxiliary state spaces on the four adjacent bonds. Finite-temperature 
and ground states are calculated by imaginary-time evolution of a 
totally mixed state along the lines of \citet{Vers04a}. 
As tensorial contractions lead (unlike in the case of the ansatz 
matrices $A[\sigma]$ appearing in one dimension) to a proliferation 
of indices on resulting tensors, a suitable truncation scheme is 
needed as described by \citet{Vers04b}.

An appealing feature of this approach is that the entropy of entanglement for a cut through 
a system of size $L\times L$ will scale, for fixed $M$, linearly in 
system size as it should generically. The errors in energy seem to 
decrease exponentially in $M$. $M$ is currently very small (up to 5), 
as the algorithm scales badly in $M$; however, as $M$ variational 
parameters are available on {\em each} bond, even a small $M$ corresponds to 
large variational spaces. 
At the time of writing, it is too early to assess the potential of 
this new approach; however, current results and its conceptual clarity make it 
seem very promising to me.
\section{BEYOND REAL SPACE LATTICES}
\label{sec:momentum}
In this section, I shall consider three groups of DMRG applications 
that seem to 
have very little in common at first sight: the study of 
translationally invariant
low-dimensional models in momentum space, 
high-precision quantum chemistry calculations for small molecules,
and studies of small grains and nuclei.
However, from a DMRG point of view, all share 
fundamental properties. Let me first discuss momentum-space DMRG, move 
on to quantum chemistry and finish by considering small grain and 
nuclear physics.
\subsection{Momentum-space DMRG}
\label{subsec:momentum}
Real-space DMRG precision dramatically deteriorates when applied 
to long-ranged interactions or hoppings. Moreover, momentum is not 
accessible as good quantum number in real-space DMRG. Momentum-space DMRG, on 
the other hand, makes momentum a good quantum number, works naturally 
with periodic boundary conditions, and allows trivial manipulation of 
the single-particle dispersion relation. Momentum-space DMRG has already yielded 
highly satisfying dispersion relations and momentum distributions 
in excellent agreement with 
analytical results \cite{Nish02}.

Consider for definiteness a Hubbard model with local
interaction $U$, but long-ranged hopping $t_{ij}$,
\begin{equation}
    \ham_{\text{LR}} = \sum_{ij\sigma} t_{ij} \ecr{i}\ean{j} + U \sum_{i}
    n_{i\uparrow} n_{i\downarrow}
    \label{eq:genhubbard}
\end{equation}    
in arbitrary dimensions; site $i$ is localized at ${\bf 
r}_{i}$ and $t_{ij}=t({\bf r}_{i}-{\bf r}_{j})$. With $L$ lattice 
sites, Fourier transformations
\begin{equation}
    \ecr{{\bf k}} = \frac{1}{\sqrt{L}} \sum_{j} e^{\imag {\bf k} \cdot 
    {\bf r}_{j}} \ecr{j}
    \label{eq:FTops}
\end{equation}
yield the band structure
\begin{equation}
    \epsilon ({\bf k}) = \sum_{{\bf r}} e^{-\imag {\bf k}\cdot{\bf r}} 
    t({\bf r})
    \label{eq:bandstructure}
\end{equation}
and the Hamiltonian
\begin{equation}
    \ham = \sum_{{\bf k},\sigma} \epsilon({\bf k}) \ecr{{\bf k}} 
    \ean{{\bf k}} + \frac{U}{L} 
    \sum_{{\bf k}_{1},{\bf k}_{2},{\bf k}_{3}}
    \eucr{{\bf k}_{1}} \edcr{{\bf k}_{2}} \edan{{\bf k}_{3}} 
    \euan{{\bf k}_{1}+{\bf k}_{2}-{\bf k}_{3}} . 
    \label{eq:momentumHubbard}
\end{equation}    
The reciprocal lattice points ${\bf k}$ are now taken to be ``sites''. 
Note that $L$ is {\em not} the linear size of the lattice, but is 
taken to be the total lattice size (because of the typical DMRG mapping to 
one dimension).

The key idea \cite{Xian96} is to arrange these ``sites'' into 
a one-dimensional chain with long-ranged interactions which is treated by the 
real-space finite-system DMRG \cite{Nish02,Lege03a}. 
In addition to the conventional good 
quantum numbers, states will also be classified by total momentum, 
which allows a further substantial segmentation of Hilbert space (by a 
factor of order $L$), hence for the same number of states kept, 
momentum-space DMRG is much faster than real-space DMRG.

To obtain an efficient implementation, several 
issues have to be addressed.

(i) In momentum space there is a huge proliferation of interaction 
terms of $O(L^3)$, that have to be generated, stored and applied to wave functions efficiently.

(ii) For an application of the finite-system DMRG algorithm, we need 
to provide blocks and operators living on these blocks for all block sizes. 
In real space, they are naturally generated during the infinite-system 
stage of the algorithm. In momentum space, another ``warm-up'' 
procedure must be found.

(iii) On a one-dimensional lattice with short-ranged interactions it 
is natural to group sites as they are arranged in real space. In 
momentum space with long-ranged interactions, there is no obvious
``site'' sequence, even in the one-dimensional case. Does the 
arrangement affect convergence properties, is there an optimal 
arrangement? 

Let us discuss these points in the following.

(i) {\em DMRG operator representation for general two-body interaction 
Hamiltonians.} In principle we should like to use DMRG to treat the generic 
Hamiltonian
\begin{equation}
\ham_{\text{two-body}}=\sum_{ij} T_{ij} \fcr{i}\fan{j} +
\sum_{ijkl} V_{ijkl} \fcr{i}\fcr{j}\fan{k}\fan{l} ,
    \label{eq:generic}
\end{equation}  
where $T_{ij}$ encodes kinetic energy and one-body potentials and 
$V_{ijkl}$ a generic two-body (Coulomb) interaction; for simplicity, 
we assume spin contained in the orbital indices. In the worst-case 
scenario, all $V_{ijkl}$ are different, although symmetries and 
model properties may yield decisive simplifications. In momentum space, 
for example, 
$V_{ijkl}\neq 0$ for one $l$ only once $ijk$ is fixed due to momentum 
conservation.

As for operators living on the same block
\begin{equation}
    \bra{m} \fcr{i}\fan{j} \ket{\tilde{m}} \neq \sum_{m'} 
    \bra{m} \fcr{i} \ket{m'}
    \bra{m'} \fan{j} \ket{\tilde{m}} ,
    \label{eq:wrongequation}
\end{equation}    
all operator pairings have to be stored separately. Leaving aside the 
simpler case where one or two of the four operators live on the 
single sites in DMRG, and assuming that they are all either in block S 
or E, this suggests that for $L$ ``sites'' of the order of $O(L^4)$ 
operators have to be stored. As their form changes for each of the 
$L$ possible block sizes, the total memory consumption would be 
$O(L^5 M^2)$ on disk for all blocks and $O(L^4 M^2)$ in RAM for the current block. 
At the same time, for each of the $L$ steps of a sweep, calculation 
time would be of order $O(L^4 M^3)$, or $O(L^5 M^3)$ for the entire 
calculation.

Memory consumption as well as the associated calculation time 
can however be reduced drastically \cite{Xian96}. Let us consider 
the three possible operator distributions on blocks S and E.

(a) Four operators in one block (4/0): Terms $V_{ijkl} \fcr{i}\fcr{j}\fan{k}\fan{l}$
are absorbed into a single block Hamiltonian operator during block 
growth. Assuming $i$, $j$, $k$ are in the previous block and site $l$ 
is added to form the current block, a representation of 
$\fcr{i}\fcr{j}\fan{k}$ in the previous block basis allows to form
$V_{ijkl} \times \fcr{i}\fcr{j}\fan{k} \times \fan{l}$ in the 
block plus site product basis, which is then transformed into the basis 
of the current block and added into the single block Hamiltonian 
operator. For $L$ blocks, $O(L^3)$ representations each of $\fcr{i}\fcr{j}\fan{k}$
are necessary. These, in turn can be compounded into {\em 
complementary operators} 
\begin{equation}
    O_{l} = \sum_{ijk} V_{ijkl} \fcr{i}\fcr{j}\fan{k} ,
    \label{eq:complementaryoperator}
\end{equation}
so that
\begin{equation}
    \sum_{ijkl} V_{ijkl} \fcr{i}\fcr{j}\fan{k}\fan{l} \rightarrow
    \sum_{l} O_{l} \fan{l} .
    \label{eq:complementarycontraction}
\end{equation}    
The complementary operators can be constructed as discussed in Sec.\ 
\ref{subsec:correlations}, assuming the knowledge 
of two-operator terms $\fcr{i}\fcr{j}$. For $L$ blocks, $O(L^2)$ of 
those exist, leading to memory consumption $O(L^3 M^2)$ on disk and
$O(L^2 M^2)$ in RAM. 

(b) Three operators in a block (3/1): One applies the strategy of (\ref{eq:complementaryoperator}) 
and (\ref{eq:complementarycontraction}), with $O_{l}$ and $\fan{l}$ acting on different 
blocks.

(c) Two operators in a block (2/2): Again, the complementary operator 
technique can be applied, with the modification that each 
complementary operator living on block S has now two matching 
operators in E. A further class of complementary operators
\begin{equation}
    O_{kl} = \sum_{ij} V_{ijkl} \fcr{i}\fcr{j} 
    \label{eq:complementaryoperator2}
\end{equation}
allows the simplification
\begin{equation}
    \sum_{ijkl} V_{ijkl} \fcr{i}\fcr{j}\fan{k}\fan{l} \rightarrow
    \sum_{kl} O_{kl} \fan{k}\fan{l} .
    \label{eq:complementarycontraction2}
\end{equation}    
Memory consumption for the second type of complementary operator is
$O(L^3 M^2)$ on disk and $O(L^2 M^2)$ in RAM. Taking all operator 
combinations together, global memory consumption is to leading order 
$O(L^3 M^2)$ on disk and $O(L^2 M^2)$ in RAM, 
which is a reduction by $L^2$ compared to the 
naive estimate. In momentum space, due to momentum conservation, 
memory consumption is reduced by another factor of $L$ to
$O(L^2 M^2)$ on disk and $O(L M^2)$ in RAM.

Using the complementary operator technique, calculation times are 
dominated by the (2/2) terms. In analogy to the 
construction of (4/0) terms ``on the fly'' by generating the new terms of the 
sum, transforming them and adding them into the operator, the 
$O_{kl}$ can be constructed at a computational expense of $O(L^3 M^2)$
for generating the $L$ new terms to be added to each of the $L^2$ 
complementary operators with $M^2$ matrix elements each and
$O(L^2 M^3)$ for transforming the $L^2$ operators into the current 
basis. Using the multiplication technique of Sec.\ \ref{subsec:large}, 
multiplying the Hamiltonian to the state vector costs 
$O(L^2 M^3)$ time at each step or $O(L^3 M^3)$ per sweep. Global 
calculation time per sweep is thus $O(L^3 M^3)+O(L^4 M^2)$, a reduction 
by $L^2$ for the dominant first term (typically, $M \gg L$ for the 
relevant DMRG applications). 

(ii) {\em Setting up finite-system blocks.} The standard practise currently 
adopted in momentum space \cite{Xian96,Nish02} is to use standard 
Wilsonian renormalization \cite{Wils75,Wils83}: for the chosen sequence of momenta, blocks 
are grown linearly (as in infinite-system DMRG), but diagonalized 
without superblock embedding and the lowest-energy states retained, 
with the important modification that it should be ensured that at 
least one or two states are retained for each relevant Hilbert space 
sector, i.e.\ all sectors within a certain spread about the average 
momentum, particle number and magnetization. Otherwise DMRG may miss 
the right ground state, as certain sectors that make an important 
contribution to the ground state may not be constructed during the 
finite-system sweep. Left and right block
sectors must be chosen such that all of them find a partner in the 
other block to combine to the desired total momentum, particle number 
and magnetization. Moreover, it is of advantage to choose 
corresponding states in different sectors such 
that obvious symmetries are not broken, e.g.\ $S^z \leftrightarrow -S^z$ if the 
final state is known to be in the $S^z_{tot}=0$ sector; it has also 
been found that choosing initial block sizes too small may result in 
much slower convergence or trapping in wrong minima \cite{Lege03a}.

(iii) {\em Determining the order of momentum ``sites''.} \citet{Nish02}
report that for short-ranged Hubbard models ordering the levels 
according to increasing distance to the Fermi energy of the non-interacting 
limit, $|\epsilon({\bf k})-\epsilon_{F}|$, thus grouping levels that 
have strong scattering together, works best, whereas for 
longer-ranged Hubbard models a simple ordering according to 
$\epsilon({\bf k})$ works best. Similar results have been 
reported by \citet{Lege03a}, who could demonstrate trapping in wrong 
minima for inadequate orderings; \citet{Lege03b} have provided 
a quantum information based method to avoid such orderings.  

\citet{Nish02} have carried out extensive convergence studies for 
Hubbard models in 1D and 2D with various hoppings.
Convergence seems to be dominated by the ratio of interaction to 
bandwidth, $U/W$ rather than some $U/t$.
The momentum-space DMRG algorithm becomes exact in the 
$U/W\rightarrow 0$ limit; convergence deteriorates drastically with 
$U/W$, but is acceptable for $U/W \leq 1$, which is $U\leq 8t$ for the 
2D Hubbard model with nearest-neighbor hopping.
Generally it seems that for momentum-space DMRG convergence depends
only weakly on the 
range of hopping as opposed to real-space DMRG. Momentum-space DMRG is worst at 
half-filling and somewhat deteriorates with dimension (if calculations 
for the same physical scale $U/W$ are compared). In two dimensions, 
for moderate $U$, momentum-space DMRG is more efficient than real 
space DMRG.

{\em Extrapolation schemes.} As can be seen from the results of 
\citet{Nish02}, even for 
values of $M$ as large as several 1000, convergence is not achieved. 
Due to the complex growth scheme, it cannot be taken for granted even 
after many sweeps that the environmental error has been eliminated; 
as in two-dimensional real-space DMRG, sudden drops in energy are 
observed. In their application, \citet{Nish02} report that for 
fixed $M$ a fit formula in $1/M$, 
\begin{equation}
    E_{fit} \left( \frac{1}{M} \right) =
    E_{\infty} + \frac{a_{1}}{M}+ \frac{a_{2}}{M^2}+ \ldots,
    \label{eq:fitform}
\end{equation}
works very well and improve results by over an order 
of magnitude even for final $M=2000$. Building on the 
proportionality between truncated weight and energy error for 
eliminated environmental error,
\citet{Lege03a} have proposed to vary $M$ to maintain a 
fixed truncated weight (``dynamical block selection scheme'').
They find that in that approach 
\begin{equation}
    \frac{E(\epsilon_\rho)-E_{\text{exact}}}{E_{\text{exact}}} = La \epsilon_\rho
    \label{eq:extrapolinrho}
\end{equation}
is very well satisfied for $a\approx 1$, where $\epsilon_{\rho}$ is the 
truncated weight. A further advantage 
of that approach is that the number of states needed for a certain 
precision varies widely in momentum-space DMRG such that for many DMRG steps 
important savings in calculation time can be realized. 

All in all, momentum-space DMRG seems a promising alternative to 
real-space DMRG, in particular for longer-ranged interactions (for 
short-ranged interactions in one dimension, real-space DMRG remains 
the method of choice). 
However, momentum-space DMRG presents additional complications (optimal ordering of 
levels, efficient encoding) that may not have been solved optimally yet.

\subsection{Quantum chemistry}
\label{subsec:chemistry}
A field where DMRG will make increasingly important 
contributions in the next few years is quantum chemistry. 
While the first quantum-chemical DMRG calculations on cyclic polyene 
by \citet{Fano98} and polyacetylene by \citet{Bend99} were still 
very much in the spirit of extended Hubbard models, more recent work 
has moved on to calculations in generic bases with arbitrary 
interactions. Let me situate this latter kind of DMRG applications in 
the general context of quantum chemistry.

Within the 
framework of the Born-Oppenheimer approximation, i.e.\ fixed nuclear 
positions, two major ways of determining the electronic properties of 
molecules are given by Hartree-Fock (HF) and post-HF calculations and 
density functional theory (DFT). DFT is computationally rather 
inexpensive and well-suited to quick and quite reliable studies of 
medium-sized molecules, but not overly precise in particular for 
small molecules. Here, Hartree-Fock calculations, incorporating Fermi 
statistics exactly and electron-electron interactions on a mean-field 
level provide good (initial) results and a starting point for 
further refinement, so-called post-HF calculations. Quantum chemistry 
DMRG is one such post-HF calculation.

For the HF and post-HF calculations, one starts from a 
first-quantized Schr\"{o}dinger equation (in atomic units) for the full molecule 
with $N$ electrons; second quantization is 
achieved by introducing some suitably chosen 
basis set $\{ \ket{\varphi_{i}} \}$. In this way, a general two-body interaction 
Hamiltonian as in Eq.\ (\ref{eq:generic}) can be derived.

A HF calculation now solves this Hamiltonian at mean-field level, and it 
can be reexpressed in terms of the HF orbitals. A closed-shell 
singlet ground state is then simply given by 
\begin{equation}
    \ket{\text{HF}} = \eucr{1}\edcr{1}\ldots\eucr{N/2}\edcr{N/2}\ket{0} .
    \label{eq:HFgroundstate}
\end{equation}

Within standard quantum chemistry post-HF calculations, one improves 
now on this ground state by a plethora of methods. Simple approaches
consist in diagonalizing in the subspace of one- or two-particle 
excited states; others choose a certain subset of electrons 
and diagonalize the Hamiltonian fully in one subset of orbitals for 
those (complete active space). Taking all electrons (excitations) and all 
orbitals into account, one obtains, within the originally chosen basis 
set, the exact many-body solution, known as the full-CI (configuration 
interaction) solution. While the approximate schemes typically scale 
as $N^5$ to $N^7$ in the number of orbitals, full-CI solutions demand exponential
effort and are available only for some very small molecules.

From the DMRG point of view, determining the ground state of the full 
second-quantized Hamiltonian is formally equivalent to an application 
of the momentum-space DMRG, with some additional complications.
As momentum conservation does not hold for $V_{ijkl}$, 
there are $O(N)$ more operators. Moreover, the basis set must be chosen 
carefully from a quantum chemist's point of view. As in momentum 
space DMRG, a good sequence of the strongly different orbitals has to be 
established for DMRG treatment such that convergence is optimized.
Also, an initial setup of blocks must be provided, including operators.

Orbital ordering in quantum chemistry turns out to be crucial to the 
performance of the method; a badly chosen ordering may lead to much 
slower convergence or even trapping in some higher energy local 
minimum. The simplest information available, the HF energy and occupation of 
the orbitals, has been exploited by 
\citet{Whit99}, \citet{Daul00b}, and \citet{Mitr01,Mitr03} 
to enforce various orderings of 
the orbitals.

\citet{Chan02} have proposed to use the symmetric reverse 
Cuthill-McKee (CMK) reordering \cite{Cuth69,Liu75}, 
where orbitals are reordered such 
that $T_{ij}$ matrix elements above a certain threshold are as 
band-diagonal as possible. This reduces effective 
interaction range, leading to faster convergence in $M$ and reduced 
sweep number, as confirmed by \citet{Lege03c}, who optimize the 
position of the Fock term $T_{ij}+\sum_{\text{k occupied}} (4V_{ikkj}-2V_{ikjk})$
and report reductions in $M$ by a third or so. 

Another approach is given by exploiting the short range of chemical 
interactions which in the basis of typically delocalized HF orbitals 
leads to a large effective range of interactions detrimental to DMRG. This may 
be reduced by changing to more localized orbitals. Orbital localization 
techniques have first been studied by \citet{Daul00b} who applied a 
localization procedure 
\cite{Pipe89} to the unoccupied orbitals to avoid increases in the 
starting configuration energy, but did not find remarkable improvement 
over orderings of the unmodified HF orbitals. 

Quantum information techniques have been used by \citet{Lege03b} both 
for momentum space and quantum chemistry DMRG to devise optimal 
orderings. Studying various {\em ad hoc} orderings, they find fastest and 
most stable convergence under sweeping for those orderings that 
maximize entanglement entropy [Eq.\ (\ref{eq:vonNeumannentropy})] 
between orbitals and the rest of the chain
for orbitals at the chain center; orderings are hence chosen such that strongly 
entangled orbitals are brought to the chain center (often those closest to the 
Fermi surface). This ordering can be 
constructed iteratively starting from a first guess as obtained by the 
CMK ordering. 

Once the orbital ordering has been selected, current applications 
follow a variety of ``warm-up'' procedures to build all sizes of 
blocks and to ensure that block states 
contain as many states as possible relevant for the final result. 
This may be done by augmenting the block 
being built by a toy environment block with some low energy states that provides 
an environment compatible with the block, i.e.\ leading to the desired 
total quantum numbers \cite{Chan02}, doing infinite-system DMRG on left 
and right blocks while targeting also excited states \cite{Mitr01}, 
and defining a minimum number of states to be kept even if the non-zero 
eigenvalues of the density matrix do not provide enough states 
\cite{Lege03a}. Block state choice can also be guided by entanglement entropy 
calculations \cite{Lege03b}.   

Various molecules have by now been studied using DMRG, both for 
equilibrium and out-of equilibrium configurations, ground states and
excitations.
H$_{2}$O has been serving as a benchmark in many quantum chemistry 
calculations and has been 
studied within various finite one-particle bases, more recently in 
the so-called DZP basis \cite{Baus86} with 8 electrons in 25 orbitals (and 2 
``frozen'' electrons in the $1s$ oxygen orbital) and the larger TZ2P basis 
\cite{Widm90} of 10 electrons in 41 orbitals. While the larger basis 
will generally yield lower energies, one can compare how close various 
approximations come to an exact diagonalization (full configuration 
interaction (CI) -- if at all possible) solution within one particular basis set.
In the smaller DZP basis, the exact ground state energy 
is at $-76,256634$H (Hartree). While the Hartree-Fock solution is several hundred mH 
above, and the single and doubles configuration interaction solution 
about ten mH too high, and various coupled cluster approximations 
\cite{Bart90} are reaching chemical accuracy of 1 mH with errors 
between 4.1 mH and 0.2 mH, DMRG results of various groups 
\cite{Whit99,Daul00b,Chan02} are better than 0.2 mH at $M\sim 400$, 
reaching an error of about 0.004 mH at $M\sim 900$ 
\cite{Chan02,Lege03a}. Moving to the larger TZ2P basis \cite{Chan03}, 
where there is 
no exact solution, accuracies of all methods are ranking similarly, 
with DMRG outperforming the best coupled cluster results with an error 
of 0.019 mH at $M\sim 3000$ and reaching an error of 0.005 mH at 
$M \sim 5000$, with the extrapolated result being $-76.314715$H. The 
comparison is even more favorable if the nuclei are arranged away 
from their equilibrium configuration.

Excellent 
performance is also observed in the extraction of dissociation energy 
curves for N$_{2}$ \cite{Mitr01} and water \cite{Mitr03}.       

Low-lying excited states have been studied for HHeH \cite{Daul00b}, 
CH$_{2}$ \cite{Lege03a} 
and LiF \cite{Lege03c}; the latter case has provided a test system to 
study how efficiently methods map out the ionic-neutral crossover of 
this alcali-halogenide for increasing bond length. With relatively 
modest numerical effort ($M$ not in excess of 500 even at the 
computationally most expensive avoided crossing) relative errors 
of $10^{-9}$, $10^{-7}$, and $10^{-3}$ for the ground state energy, 
first excited state energy and the dipole moment compared to the full CI solution
\cite{Baus88} have been obtained. The dipole moment precision 
increases by orders of magnitude away from the avoided crossing.

To assess the further potential of quantum chemistry DMRG for larger 
molecules and $N$ orbital bases, it should 
be kept in mind that the generic $N^4$ scaling of the algorithm - 
which does compare extremely favorably with competing methods that 
scale as $N^5$ to $N^7$ - is modified by several factors. On the one 
hand, for a given desired error in energy, various authors report, as 
for conventional DMRG, that $\delta E \sim \epsilon_{\rho}$, the 
truncated weight, which scales (Sec.\ \ref{subsec:density})
as $\epsilon_{\rho} = \exp (-k \ln^2 M)$. $k$, 
however, shrinks with size in a model-dependent fashion, the most 
favorable case being that orbitals have essentially chain-like 
interactions with a few ``nearest neighbor'' orbitals only. 
It may be more realistic to think about molecular orbitals 
as arranged on a quasi one-dimensional strip of some width $L$ 
related to the number of locally interacting orbitals; in 
that case, for standard strip geometries, the constant has been 
found to scale as $L^{-1}$. So $N^4$ scaling is in my view overly 
optimistic. On the other hand, on larger molecules orbital localization 
techniques may be more powerful than on the small molecules studied so 
far, such that orbital interactions become much more sparse and 
scaling may actually improve. 

One other possible way of improving the scaling properties of quantum 
chemistry DMRG might be the {\em canonical diagonalization approach} 
proposed by \citet{Whit02a}, which attempts to transform away 
numerically by a sequence of canonical basis transformations 
as many of the non-diagonal matrix elements of the 
second-quantized Hamiltonian $\ham$ of Eq.\ (\ref{eq:generic}) as 
possible such that entire orbitals can be removed from $\ham$, 
resulting in a new, smaller quantum chemistry problem in a different 
basis set, which may then be attacked by some DMRG technique such as 
those outlined above. The removed orbitals, which are no 
longer identical with the original ones, are typically strongly 
overlapping with the energetically very high-lying and low-lying orbitals of 
the original problem that are almost filled or empty. 
Of course, these transformations cannot be 
carried out for the exponentially large number of Hilbert space 
states. \citet{Whit02a} moves to the HF basis and carries out a 
particle-hole transformation; the canonical transformations are then 
carried out in the much smaller space of states not annihilated by 
one of the $V_{ijkl}$ {\em and} formed by the minimum of creation 
operators from the HF vacuum (ground state). For example, the term
$V_{ijkl} d^{\dagger}_{i}d_{j}d_{k}d_{l}$, where the $d^{\dagger}_{i},d_{i}$ are 
particle-hole transformed fermionic operators, implies the 
consideration of the ``left'' state $d^{\dagger}_{i}\ket{0}$ with HF 
energy $E_{l}$ and the ``right'' state  $d^{\dagger}_{j}d^{\dagger}_{k}d^{\dagger}_{m}\ket{0}$
with HF energy $E_{r}$.
In this smaller state space, a sequence of canonical basis transformations 
may be implemented as a differential equation as originally proposed by 
\citet{Wegn94} as {\em flow equation method} and \citet{Glaz94}
as {\em similarity renormalization}: with $A$ some antihermitian 
operator, a formal time-dependence is introduced to the unitary 
transformation $\ham(t)=\exp(tA)\ham(0)\exp(-tA)$, where $\ham(0)$ is 
the original Hamiltonian of Eq.\ (\ref{eq:generic}) expressed in the 
HF basis. The 
corresponding differential equation
\begin{equation}
    \frac{d\ham(t)}{dt} = [A,\ham(t)]
    \label{eq:conttrafo}
\end{equation}
is modified by making $A$ time-dependent itself. One now expands
\begin{equation}
    \ham(t)=\sum_{\alpha} a_{\alpha}(t)h_{\alpha}
    \label{eq:hamexpan}
\end{equation}    
and 
\begin{equation}
    A(t) = \sum_{\alpha} s_{\alpha}a_{\alpha}(t)h_{\alpha},
\end{equation}
where each $h_{\alpha}$ is the product of creation and annihilation 
operators and each $s_{\alpha}$ some constant yet to be chosen under 
the constraint of the anti-hermiticity of $A$. 
To avoid operator algebra, White introduces 
\begin{equation}
    [h_{\alpha},h_{\beta}] = \sum_{\gamma} B^\gamma_{\alpha\beta} 
    h_{\gamma},
\end{equation} 
where additional contributions to the commutator that cannot 
be expressed by one of the operator terms in Eq.\ (\ref{eq:hamexpan}) 
are suppressed; this eliminates three-body interactions generically generated by 
the transformation and not present in the original Hamiltonian. 
Then Eq.\ (\ref{eq:conttrafo}), with $A$ time-dependent, reduces 
to a set of differential equations 
\begin{equation}
    \frac{da_{\gamma}(t)}{dt} = \sum_{\alpha\beta} 
    B^\gamma_{\alpha\beta} s_{\alpha} a_{\alpha}(t) a_{\beta}(t) ,
\end{equation}
that can now be integrated numerically up to some time $t$. It can now 
be shown that the goal of diminishing $a_{\alpha}(t)$ for all 
non-diagonal contributions to $\ham$ can be achieved efficiently by setting
\begin{equation}
    s_{\alpha} = (E_{l}-E_{r})^{-1},
\end{equation}
where $E_{l}$ and $E_{r}$ are the HF energies of the left and right 
HF orbitals in $h_{\alpha}$. Stopping after some finite time $t$, one 
can now remove orbitals (e.g.\ with the lowest occupancy, i.e.\ 
almost full or empty before the particle-hole transformation) 
and use DMRG for the full many-body problem in the reduced orbital 
space. \citet{Whit02a} finds 
this approach to yield very good results as compared to full-CI 
calculations for a stretched water molecule, but as of now the 
full potential of the method remains unexplored.

To conclude, in my opinion, DMRG has been established as a 
competitive, essentially full-CI quality method in quantum 
chemistry which should be taken seriously. 
However, efficient geometrical optimization techniques to 
determine lowest-energy molecular geometries are still lacking.

\subsection{DMRG for small grains and nuclear physics}
\label{subsec:energybasics}
Let us reconsider the generic Hamiltonian of Eq.\ (\ref{eq:generic}),
where energy levels above and below the Fermi energy 
are ordered in ascending fashion and where a 
simple model interaction is chosen. This Hamiltonian can then be 
treated in the following so-called {\em particle-hole} reformulation of 
{\em infinite-system} real-space DMRG \cite{Duke99a,Duke00}: 
Starting from two initial blocks from a small number of states, one (``hole 
block'') from the lowest 
unoccupied one-particle levels above the Fermi energy and the other 
(``particle block'') from the highest 
occupied one-particle levels below, we iteratively 
add further states to particle and hole block, moving away from the 
Fermi surface, determine the 
ground state, and in the usual DMRG procedure calculate density 
matrices and the reduced basis transformations 
by tracing out particle and hole states respectively.
This approach may work if there is a clear physical reason that states 
far away from the Fermi surface will
hardly influence low energy physics and if interactions show ``translational invariance'' in energy 
space; otherwise the infinite-system algorithm should fail. The 
advantage of such simple models is that much larger systems can be 
treated than in quantum chemistry.

A model Hamiltonian which happens to combine both features is the reduced 
BCS Hamiltonian
\begin{equation}
\ham_{BCS}=\sum_{j\sigma}(\epsilon_{j}-\mu)\fcr{j\sigma}\fan{j\sigma} - 
\lambda d \sum_{ij} 
\fcr{i\uparrow}\fcr{i\downarrow}\fan{j\downarrow}\fan{j\uparrow},
\end{equation}
where DMRG has been able to definitely settle 
longstanding questions \cite{Duke99a,Duke00,Duke01} on the nature of 
the breakdown of BCS superconductivity in small grains expected when
the level spacing $d$ in the finite system becomes of the order of the BCS 
gap $\Delta$ \cite{Ande59}. DMRG has conclusively shown that there is a smooth crossover between a 
normal and a superconducting regime in the pairing order parameter and 
other quantities, in contradiction to analytical approaches 
indicating a sharp crossover. 
This approach to DMRG has been extended to the study of the Josephson 
effect between superconducting nanograins with discrete energy levels 
\cite{Gobe03a} and to the observation and calculation 
of well-defined quasi particle excitations in small interacting 
disordered systems with high dimensionless conductance $g$ \cite{Gobe03b}.  

Particle-hole DMRG has also been applied successfully in nuclear 
physics \cite{Duke01,Dimi02,Duke02} in the framework of the nuclear 
shell model, where a nucleus 
is modeled by core orbitals completely filled with neutrons and
protons (``doubly magic core'') and valence orbitals partially filled 
with nucleons. The core is considered inert and, starting from some 
Hartree-Fock level orbital configuration for the ``valence'' nucleons, 
a two-body Hamiltonian for these nucleons is solved using the 
particle-hole method. This is conceptually very similar to the 
post-HF approaches of quantum chemistry, with model interactions for 
the nucleons as they are not so precisely known, such as pairing 
interactions, quadrupolar interactions, and the like.

This approach has been very successful for nucleons in very large 
angular momentum shells interacting through a pairing and a 
quadrupolar force in an oblate nucleus \cite{Duke01}, with up to 20 
particles of angular momentum $j=55/2$, obtaining energies converged 
up to $10^{-6}$ for $M=60$; for 40 nucleons of $j=99/2$, 4 digit 
precision was possible; in this case, 38 \% of energy was contained in the 
correlations \cite{Duke02}. For more realistic calculations of the 
nucleus $^{24}$Mg, with 4 neutrons and 4 protons outside the doubly 
magic $^{16}$O core, convergence in $M$ was so slow that almost the 
complete Hilbert space had to be considered for good convergence 
\cite{Dimi02}. The fundamental drawback of the particle-hole approach 
is that it does not allow for an easy implementation of the finite-system 
algorithm (levels far away from the Fermi surface hardly couple 
to the system, giving little relevant information via the density 
matrix) and that angular momentum conservation is not exploited. 
\citet{Pitt03} are currently aiming at implementing an algorithm for 
nuclei using
angular momentum, circumventing these difficulties (see also 
\citet{Duke04}).

\section{TRANSFER MATRIX DMRG}
\label{sec:transfer}
Conventional DMRG is essentially restricted to $T=0$ 
calculations, with some computationally expensive forays into the very-low 
temperature regimes possible \cite{Mouk96a,Bati98,Hall99}. 
Decisive progress was made by \citet{Nish95} who realized that the DMRG idea
could also be used for the decimation of transfer matrices, leading 
to the name of {\em transfer-matrix renormalization group} (TMRG). This opened 
the way to DMRG studies of classical statistical mechanics at finite 
temperature for systems on two-dimensional $L\times\infty$ strips.  
If one applies the generic mapping of $d$-dimensional quantum 
systems at finite temperature to $(d+1)$-dimensional classical systems,
TMRG also permits study of thermodynamic properties of one-dimensional quantum 
systems at finite temperature.
\subsection{Classical 2D transfer matrix DMRG: TMRG}
\label{subsec:TMRG}
Consider the textbook transfer matrix method for a one-dimensional
classical system with Hamiltonian $\ham$ and $N_{\text{site}}$ states 
$|\sigma_{i}\rangle$ on each site $i$, e.g.\ the Ising model. Assuming
nearest-neighbor interactions, one performs a
local decomposition of the Hamiltonian $\ham=\sum \hat{h}_{i}$, where
each $\hat{h}_i$ contains the interaction between sites $i$ and $i+1$.
Taking the partition function of a system of length $N$ with 
periodic boundary conditions,
\begin{equation} Z_{L} = \mbox{Tr\ } e^{-\beta {\cal H}} =  
\mbox{Tr\ } e^{-\sum_{i=1}^{L}\beta \hat{h}_i},
\end{equation}
and inserting identities $\sum_{\sigma_{i}} |\sigma_{i}\rangle\langle \sigma_{i}|$, 
one obtains for a translationally invariant system
$Z_{L}=\mbox{Tr\ } {\cal T}^{L}$, where the 
$N_{\text{site}} \times N_{\text{site}}$ {\em transfer matrix} ${\cal T}$ 
reads 
\[ \langle \sigma_{i} |{\cal T}|\sigma_{i+1} \rangle 
=\langle \sigma_{i} | e^{-\beta h_{i}} | \sigma_{i+1} \rangle . \]
From this one deduces for $N \rightarrow \infty$ the
free energy per site $f=-k_{B}T \ln \lambda_{1}[{\cal T}]$,
where $\lambda_1$ is the largest eigenvalue of ${\cal T}$, from which 
all thermodynamic properties follow.

\begin{figure}
\centering\epsfig{file=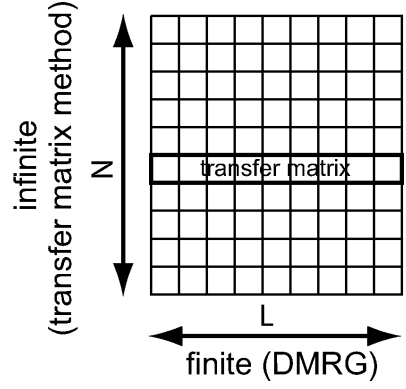,width=0.6\linewidth}
\caption{Strategy for the DMRG treatment of two-dimensional classical systems.}
\label{fig:classical_strategy}
\end{figure}

Moving now to a two-dimensional geometry, on a strip of dimension 
$L\times N$ with a short-ranged Hamiltonian, the transfer matrix method
may be applied in the limit $N\rightarrow\infty$ for some finite 
width $L\ll N$, treating a row of $L$ sites as one big site; transfer matrix
size then grows 
as $N_{\text{site}}^L$, strongly limiting this approach in complete analogy 
to the exact diagonalization of quantum Hamiltonians.  Hence (Fig.\ 
\ref{fig:classical_strategy})
the first dimension, with {\em finite} system length $L$, is attacked
by DMRG applied to the transfer instead of the Hamiltonian 
matrix, to keep transfer matrix size
finite and sufficiently small by decimation, 
while retaining the most
important information.
Results are extrapolated to infinite size. The second dimension, 
with {\em infinite} system length, is accounted for by the 
one-dimensional transfer matrix method.  

To prove that this concept works, i.e.\ that an optimal decimation 
principle can be set up for transfer matrices in analogy to the case made for 
Hamiltonians, consider a short-ranged classical Hamiltonian at $T>0$ 
on a $L \times 
N$ strip, where $N \rightarrow\infty$. We now define an unnormalized 
density-matrix $\dm_{u}=e^{-\beta\ham}$ (in reality an evolution operator) by
\begin{equation}
    \bra{\fat{\tilde{\sigma}}} \dm_{u} \ket{\fat{\sigma}} =
    \bra{\fat{\tilde{\sigma}}} [{\cal T}^{(L)}]^N \ket{\fat{\sigma}} ,
    \label{eq:totaldm}
\end{equation}
where $\fat{\tilde{\sigma}}$ and $\fat{\sigma}$ label states of the $L$ 
sites on top and bottom of the strip. ${\cal T}^{(L)}$ is the band 
transfer matrix of Fig.\  \ref{fig:classical_strategy}. 
The partition function $Z =\text{Tr } \dm_{u}$ is then given as
\begin{equation}
    Z = \sum_{i=1}^{N_{\text{site}}^L} \lambda_{i}^N,
    \label{eq:partitionfortransfer}
\end{equation}
where $\lambda_{1} > \lambda_{2} \geq \ldots$ are the eigenvalues of 
${\cal T}^{(L)}$; the largest being positive and non-degenerate for 
positive Boltzmann weights, 
partition function and unnormalized density-matrix simplify in the thermodynamic 
limit $N\rightarrow\infty$ to
\begin{equation}
    Z = \lambda_{1}^{N} 
    \label{eq:partitioninfinity}
\end{equation}
and
\begin{equation}
    \dm_{u} = \lambda_{1}^{N}\ket{\lambda_{1}}\bra{\lambda_{1}} ,
    \label{eq:densityinfinity}
\end{equation}    
where $\ket{\lambda_{1}}$ is the normalized eigenvector of the largest transfer 
matrix eigenvalue. 

\begin{figure}
\centering\epsfig{file=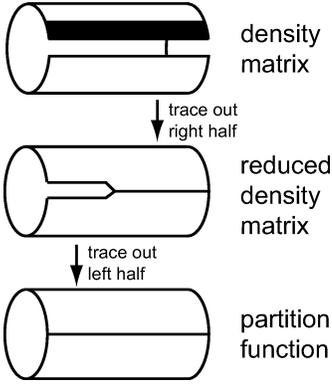,width=0.5\linewidth}
\caption{Pictorial representation of the partition function, the 
(unnormalized) density matrix and the reduced density-matrix for a two-dimensional 
classical system on a strip geometry. The black stripe represents a 
band transfer matrix. Adapted from Nishino in \citet{Pesc99}.}
\label{fig:nishinoontransfer}
\end{figure}

Consider now Fig.\ \ref{fig:nishinoontransfer}. The unnormalized 
density-matrix has
exactly the same form, up to the prefactor, 
of the pure state projector of standard DMRG, with  $\ket{\lambda_{1}}$
assuming the role of the target state there. Tracing out the 
right ``environment'' half, the left ``system'' unnormalized reduced density-matrix is given
by
\begin{equation}
    \dm_{uS}= 
    \text{Tr}_{E}\lambda_{1}^{N}\ket{\lambda_{1}}\bra{\lambda_{1}} = \lambda_{1}^{N} \sum_{\alpha=1}^{N_{\text{site}}^{L/2}} w_{\alpha} 
    \ket{w_{\alpha}}\bra{w_{\alpha}},
    \label{eq:reddmTMRG}
\end{equation}    
where $w_{\alpha}$ are the positive eigenvalues to the normalized eigenvectors
$\ket{w_{\alpha}}$ of the  reduced ``system'' density matrix; 
$\sum_{\alpha}w_{\alpha}=1$. 
Completing the partial trace, one finds
\begin{equation}
    Z = \text{Tr}_{S} \dm_{uS} =  \lambda_{1}^{N} \sum_{\alpha=1}^{N_{\text{site}}^{L/2}} 
    w_{\alpha} ,
    \label{eq:newpartition}
\end{equation}    
such that indeed the best approximation to $Z$ is obtained by 
retaining in a reduced basis the eigenvectors to the largest 
eigenvalues $w_{\alpha}$ as in conventional DMRG.

\begin{figure}
\centering\epsfig{file=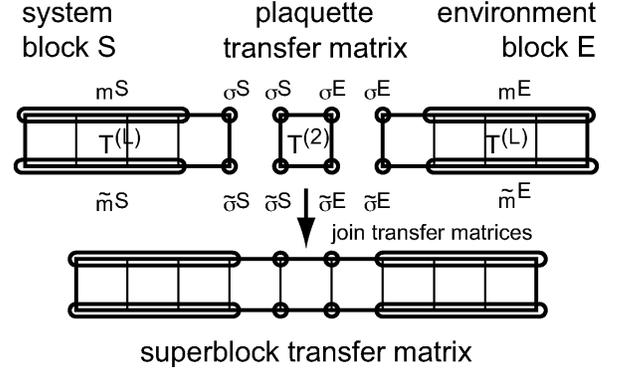,width=0.9\linewidth}
\caption{Transfer matrix DMRG step.}
\label{fig:trans_DMRG_step}
\end{figure}

Let us explain the DMRG transfer-matrix renormalization \cite{Nish95} 
in more detail. Because the transfer matrix factorizes into 
local transfer matrices, some minor modifications of the standard 
DMRG approach are convenient. 
Assume as given a $M \times M$ transfer matrix of a system
of length $L$,
\begin{equation} 
\langle m \sigma | {\cal T}^{(L)} | \tilde{m} \tilde{\sigma} \rangle , 
\end{equation}
where the sites at the active (growth) end are kept explicitly as opposed to 
standard DMRG, and $\ket{m},\ket{\tilde{m}}$ are block states (Fig.\ 
\ref{fig:trans_DMRG_step}). 
Considering now the 
superblock of length $2L$, the transfer matrix reads
\begin{eqnarray}
    & & \bra{m^{S}\sigma^{S}\sigma^{E}m^{E}}
    T^{(2L)} 
    \ket{\tilde{m}^{S}\tilde{\sigma}^{S}\tilde{\sigma}^{E}\tilde{m}^{E}}
    = \nonumber \\
    & & \langle m^{S} \sigma^{S} | {\cal T}^{(L)} | \tilde{m}^{S} 
    \tilde{\sigma}^{S} \rangle
    \times \\
    & & \langle \sigma^{S} \sigma^{E} | {\cal T}^{(2)} |  
    \tilde{\sigma}^{S} \tilde{\sigma}^{E} \rangle
    \langle m^{E} \sigma^{E} | {\cal T}^{(L)} | \tilde{m}^{E} 
    \tilde{\sigma}^{E} 
    \rangle . \nonumber 
\end{eqnarray}

Lanczos diagonalization yields the eigenvector of ${\cal T}^{(2L)}$ to the 
maximum eigenvalue $\lambda_{1}$. Again in some analogy to conventional 
DMRG, the fact that the transfer matrix is a product can be used to 
speed up calculations by decomposing $\ket{\phi} = {\cal T}^{(2L)} 
\ket{\psi}$ into three successive multiplications,
$\ket{\phi} = {\cal T}^{(L)}[{\cal T}^{(2)}({\cal T}^{(L)} 
\ket{\psi})]$.

\begin{figure}
\centering\epsfig{file=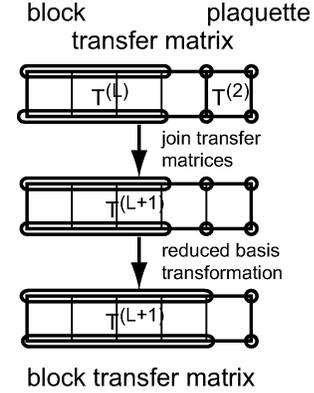,width=0.45\linewidth}
\caption{Construction of the enlarged transfer matrix.}
\label{fig:transmat}
\end{figure}

The obtained eigenvector now yields the reduced density matrices and the 
reduced basis transformations.
The last step is to build the $MN_{\text{site}} \times MN_{\text{site}}$ 
transfer matrix for a system of length $L+1$ 
(Fig.\ \ref{fig:transmat}):
\begin{eqnarray}
& & \langle m\, \sigma | {\cal T}^{(L+1)} | \tilde{m} \tilde{\sigma} \rangle = \\
& & \sum_{n\tilde{n}\tau\tilde{\tau}}
\langle m | n \tau \rangle 
 \langle n\tau | {\cal T}^{(L)} | \tilde{n}\tilde{\tau} \rangle 
\langle \tau\sigma | {\cal T}^{(2)}| \tilde{\tau}\tilde{\sigma} \rangle
\langle \tilde{n}\tilde{\tau} | \tilde{m} \rangle \nonumber .
\end{eqnarray} 

The TMRG procedure is repeated up to the desired final length, 
the free energy then obtained from $\lambda_{1}$; this allows the
calculation of all thermodynamic quantities through numerical 
differentiation of the free energy. To avoid numerically unstable 
second derivatives for specific heat and susceptibility, it is 
convenient to consider the first order derivative of the average 
energy extracted from expectation values such as $\langle \sigma_{i} 
\sigma_{j} \rangle = \bra{\lambda_{1}} \sigma_{i} 
\sigma_{j} \ket{\lambda_{1}}$ obtained by replacements
$e^{-\beta\hat{h}_{i}}\rightarrow \sigma_{i}e^{-\beta\hat{h}_{i}}$
in the above calculations; at the same time, 
correlation lengths may be extracted both from these expectation 
values or from the two leading transfer matrix eigenvalues,
\begin{equation}
    \xi = -1/\ln \repart (\lambda_{2}/\lambda_{1}) .
\label{eq:corrfromlambda}
\end{equation}
To refine results, 
a finite-system calculation can be set up as in conventional DMRG.
A main technical difference between 
conventional DMRG and classical transfer matrix DMRG is the
{\em absence of good quantum numbers}, which simplifies the algorithm, 
but is hard on computational resources.

A large body of works has emerged that actually strongly relies on 
the finite strip-width provided by the TMRG, studying competing bulk 
and surface effects both in two-dimensional Ising models on a 
$L\times\infty$ strip. The confined Ising model has been used to model two 
coexisting phases in the presence of bulk 
and surface fields as well as gravity, in order to model wetting and 
coexistence phenomena and the competition of bulk and surface effects 
in finite geometries. In the case of {\em opposing} surface fields 
favoring phase coexistence at zero bulk field up to some 
temperature that (unintuitively) goes to the wetting temperature for
$L\rightarrow\infty$ \cite{Parr90}, it could be shown that gravity 
along the finite width direction suppresses fluctuations such that it restores 
the bulk critical temperature as limiting temperature for coexistence
\cite{Carl97,Carl98d}. These studies were extended to the competition of surface 
and bulk fields \cite{Drze98}. Further studies elucidated finite-size corrections 
to the Kelvin equation at complete wetting for parallel surface 
and bulk fields \cite{Carl98b}, the nature of coexisting excited 
states \cite{Drze00a}, the scaling of cumulant ratios \cite{Drze00b}, 
and an analysis of the confined Ising model at and near criticality for 
competing surface and bulk fields \cite{Maci01a,Maci01b}. In the case 
of the Potts model with $Q>4$, where there is a first-order phase 
transition in the bulk \cite{Baxt82}, but a second-order phase 
transition at the surface \cite{Lipo82}, TMRG permitted to extract 
surface critical exponents that demonstrate that the universality class of the 
surface transition is $Q$-independent \cite{Iglo99}.

TMRG has also been used to study critical properties of truly 
two-dimensional classical systems 
\cite{Hond97,Tsus97,Carl99a,Sato00}, such as the spin-$3/2$ Ising 
model on a square lattice \cite{Tsus97}, the Ising model with line 
defects \cite{Chun00a}, the three-state chiral clock model as 
example of a commensurate-incommensurate phase transition 
\cite{Sato00}, the 19-vertex model in two dimensions as a discrete 
version of the continuous XY model to study the Kosterlitz-Thouless 
phase transition \cite{Hond97}, and the random exchange coupling 
classical $Q$-state Potts model which was demonstrated to have 
critical properties independent of $Q$ \cite{Carl99a}. \citet{Lay02} 
have used the TMRG to study a continuous Ginzburg-Landau field-theory 
formulation of the two-dimensional Ising model by retaining the 
largest eigenvalue eigenfunctions of bond transfer matrices as state space.

\citet{Gend00} have extended the TMRG to periodic 
boundary conditions to obtain, at the expense of the expected lower 
DMRG precision, much better thermodynamic limit scaling behavior 
for $L\rightarrow\infty$. Critical temperature, 
thermal and magnetic critical exponent are all extracted with modest 
computational effort at a staggering 6 to 7 digit precision compared 
to the Onsager solution.

\subsection{Corner transfer-matrix DMRG}
\label{subsec:CTMRG}
\begin{figure}
\centering\epsfig{file=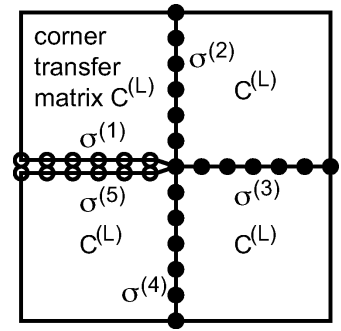,width=0.5\linewidth}
\caption{Corner transfer matrix setup for the calculation of reduced 
density matrices and partition functions in two-dimensional classical 
models.}
\label{fig:ctmsetup}
\end{figure}

Following \citet{Baxt82}, one may conclude that the essential aspect 
of the reduced density-matrix is that it represents a half-cut in the 
setup of Fig.\ \ref{fig:nishinoontransfer} whose spatial extension is to be 
taken to the thermodynamic limit. The same setup can be obtained 
considering the geometry of Fig.\ \ref{fig:ctmsetup}, where four 
{\em corner transfer-matrices (CTM)} are defined as
\begin{equation}
   \bra{\fat{\sigma}'} C^{(L)} \ket{\fat{\sigma}}
   = \sum_{\fat{\sigma}_{\text{int}}} \prod_{\langle ijkl \rangle} 
   \bra{\sigma_{i}\sigma_{j}} {\cal T}^{(2)} 
   \ket{\sigma_{k}\sigma_{l}},
   \label{eq:ctmdefine}
\end{equation}
where the product runs over all site plaquettes and the sum over all 
internal site configurations, i.e.\ the states on the sites linking to
neighboring CTMs are kept fixed. The corner site state is invariant 
under application of the CTM. The unnormalized reduced density-matrix is then 
given by
\begin{eqnarray}
   & & \bra{\fat{\sigma}^{(5)}} \dm_{u} \ket{\fat{\sigma}^{(1)}} =
   \sum_{\fat{\sigma}^{(4)}\fat{\sigma}^{(3)}\fat{\sigma}^{(2)}} 
   \nonumber \\
   & &    \bra{\fat{\sigma}^{(5)}} C^{(L)} \ket{\fat{\sigma}^{(4)}}
          \bra{\fat{\sigma}^{(4)}} C^{(L)} \ket{\fat{\sigma}^{(3)}} \\
   & &    \bra{\fat{\sigma}^{(3)}} C^{(L)} \ket{\fat{\sigma}^{(2)}}
          \bra{\fat{\sigma}^{(2)}} C^{(L)} \ket{\fat{\sigma}^{(1)}} 
          \nonumber
\end{eqnarray}
with the center site fixed. Diagonalizing $C^{(L)}$, 
and obtaining eigenvalues $\lambda_{i}$ and eigenvectors 
$\ket{\lambda_{i}}$, 
the reduced density-matrix $\dm_{u}(L)$ then reads
\begin{equation}
    \dm_{u}(L)=\sum_{i}\lambda_{i}^4 \ket{\lambda_{i}}\bra{\lambda_{i}}
    \label{eq:ctmdensity}
\end{equation}
and the partition function, obtained by tracing out the reduced density 
matrix, is
\begin{equation}
    Z(L) = \sum_{i} \lambda_{i}^4 .
    \label{eq:ctmpartition}
\end{equation}    

\begin{figure}
\centering\epsfig{file=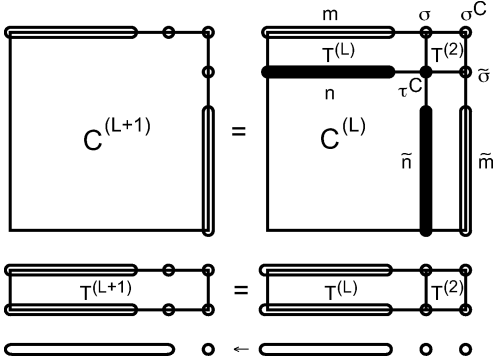,width=0.75\linewidth}
\caption{Corner transfer matrix growth using band and plaquette 
transfer matrices. Solid states are summed over.}
\label{fig:ctmgrowth}
\end{figure}

Following the argument for classical TMRG, \citet{Nish96a} have 
introduced the CTMRG or {\em corner transfer-matrix renormalization 
group}. A sequence of increasingly 
large corner transfer matrices is built (Fig.\ \ref{fig:ctmgrowth})
as
\begin{eqnarray}
    & & \bra{m \sigma \sigma^{C}} C^{(L+1)} 
    \ket{\tilde{m} \tilde{\sigma} \sigma^{C}} = 
    \sum_{n \tilde{n} \tau^{C}} \nonumber \\
    & & \bra{\sigma\sigma^{C}} {\cal T}^{(2)} \ket{\tau^{C}\tilde{\sigma}}  
    \bra{n \tau^{C}} C^{(L)} \ket{\tilde{n} \tau^{C}}\label{eq:ctmgrowth} \\
    & & \bra{m\sigma} {\cal T}^{(L)} \ket{n\tau^{C}} 
    \bra{\tilde{n}\tau^{C}} {\cal T}^{(L)} 
    \ket{\tilde{m}\tilde{\sigma}} \nonumber ,
\end{eqnarray}
where previous reduced basis transformations are supposed to have 
taken place for $C^{(L)}$ and ${\cal T}^{(L)}$. Similarly, ${\cal 
T}^{(L+1)}$ is built as in classical TMRG. Diagonalizing
$C^{(L+1)}$ yields eigenvalues $\lambda_{i}$ and associated 
eigenvectors, the $M$ most important of which define the reduced 
basis transformation. This reduced basis transformation is then 
carried out on ${\cal T}^{(L+1)}$ and $C^{(L+1)}$, completing one 
CTMRG step. The crucial advantage of this algorithm is that it 
completely avoids the diagonalization of some large sparse matrix. 

CTMRG allows to calculate similar quantities as TMRG, all 
thermodynamic variables as some derivative of the free energy or 
using local expectation values. For the two-dimensional Ising model 
\citet{Nish97a} obtained Ising critical exponents at 4 
digit precision, using systems of up to $L=20000$. For the $Q=5$ 
two-dimensional Potts model, which has a very weak first-order 
transition they could determine the latent heat ${\cal L} \approx 0.027$ 
compared to an exact ${\cal L}=0.0265$ \cite{Baxt82}, which is hard to see in 
Monte Carlo simulations due to metastability problems. 
Other applications have considered the 
spin-$3/2$ Ising model \cite{Tsus98} and a vertex model with 7 vertex 
configurations, modeling an order-disorder transition \cite{Taka01}. 
More recent extensions study self-avoiding walk models in two 
dimensions \cite{Fors03a,Fors03b}.

One may also generalize the CTMRG to three-dimensional classical 
models \cite{Nish98a}. While the implementation is cumbersome, the 
idea is simple: the semi-infinite cut line in the plane leading to 4 corner 
matrices gives way to a semi-infinite cut plane in some volume leading 
to 8 corner tensors two of which have an ``open'' side. Growing these 
corner tensors involves concurrent growing of corner matrices, band 
transfer matrices, and plaquette transfer matrices. Numerical results, however, 
indicate clear practical limitations of the concept.

\subsection{Quantum transfer matrix DMRG in 1D}
\label{subsec:quanttransfer}
It is now straightforward -- at least in 
principle -- to extend classical TMRG to the calculation of the 
thermodynamics of one-dimensional quantum 
systems due to the general relationship between $d$-dimensional quantum and 
$(d+1)$-dimensional classical problems. This approach was first 
used by \citet{Burs96} and fully developed by \citet{Wang97} as 
well as by \citet{Shib97b}.
To appreciate the following, it is best to visualize
quantum TMRG as an extension of the quantum transfer-matrix 
method \cite{Bets84,Bets85} in the same
way conventional DMRG extends exact diagonalization.

Consider the mapping of one-dimensional {\em quantum} to two-dimensional
{\em classical} systems, as used in the quantum transfer matrix 
and quantum Monte Carlo \cite{Suzu76} methods: Assume a 
Hamiltonian that contains nearest-neighbor interactions only and 
is invariant under translations 
by an even number of sites.
We now introduce the well-known checkerboard decomposition \cite{Suzu76} 
$\ham=\hat{H}_{1}+\hat{H}_{2}$.
$\hat{H}_{1}=\sum_{i=1}^{N/2} \hat{h}_{2i-1}$ and 
$\hat{H}_{2}=\sum_{i=1}^{N/2} \hat{h}_{2i}$. $\hat{h}_{i}$ is the 
local Hamiltonian 
linking sites $i$ and $i+1$; neighboring local Hamiltonians will in general not 
commute, hence $[\hat{H}_{1},\hat{H}_{2}]\neq 0$. However, all terms 
in $\hat{H}_{1}$ or $\hat{H}_{2}$ commute.
$N$ is the system size in real space (which we will take to infinity).

Following \citet{Trot59}, we consider 
a sequence of approximate partition functions 
\begin{equation}
Z_{L} := \mbox{Tr\ } (e^{-\beta H_{1}/L}e^{-\beta H_{2}/L})^{L},
\label{eq:approxpartition}
\end{equation}
with Trotter number $L$. Then
\begin{equation} 
Z = \lim_{L\rightarrow\infty} Z_{L}
\label{eq:trotter} 
\end{equation}
holds: quantum effects are suppressed as neglected
commutators between local Hamiltonians scale as $(\beta/L)^{2}$.

\begin{figure}
\centering\epsfig{file=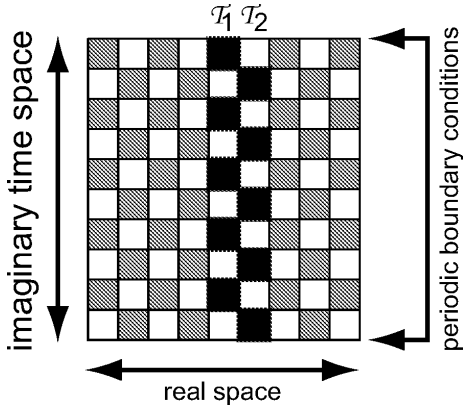,width=0.7\linewidth}
\caption{Checkerboard decomposition: active vs.\ inactive plaquettes 
of the two-dimensional effective classical model.}
\label{fig:checkerboard}
\end{figure}

One now expands $\hat{H}_{1}$ and $\hat{H}_{2}$ in the exponentials of $Z_{L}$ 
in Eq.\ (\ref{eq:approxpartition})
into the sums of local Hamiltonians and inserts $2L+1$ times 
the identity decomposition
\begin{equation}
    \id = \prod_{i=1}^N \sum_{\sigma_{i}} 
    \ket{\sigma_{i}}\bra{\sigma_{i}} ,
    \label{eq:tmrgid}
\end{equation}    
sandwiching each exponential. Introducing an additional label $j$ for the 
identity 
decompositions, we find $(2L+1)\times N$ ``sites'', where each site carries two 
labels, corresponding to two dimensions,
the real space coordinate (subscript) $i$ and the Trotter space (imaginary time)
coordinate (superscript) $j$. Thinking of both coordinates as spatial,
one obtains a two dimensional lattice of dimension $(2L+1) \times N$, 
for which the 
approximate partition function $Z_{L}$ reads:
\begin{eqnarray*}
Z_{L} = \mbox{Tr\ } \prod_{i=1}^{N/2} & & \prod_{j=1}^{L} 
\langle \sigma_{2i-1}^{2j+1} \sigma_{2i}^{2j+1} | e^{-\beta h_{2i-1}/L} |
\sigma_{2i-1}^{2j} \sigma_{2i}^{2j} \rangle  \\
 & &  
\langle \sigma_{2i}^{2j} \sigma_{2i+1}^{2j} | e^{-\beta h_{2i}/L} |
\sigma_{2i}^{2j-1} \sigma_{2i+1}^{2j-1} \rangle 
\label{eq:checkerZ}
\end{eqnarray*}
 Figure \ref{fig:checkerboard} shows that as 
in a checkerboard only every second plaquette of the two-dimensional 
lattice is active (i.e.\ black). To evaluate $Z_{L}$ the trace is 
taken over all local states of the $(2L+1)\times N$ sites, while noting
that the trace in 
Eq.\ (\ref{eq:approxpartition}) enforces
{\em periodic boundary conditions} along the imaginary time 
direction, $\ket{\sigma_{i}^{1}}=\ket{\sigma_{i}^{2L+1}}$. Nothing 
specific needs to be assumed about boundary conditions along the real 
space direction. Note that the orientation of the transfer matrix has 
changed compared to Fig.\ \ref{fig:classical_strategy}.

Working the way backward from the partition function $Z_{L}$, we can now 
identify transfer and density matrices.
Introducing a local transfer-matrix as $\hat{\tau}_{k}=
\exp (-\beta h_{k}/L)$, and exploiting the assumed restricted 
translational invariance, the global transfer matrices
\begin{eqnarray*}
& &\bra{\sigma^{1}\ldots \sigma^{2L+1}}{\cal T}_{1}^{(2L+1)}\ket{\nu^{1}\ldots 
\nu^{2L+1}} = \\ 
& & \prod_{j=1}^{L} \langle \sigma^{2j+1}\nu^{2j+1} | \hat{\tau}_{1} | 
\sigma^{2j}\nu^{2j} \rangle \\
& & \bra{\nu^{1}\ldots \nu^{2L+1}}{\cal T}_{2}^{(2L+1)} \ket{\tau^{1}\ldots 
\tau^{2L+1}} = \\ 
& & \prod_{j=1}^{L} \langle \nu^{2j}\tau^{2j} | \hat{\tau}_{2} | 
\nu^{2j-1}\tau^{2j-1} \rangle 
\end{eqnarray*}
can be defined (see Fig.\ \ref{fig:checkerboard}), representing
odd and even bonds on the chain because of the alternating checkerboard 
decomposition. The local transfer matrices are linking the states of 
two sites at different Trotter times and are evaluated using the 
Trotter-time independent eigenbasis of the local Hamiltonian, 
$\hat{h}\ket{i}=E_{i}\ket{i}$:
\begin{equation}
    e^{-\beta \hat{h}/L} = \sum_{i} e^{-\beta E_{i}/L} \ket{i}\bra{i} .
    \label{eq:localeigenbasis}
\end{equation}    
Summing over all internal degrees of freedom $\ket{\nu^i}$, we obtain matrix 
elements
\begin{equation}
    \bra{\sigma^{1}\ldots \sigma^{2L+1}} {\cal T}_{1}{\cal T}_{2} 
    \ket{\tau^{1}\ldots \tau^{2L+1}} .
\end{equation}    
Using the global transfer matrices, the unnormalized density-matrix reads
\begin{equation}
\dm_{uL}= [{\cal T}_{1}{\cal T}_{2}]^{N/2}
\end{equation}
and
\begin{equation} 
Z_{L}= \mbox{Tr\ } [{\cal T}_{1}{\cal T}_{2}]^{N/2}, 
\end{equation}
somewhat modified from the classical TMRG
expression.

As $Z_{L} = \lambda_{1}^{N/2} + \lambda_{2}^{N/2} + \ldots$, where 
$\lambda_{i}$ are the eigenvalues of ${\cal T}_{1}{\cal T}_{2}$,
the largest eigenvalue $\lambda_{1}$ of ${\cal T}_{1}{\cal T}_{2}$ 
dominates in the thermodynamic limit $N\rightarrow\infty$. The density 
matrix simplifies to
\begin{equation}
    \dm_{uL} = \lambda_{1}^{N/2} \ket{\psi_{R}}\bra{\psi_{L}} ,
\end{equation}    
where $\bra{\psi_{L}}$ and $\ket{\psi_{R}}$ are the left and right 
eigenvectors to eigenvalue $\lambda_{1}$. Due to the checkerboard 
decomposition, the transfer matrix ${\cal T}_{1}{\cal T}_{2}$ is 
non-symmetric, such that left and right eigenvectors are not identical.
Normalization to $\langle \psi_{L}| \psi_{R} \rangle =1$ is assumed.
The free energy per site is
given as
\begin{equation}
    f_{L} = -\frac{1}{2} k_{B}T \ln \lambda_{1} .
\end{equation}    
If $Z_{L}$ (i.e.\ the eigenvalue $\lambda_{1}$) is calculated exactly, 
computational effort scales 
exponentially with $L$ as in exact diagonalization. As the errors 
scale as $(\beta/L)^2$, reliable calculations 
are restricted to small $\beta$, and the interesting low-temperature 
limit is inaccessible. Instead, one can adopt the classical TMRG on the 
checkerboard transfer matrix to access large $L$, hence low 
temperatures. 

Conceptually, the
main steps of the algorithm are unchanged from the
classical case and the argument for the optimality of the decimation procedure is 
unchanged. 
Explicit construction and decimation formulae are however slightly
changed from the classical case and much more complicated notationally 
because of the checkerboard decomposition; see
\citet{Wang97} and \citet{Shib97b,Shib03}.

\begin{figure}
\centering\epsfig{file=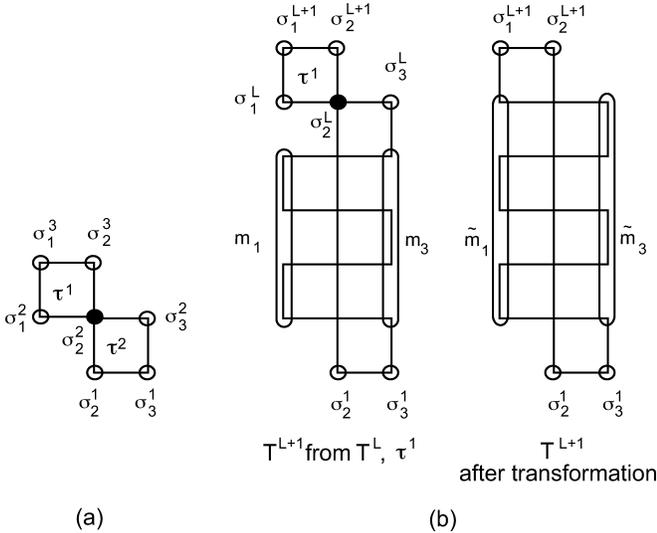,width=\linewidth}
\caption{Two plaquette transfer matrix ${\cal T}^{(3)}$ 
and transfer matrix growth. Solid circles are summed over.}
\label{fig:mat1}
\end{figure}

At the start, the transfer matrix ${\cal T}^{(3)}$ of two plaquettes is given 
by (Fig.\ \ref{fig:mat1})
\begin{eqnarray} & & {\cal T}^{(3)}
(\sigma^{3}_{1}\sigma^{3}_{2};
 \sigma^{2}_{1}\sigma^{2}_{3};
 \sigma^{1}_{2}\sigma^{1}_{3}) = 
\nonumber \\
& & \sum_{\sigma^{2}_{2}} 
\langle \sigma^{3}_{1} \sigma^{3}_{2} | \hat{\tau}_{1} | 
        \sigma^{2}_{1} \sigma^{2}_{2} \rangle 
\langle \sigma^{2}_{2} \sigma^{2}_{3} | \hat{\tau}_{2} | 
        \sigma^{1}_{2} \sigma^{1}_{3} \rangle .  
\end{eqnarray}

The addition of plaquettes follows a zig-zag pattern. Let me 
just discuss the case where the transfer matrix grows to comprise an 
{\em even} number of plaquettes. This number be $L$. Then the growth
formula reads (Fig.\ \ref{fig:mat1}):
\begin{eqnarray*}
& & {\cal T}^{(L+1)}
(\sigma^{L+1}_{1}\sigma^{L+1}_{2};
m_{1}\sigma^{L}_{1} m_{3}\sigma^{L}_{3};
\sigma^{1}_{2}\sigma^{1}_{3}) = \\
& &
\sum_{\sigma^{L}_{2}} 
\langle \sigma^{L+1}_{1} \sigma^{L+1}_{2} | \hat{\tau}_{1} | 
        \sigma^{L}_{1} \sigma^{L}_{2} \rangle
 {\cal T}^{(L)}
(\sigma^{L}_{2}\sigma^{L}_{3};
m_{1} m_{3};
\sigma^{1}_{2}\sigma^{1}_{3}).
\end{eqnarray*}
I distinguish between inner and outer states, the outer states being 
those of the first and last row of the transfer matrix. For the inner 
states, the compound notation $m_{1}$ and $m_{3}$ indicates that they 
may have been subject to a reduced basis transformation beforehand. If 
the number of states $m_{1}\sigma^{L}_{1}$ and $m_{3}\sigma^{L}_{3}$ 
is in excess of the number of states to be kept, a reduced basis 
transformation is carried out, $m_{1}\sigma^{L}_{1}\rightarrow 
\tilde{m}_1$ and $m_{3}\sigma^{L}_{3}\rightarrow\tilde{m}_3$, using 
the reduced basis transformation as obtained in the {\em previous} 
superblock diagonalization.

\begin{figure}
\centering\epsfig{file=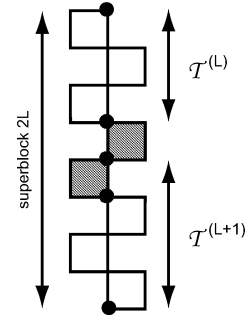,width=0.35\linewidth}
\caption{Quantum transfer matrix DMRG step. Solid circles are summed 
over; note the periodic boundary conditions.}
\label{fig:quantum2}
\end{figure}

The superblock transfer matrix of $2L$ plaquettes is now formed from the product 
of two such ${\cal T}^{(L+1)}$, summing over states $\sigma_{2}^1 
\equiv \sigma_{2}^{2L+1}$ and $\sigma^{L+1}_{2}$  (Fig.\ \ref{fig:quantum2}). 
This superblock 
transfer matrix is
diagonalized using some large sparse eigensolver to get
the maximum eigenvalue  $\lambda_{1}$ and the right and left 
eigenvectors $\bra{\psi_{L}}$ and $\ket{\psi_{R}}$, which may be chosen mutually 
biorthonormal; $\langle \psi_{L}|\psi_{R}\rangle=1$. In view of the 
possible need for a reduced basis transformation in the next step, 
non-symmetric reduced density matrices are formed,
\begin{equation}
    \dm=\text{Tr } \ket{\psi_{R}} \bra{\psi_{L}},
    \label{eq:QTMRGdensity}
\end{equation}    
where trace is over all states in columns 1 and 3 
except $\tilde{m}^1$, $\tilde{m}^3$, $\sigma^{L+1}_{1}$ and 
$\sigma^{L+1}_{3}$ as they are the states subject to a subsequent 
reduced basis transformation. 
Row and column matrices of the $M$ left and 
right eigenvectors with highest eigenvalue weight 
yield the transformation matrices for left and 
right basis states, i.e.\ in columns 1 and 3. This construction implies that 
basis vectors $\bra{m_{1}}$ and $\ket{m_{3}}$ will also be 
biorthonormal. 
The procedure is repeated until the system has reached the desired 
final size of $2L_{\text{max}}$ plaquettes.

Several technical issues have to be discussed for quantum TMRG.
An important reduction of the computational load comes from the
existence of modified good quantum numbers derived from conservation 
laws of the underlying Hamiltonian. Their existence was already observed 
in the quantum transfer matrix method \cite{Nomu91}:
The action of a local magnetization-conserving Hamiltonian 
in Trotter direction between imaginary times $j$ and $j+1$ obeys 
$[S_{1}^{j+1}]^z + [S_{2}^{j+1}]^z =  
[S_{1}^{j}]^z + [S_{2}^{j}]^z $, or,
$ [S_{1}^{j+1}]^z - [S_{1}^{j}]^z = 
-[S_{2}^{j+1}]^z + [S_{2}^{j}]^z$. This generalizes to
$ \sum_{j} (-1)^{j} [S^{j}_{1}]^{z} = - \sum_{j} (-1)^{j} 
[S^{j}_{2}]^{z}$.
Thus the staggered magnetization, summed in Trotter direction, is a conserved
quantity, i.e.\ constant along the real-space axis: 
\begin{equation}
Q = 
\sum_{j=1}^{2L} (-1)^{i+j} [S_{i}^{j}]^z = \mbox{cst.}
\end{equation}
In fermionic systems, particle number conservation implies that
$ P = \sum_{j=1}^{2L} (-1)^{i+j} n_{i}^{j}$
is an additional good quantum number \cite{Shib99a}.
As in $T=0$ DMRG the
``good quantum numbers'' are conserved by the DMRG decimation process. 
Moreover, the eigenstate with the maximum eigenvalue is always in the
subspace $Q=0$ (or $P=0$, respectively): Thermodynamic quantities are 
independent of boundary conditions in high-temperature phases, which are found
for all finite temperatures in one dimension. For open boundary 
conditions in real space,
the lack of an update at the open ends at every second Trotter time 
step implies the subtraction of equal
quantities when forming $Q$, leading to $Q=0$.


%
An important technical complication occurs because the transfer matrices 
to be diagonalized are asymmetric due to the checkerboard decomposition. 
We therefore have to distinguish between left and right eigenvectors 
$\bra{\psi_{L}}$ and $\ket{\psi_{R}}$. Nonhermitian diagonalization 
routines lack the stabilizing variational principle of the hermitian 
case and have to be monitored extremely carefully. The simplest 
choice, very slow but stable, is the power method applied to the transfer matrix 
and its transpose. Faster algorithms that have been used to obtain 
stable results are the Arnoldi and unsymmetric Lanczos methods 
\cite{Golu97}, 
combined with rebiorthogonalization. Once the eigenvectors have been 
obtained, the unsymmetric density-matrix [Eq.\ (\ref{eq:QTMRGdensity})] has 
to be fully diagonalized. Here, two numerical problems typically occur, 
appearance of complex conjugate eigenvalue pairs with spurious 
imaginary parts and loss of biorthonormality. The former is dealt with 
by discarding the spurious imaginary parts and by taking the real and 
imaginary part of one of the eigenvectors as two real-valued 
eigenvectors. The latter, less frequent one, is dealt with by iterative 
reorthogonalization \cite{Ammo99}. 

Both left and right eigenvectors $\bra{\psi_{L}}$ and $\ket{\psi_{R}}$
are needed to construct the density matrix as in Eq.\ 
(\ref{eq:QTMRGdensity}). Having both eigenvectors, a symmetric choice 
(with $\bra{\psi_{R}} = (\ket{\psi_{R}})^\dagger$)
\begin{equation}
    \dm_{\text{symm}}= \text{Tr} \frac{1}{2} 
    [\ket{\psi_{R}}\bra{\psi_{R}}+\ket{\psi_{L}}\bra{\psi_{L}}]
    \label{eq:symmdensity}
\end{equation}
for the density matrix is also conceivable, if one argues that this 
amounts to trying to optimally represent both the left and right 
eigenvectors at the same time.
This choice is much more 
stable numerically, as no complex eigenvalue problem appears and 
diagonalization routines are much more stable, and it somehow also accounts
for the information carried by both eigenvectors.
A detailed study by \citet{Nish99b} clearly favors the 
asymmetric choice. They showed that if the asymmetric density-matrix has 
real eigenvalues, it is for $M$ retained states as precise as the 
symmetric choice for $2M$ states; if they are complex, there is no 
such advantage. 

For the {\em extraction of physical quantities} it is now of advantage 
not to fix the inverse temperature $\beta$ and vary the Trotter 
number $L$, but to replace $\beta/L$ by a fixed initial inverse 
temperature $\beta_{0}\ll 1$ in 
Eq.\ (\ref{eq:approxpartition}). Quantum TMRG growth then reaches for 
$2L$ plaquettes all inverse temperatures $\beta_{0},2\beta_{0},\ldots,L\beta_0$; 
at each step we obtain the
free energy per site $f(T)$ and thus all thermodynamic quantities, such as
the internal energy $u$, magnetization $m$, specific heat at constant 
volume $c_{v}$, and magnetic susceptibility $\chi$ 
by numerical differentiation. Temperatures less than a hundredth of 
the main energy scale of the Hamiltonian can be reached. Convergence 
must be checked both in $M\rightarrow\infty$ and in $\beta_{0}\rightarrow 0$;
the latter convergence is in $\beta_{0}^2$. For a fixed 
number of plaquettes, finite-system DMRG runs can be carried out.

Severe numerical problems may occur for the second
order derivatives $c_v$ and $\chi$. A {\em direct} calculation of $u$ and 
$m$, which are expectation values, reduces the number of differentiations to
one. 

To do this, consider the internal energy $\langle u \rangle = Z^{-1} \mbox{Tr\ } 
[\hat{h}_{1}e^{-\beta \hat{H}}]$.
In the Trotter decomposition ${\cal T}_{1}$ changes to:
\begin{eqnarray}
& & \bra{\sigma^{1}\ldots \sigma^{2L+1}}{\cal T}_{u}^{(2L+1)}\ket{\nu^{1}\ldots 
\nu^{2L+1}} = \label{eq:inserttransfer} \\
& & \langle \sigma^{3}\nu^{3} | \hat{h}_{1} e^{-\beta \hat{h}_{1}} | 
\sigma^{2} 
\nu^{2} 
\rangle 
\prod_{j=2}^{L} \langle \sigma^{2j+1}\nu^{2j+1} | \hat{\tau}_{1} | 
\sigma^{2j}\nu^{2j} \rangle 
\nonumber
\end{eqnarray}
and therefore ($\langle \psi_{L}|$ and $|\psi_{R} \rangle$ are the left (right)
eigenvectors of $\lambda_1$) 
\begin{equation}
\langle u \rangle = 
\frac{\mbox{Tr\ } [\hat{h}_{1}e^{-\beta \hat{H}}]}{\mbox{Tr\ }[e^{-\beta 
\hat{H}}]} =
\frac{\langle \Psi_{L}| {\cal T}_{u}{\cal T}_{2}| 
\Psi_{R} \rangle}{\lambda_{1}}. 
\end{equation}
A direct calculation of $c_v$ from energy fluctuations is less
accurate.
The susceptibility can be obtained from calculating the 
magnetization $m$
at two infinitesimally different fields.

In the beginning, applications of the quantum TMRG focused on 
the thermodynamics of simple Heisenberg spin chains \cite{Wang97,Shib97b,Xian98};
but modified (e.g., frustrated and dimerized) spin chains are also easily accessible 
\cite{Mais98,Klum99,Klum00,Maes00,Shib01a,Wang00c,John00,Kara01}; for frustrated chains, effective 
sites of two original sites are formed to ensure nearest-neighbor 
interactions and the applicability. Quantum TMRG studies of 
ferrimagnetic spin chains have revealed an interesting combination of 
thermodynamics typical of ferromagnets and antiferromagnets
\cite{Yama98a,Mais98a,Kole99}. In fact, the results are remarkably 
similar to those obtained for spin ladders \cite{Hagi00,Wang00b}.

Electronic degrees of freedom have been studied for the
$t$-$J$ model by \citet{Ammo99}, and \citet{Sirk02a,Sirk02b} and for
the Kondo lattice 
\cite{Shib98,Shib99b}. Lately, quantum TMRG has also been applied to 
the algorithmically very similar spin-orbit chain by \citet{Sirk03}.

\citet{Romm99} studied one localized impurity linking two spin chains 
in the thermodynamic limit; there has also been strong interest in the 
case of multiple mobile impurities \cite{Ammo00a,Ammo00b,Ammo01}.

Dynamic properties at finite temperature are among the most frequently 
available experimental data. Quantum TMRG offers, in close analogy to
quantum Monte Carlo calculations, a way to calculate these properties, 
albeit with far less than the usual precision. It has been applied to
anisotropic spin chains \cite{Naef99}, to the one-dimensional Kondo 
chain \cite{Muto98a,Muto99,Shib99a} and has been used to calculate
nuclear relaxation rates \cite{Naef00}. One starts by calculating  
imaginary-time correlations $G(\tau)$ by inserting two operators at 
different imaginary times and time distance $\tau$ into the transfer 
matrix analogous to Eq.\ (\ref{eq:inserttransfer}). 
The spectral function $A(\omega)$ can now be extracted from the 
well-known relationship to $G(\tau)$,
\begin{equation}
    G(\tau)=\frac{1}{2\pi} \int_{0}^{\infty} d\omega\, K(\tau,\omega) A(\omega),
    \label{eq:spectralfunction}
\end{equation}
where the kernel $K(\tau,\omega)$ is 
\begin{equation}
    K(\tau,\omega)= e^{-\tau\omega}+e^{-\beta\omega+\tau\omega}
    \label{eq:kernel}
\end{equation}
and the extension to negative frequencies obtained through $A(-\omega)
= e^{-\beta\omega} A(\omega)$. This decomposition is not unique; 
authors also use related expressions \cite{Naef99}, but the essential 
difficulty in using these expressions is invariant: If we introduce a
finite number of
discrete imaginary times $\tau_{i}$ and frequencies $\omega_{j}$, and
a set $K_{ij}=K(\tau_{i},\omega_{j})$, $G_{i}=G(\tau_{i})$,
$A_{j}=A(\omega_{j})$ the spectral function $A(\omega)$ is 
in principle obtained from inverting (by carrying out a 
singular value decomposition of $K_{ij}$)
\begin{equation}
    G_{i}= \sum_{j} K_{ij} A_{j} ,
    \label{eq:matrixspectral}
\end{equation}
which, as has been known to practitioners of quantum Monte Carlo for a 
long time, is a numerically unstable procedure, because $K_{ij}$ is 
very badly conditioned. The ratio of its largest eigenvalues to its 
smallest eigenvalues is normally so big that it is not encoded 
properly in computer number representations. \citet{Naef99} have 
argued, as is done in quantum Monte Carlo, that given the 
algorithm-based imprecisions in the imaginary-time data, one may 
merely try to find that spectral function that is (in a probabilistic 
sense) most compatible with the raw data. This {\em maximum entropy}
procedure leads to finding that set $\{ A(\omega_{j}) \}$ that 
maximizes
\begin{equation}
    \alpha S[A] - \frac{1}{2} \chi^2[A],
    \label{eq:maxentfunctional}
\end{equation}
where $S[A]$ is the relative entropy
\begin{equation}
    S[A] = \sum_{j} (A_{j} - m_{j} - A_{j} \ln 
    (A_{j}/m_{j}))(1+e^{-\beta\omega})
    \label{eq:relentropy}
\end{equation}
and $\chi^2[A]$ accommodates the noisiness of the raw data,
\begin{equation}
    \chi^2[A] = \sum_{i} (G_{i}-\sum_{j} K_{ij}A_{j})^2/\sigma_{i}^2 .
    \label{eq:chisquared}
\end{equation}    
In the relative entropy, $m_{j}$ is some spectral function that 
embodies previous knowledge relative to which information of 
$A(\omega_{j})$ is measured. It may simply be assumed flat. The
factor $1+e^{-\beta\omega}$ ensures that contributions for $\omega<0$ 
are considered correctly. $\sigma_{i}^2$ measures the estimated error of the 
raw data and is in TMRG of the order of $10^{-6}$. It is found by looking 
for that order of magnitude of $\sigma_{i}^2$ where  
results change least under varying it.  
$\alpha$ is determined self-consistently such that of all solutions 
that can be generated for various $\alpha$ the one most probable 
in the light of the input data is 
chosen \cite{Jarr96}.

Another way of calculating $A(\omega)$ is given by Fourier 
transforming the raw imaginary time data to Matsubara frequencies on 
the imaginary axis, $\omega_{n}=(2n+1)/\pi$ for fermions and
$\omega_{n}=2n/\pi$ for bosons, using $\tau_{i}=i\beta/L$ and
\begin{equation}
    G(\imag\omega_{n}) = \frac{\beta}{L} \sum_{i=0}^{L} 
    e^{\imag\omega_{n}\tau_{i}} G(\tau_{i}) .
    \label{eq:matsubara}
\end{equation}    
$G(\imag\omega_{n})$ is written in some Pad\'{e} approximation and then 
analytically continued from $\imag\omega_{n}$ to infinitesimally above 
the real frequency axis, $\omega+\imag\eta$, $\eta=0^+$. 

The precision of quantum TMRG dynamics is far lower than that of $T=0$ 
dynamics. Essential spectral features are captured, but results are 
not very reliable quantitatively. A serious alternative is currently 
emerging in time-dependent DMRG methods at finite temperature (Sec.\ 
\ref{subsec:time}), but their potential in this field has not been 
demonstrated yet.

\section{SYSTEMS OUT OF EQUILIBRIUM}
\label{sec:nonequilibrium}
The study of strongly correlated electronic systems, from
which DMRG originated, is dominated by attempts to understand 
equilibrium properties of these systems. Hence, the theoretical 
framework underlying all previous considerations has been that of 
equilibrium statistical mechanics, which has been extremely well 
established for a long time. Much less understanding and in particular 
no unifying framework is available so far for non-equilibrium physical 
systems, the main difficulty being the absence of a canonical 
(Gibbs-Boltzmann) ensemble. Physical 
questions to be studied range from transport through quantum dots 
with strong voltage bias in the leads far from linear response, to 
reaction-diffusion processes in chemistry, to ion transport in 
biological systems, to traffic flow on multilane systems. Quite 
generically these situations may be described by time-evolution rules 
that, if cast in some operator language, lead to non-hermitian 
operators \cite{Glau63}. 

The application of DMRG to such problems in one spatial dimension was 
pioneered by \citet{Hiei98}, who studied, using a transfer-matrix approach, 
the discrete-time asymmetric exclusion process (ASEP), a biased 
hopping of hard-core particles on a chain with injection and removal of 
particles at both ends \cite{Derr93}. Excellent agreement with 
exact solutions for particular parameter sets (which correspond to 
matrix-product ansatzes) and with other numerics was found.  
\citet{Kaul98} studied the $q$-symmetric Heisenberg model out of equilibrium.

\subsection{Transition matrices}
\label{subsec:transition}
In the field of non-equilibrium statistical mechanics 
there has been special interest in those one-dimensional systems with 
transitions from active to absorbing steady states under a change of
parameters. Active steady states show internal particle dynamics, 
whereas absorbing steady states are truly frozen (e.g.\ a vacuum state).
These transitions are 
characterized by universal critical exponents, quite analogously to 
equilibrium phase transitions; see \citet{Hinr00} for a review.
\citet{Carl99b}
have shown that DMRG is able to provide reliable estimates of 
critical exponents for
non-equilibrium phase transitions in one-dimensional reaction-diffusion models. 
To this end they applied DMRG to a (non-hermitian) 
master equation of a pseudo-Schr\"{o}dinger form. This approach has been at 
the basis of most later DMRG work on out-of-equilibrium phenomena. 
Considering a chain of length $L$, in which each site is either occupied 
by a particle ($A$) or empty ($\emptyset$), the time evolution of the 
system is given in terms of microscopic rules involving only 
neighboring sites. Typical rules of evolution are site-to-site 
hopping (diffusion) $A\emptyset \leftrightarrow \emptyset A$ or the 
annihilation of neighboring particles $AA \rightarrow  \emptyset 
\emptyset$, all with some fixed rates. Non-equilibrium phase 
transitions in the steady state may arise for competing reactions, 
such as $AA \rightarrow  \emptyset \emptyset$ and $A\emptyset,\emptyset A \rightarrow AA$
simultaneously present. The last model is referred to as the contact 
process. Its steady-state transition between steady states of 
finite and vanishing density in the thermodynamic limit 
is in the universality class 
of directed percolation.

Once the reaction rates are given, the stochastic evolution follows from
a master equation, which can be written as 
\begin{equation}
\frac{d | P(t) \rangle }{d t} = - \ham | P(t) \rangle ,
\label{eq:master}
\end{equation}
where $| P(t) \rangle$ is the state vector containing the probabilities 
of all configurations, as opposed to the true quantum case, where 
the coefficients are probability amplitudes. The elements of the 
``Hamiltonian'' $\ham$ (which is rather a transition matrix) 
are given by
\BEA
\langle \sigma | \ham | \tau \rangle &=& - w(\tau \rightarrow \sigma) \;\; ; \;\;
\sigma\neq\tau \nonumber \\
\langle \sigma | \ham | \sigma \rangle &=& \sum_{\tau \neq \sigma}
w( \sigma \rightarrow \tau) ,
\EEA
where $| \sigma \rangle$, $| \tau \rangle$ are the state vectors of two
particle configurations and
$w(\tau \rightarrow \sigma)$ denotes the transition rates
between the two states and is constructed from the rates 
of the elementary processes. $\ham$ will in 
general be non-hermitian.
Since $\ham$ is a
stochastic matrix, its columns add to zero. The left ground state $\langle 
\psi_{L}|$ is hence given by
\begin{equation}
\langle \psi_{L}| = \sum_{\sigma} \langle \sigma | ,
\end{equation}
with ground state energy $E_0=0$, since $\langle \psi_{L}|\ham =0$; the right 
ground state $\ket{\psi_{R}}$ is problem-dependent. All other eigenvalues 
$E_i$ of $\ham$ have a non-negative real part $\repart E_i \geq 0$ 
\cite{Alca94,Kamp97}. 

Since the formal solution of Eq.\ (\ref{eq:master}) is
\begin{equation}
|P(t)\rangle = e^{-\ham t} |P(0)\rangle
\label{eq:timeevolution}
\end{equation}
the system evolves towards its steady state $|P(\infty)\rangle$. Let
$\Gamma := \inf_{i} \repart E_i$ for $i\neq 0$. 
If $\Gamma>0$, the approach towards the steady state is characterized by a 
{\em finite}\ relaxation time $\tau=1/\Gamma$, but if $\Gamma=0$, that
approach is algebraic. This situation is quite analogous to non-critical
phases ($\tau\neq 0$) and critical points ($\tau=\infty$), 
respectively, which may arise in equilibrium quantum Hamiltonians. 
The big advantage of DMRG is that the steady-state behavior is 
adressed directly and that {\em all} initial configurations are inherently 
averaged over in this approach; the price to be paid are restrictions 
to relatively small system sizes of currently $L\sim 100$ due to inherent numerical instabilities in 
nonhermitian large sparse eigenvalue solvers.  

Standard DMRG as in one-dimensional $T=0$ quantum problems is now used 
to calculate both left and right lowest eigenstates. Steady state 
expectation values such as density profiles or steady state two-point 
correlations are given by expressions
\begin{equation}
\langle n_{i} \rangle = \bra{ \psi_{L}} n_{i} \ket{ \psi_{R}} / 
\langle \psi_{L} | \psi_{R} \rangle ,
\label{eq:expvalues}
\end{equation}
where the subscripts $L$, $R$ serve to remind that two distinct left 
and right ground states are employed in the calculation.
The gap 
allows to extract the relaxation time on a finite lattice and using 
standard finite-size scaling theory. The Bulirsch-Stoer 
transformation (BST) extrapolation scheme
\cite{Henk88} has been found very useful to extract critical exponents 
from the finite-size scaling functions for density profiles and gaps. In 
this way 
\citet{Carl99b} determined both bulk and surface critical exponents of 
the contact process and found their values to be compatible with the 
directed percolation class, as expected.

In cases where the right ground state is also trivially known (e.g.\ 
the empty state), DMRG calculations can be made more stable by 
shifting this trivially known eigenstate pair to some high ``energy'' 
in Hilbert space by adding a term 
$E_{\text{shift}}\ket{\psi_{R}}\bra{\psi_{L}}$ to 
the Hamiltonian, turning the non-trivial first excitation into the 
more easily accessible ground state.

For a correct choice of the density matrix it is crucial always to target 
the ground state, even if it shifted away; otherwise it will reappear 
due to numerical inaccuracies in the representation of the shift 
operator \cite{Carl01b}. Moreover, the trivial left ground state must be targeted, 
although it contains no information, to get a good representation of
the right eigenstates that are joined in a biorthonormality relation 
\cite{Carl99b}. For 
nonhermitean Hamiltonians, there is also a choice between symmetric 
and nonsymmetric density matrices. It was found empirically \cite{Carl99b} that
the symmetric choice
\begin{equation}
    \dm = \text{Tr}_{E} \frac{1}{2} \left( \ket{\psi_{R}}\bra{\psi_{R}} + 
    \ket{\psi_{L}}\bra{\psi_{L}} \right)
    \label{eq:symmdens}
\end{equation}
was most efficient, as opposed to quantum TMRG, where the 
nonsymmetric choice is to be preferred. This is probably due to 
numerical stability: a nonsymmetric density-matrix makes this 
numerically subtle DMRG variant even less stable. The same observation 
was made by \citet{Sent99} in the context of SUSY chains.

The method has been applied to the ASEP model by \citet{Nagy02}.
A series of papers has been able to produce impressive results on 
reptating polymers exposed to the drag of an external field  such as 
in electrophoresis in the 
framework of the Rubinstein-Duke model 
\cite{Carl01a,Carl02,Carl03,Paes02,Paes03}. 
DMRG has been 
able to show for the renewal time (longest relaxation time) $\tau$ a crossover 
from  
an (experimentally observed) behavior $\tau \sim N^{3.3\pm 0.1}$ for short 
polymers to the theoretically predicted $\tau \sim N^{3}$ for long 
polymers.

Another field of application which is under active debate are the 
properties of the one-dimensional pair-contact process with 
single-particle diffusion (PCPD), which is characterized by the 
reactions $AA \rightarrow \emptyset\emptyset$, 
$\emptyset AA, AA\emptyset \rightarrow AAA$. While early 
field-theoretic studies \cite{Howa97} showed that its steady-state 
transition is not in the directed percolation class, the first 
quantitative data came from a DMRG study \cite{Carl01b} and suggested 
again a universality class different from directed percolation. At 
present, despite further extensive DMRG studies \cite{Henk01,Carl03}
and additional simulational and analytical studies, no consensus on 
the critical behavior of the one-dimensional PCPD is reached yet.
It may be safely stated that DMRG has 
sparked a major controversial debate in this field. While the raw data 
as such are not disputed, it is the extrapolations to the 
infinite-size limit that are highly controversial, 
such that the answer to this issue is well outside the scope of this 
review. 

\subsection{Stochastic transfer matrices}
\label{subsec:stochastic}
\citet{Kemp01} have made use of the pseudo-Hamiltonian formulation of 
stochastic systems as outlined above to apply the formalism of quantum 
TMRG. The time evolution of Eq.\ (\ref{eq:timeevolution}) and 
the Boltzmann operator $e^{-\beta\ham}$ are formally identical, such 
that in both cases the transfer matrix construction outlined in
Sec.\ \ref{subsec:quanttransfer} can be carried out. In the 
thermodynamic limit $N\rightarrow\infty$ all physical information is 
encoded in the left and right eigenvectors to the largest eigenvalue, 
which are approximately represented using the TMRG machinery.  

In the stochastic TMRG the local transfer matrices are direct 
evolutions in a time interval $\Delta t$ obtained from the discretized 
time evolution,
\begin{equation}
    \ket{P(t+\Delta t)}=e^{-\ham \Delta t} \ket{P(t)} .
    \label{eq:discretetimeevolution}
\end{equation}
Otherwise, the formalism outlined in Sec. \ref{subsec:quanttransfer} 
carries over identically, with some minor, but important 
modifications. Boundary conditions in time are open, since periodic 
boundary conditions 
would set the future equal to the past. Moreover, stochastic 
TMRG only works with a symmetric density-matrix constructed from left 
and right eigenvectors: \citet{Enss01} have shown that the open 
boundary conditions imply that the unsymmetric choice of the density 
matrix, $\dm=\text{Tr}_{E} \ket{\psi_{R}} \bra{\psi_{L}}$, which was 
superior in quantum TMRG, has no meaningful information because it 
has only one non-zero eigenvalue and that for 
stochastic TMRG the use of the symmetric choice, $\dm=(1/2) 
\text{Tr}_{E} [\ket{\psi_{L}} \bra{\psi_{L}}+ \ket{\psi_{R}} \bra{\psi_{R}}]$, 
is mandatory. 

The advantage of the stochastic TMRG is that it works in the thermodynamic 
limit and that, although it can only deal with comparatively short 
time scales of some hundred time steps, it automatically averages 
over all initial conditions and is hence bias-free; it can be seen as 
complementary to the method of the last section. 
Unfortunately, it has also been shown by \citet{Enss01} that there is a 
critical loss of precision strongly limiting the number $L$ of possible time steps;
it is a property of probability-conserving stochastic transfer matrices
that the norm of the left and right eigenvectors to the largest 
eigenvalue diverges exponentially 
in $L$, while biorthonormality $\langle \psi_{L} | \psi_{R} \rangle = 1$ 
should hold exactly as well, which cannot be ensured by finite computer
precision.
        
\begin{figure}
\centering\epsfig{file=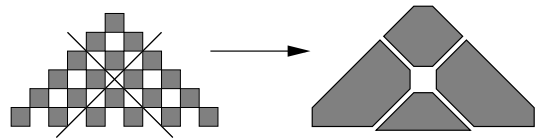,scale=0.5}
\vspace{0.3truecm}
\caption{TMRG applied to the causal light cone of a stochastic 
Hamiltonian evolving in real time: four corner transfer matrices make 
up the light cone. From \citet{Kemp03}. Reprinted with permission.}
\label{fig:lightcone}
\end{figure}

Building on the observation by \citet{Enss01} that the local physics at 
some place and at time $t$ is determined by past events in a 
finite-sized light-cone, \citet{Kemp03} have applied a variant of the 
corner transfer matrix algorithm to the light-cone (Fig.\ 
\ref{fig:lightcone}), reporting that the numerical loss of precision 
problem now occurs at times several orders of magnitude larger, 
greatly enhancing the applicability of stochastic TMRG. This method 
has been applied by \citet{Enss04} to study scaling functions for 
ageing phenomena in systems without detailed balance.

\subsection{Time-dependent DMRG}
\label{subsec:time}

So far, all physical properties discussed in this review and obtained 
via DMRG have been either true equilibrium quantities, static or 
dynamic, or steady-state quantities. 
However, time-dependent phenomena in strongly correlated 
systems are more and more coming to the forefront of interest. On the 
one hand, in technological applications such as envisaged in 
nanoelectronics, it will be of great interest to fully understand the 
time-dependent response of quantum many-body systems to external 
time-dependent perturbations and to calculate transport far from 
equilibrium. On the other hand, the recent mastery of 
storing ultracold bosonic atoms in a magnetic trap superimposed by an 
optical lattice has allowed to drive, at will, by time-dependent 
variations of the optical lattice strength, quantum phase transitions 
from the superfluid (metallic) to Mott insulating regime, which is one 
of the key phase transitions in strongly correlated electron systems 
\cite{Grei02}. 

The fundamental difficulty can be seen considering the time-evolution 
of a quantum state $\ket{\psi(t=0)}$ under the action of some 
time-independent Hamiltonian $\ham\ket{\psi_{n}}=E_{n}\ket{\psi_{n}}$.
If the eigenstates $\ket{\psi_{n}}$ are known, expanding $\ket{\psi(t=0)}=
\sum_{n}c_{n}\ket{\psi_{n}}$ leads to the well-known time evolution
\begin{equation}
    \ket{\psi(t)}=\sum_{n}c_{n}\exp (-\imag E_{n}t) \ket{\psi_{n}} , 
\label{eq:simpletime}
\end{equation}
where the modulus of the expansion coefficients of $\ket{\psi(t)}$ is 
time-independent. A sensible Hilbert space truncation is then given by a 
projection onto the large-modulus eigenstates. In strongly correlated 
systems, however, we usually have no good knowledge of the 
eigenstates. Instead, one uses some orthonormal basis with unknown 
eigenbasis expansion, $\ket{m}=\sum_{n} a_{mn} \ket{\psi_{n}}$. The 
time evolution of the state $\ket{\psi(t=0)}=
\sum_{m}d_{m}(0)\ket{m}$ then reads
\begin{equation}
    \ket{\psi(t)}=\sum_{m}\left( \sum_{n} d_{m}(0) a_{mn} e^{-\imag 
    E_{n}t} \right) \ket{m} \equiv \sum_{m} d_{m}(t) \ket{m} , 
\label{eq:complicatedtime}
\end{equation}
where the modulus of the expansion coefficients is {\em 
time-dependent}. For a general orthonormal basis, Hilbert space truncation at one 
fixed time (i.e. $t=0$) will therefore not ensure a reliable 
approximation of the time evolution. Also, 
energy {\em differences} matter in time evolution. 
The sometimes justified hope that
DMRG yields a good approximation to the low-energy Hamiltonian is 
hence of limited use. 

All time-evolution schemes for DMRG so far follow one of two different 
strategies. {\em Static} Hilbert space DMRG methods try to enlarge 
the truncated Hilbert space optimally to approximate $\ket{\psi(t=0)}$ 
such that it is big enough to 
accommodate (i.e.\ maintain large overlap with the exact result) 
$\ket{\psi(t)}$ for a sufficiently long time to a very good 
approximation. More recently, {\em adaptive} Hilbert space DMRG methods 
keep the size 
of the truncated Hilbert space fixed, but try to change it as time 
evolves such that it also accommodates $\ket{\psi(t)}$ to a very good 
approximation.

{\em Static time-dependent DMRG.} \citet{Caza02} were the first to exploit DMRG to
systematically calculate
quantum many-body effects out of equilibrium.
After applying a standard DMRG calculation to the Hamiltonian 
$\ham(t=0)$, the time-dependent Schr\"{o}dinger equation is 
numerically integrated forward in time, building an effective 
$\ham_{\text{eff}}(t)=\ham_{\text{eff}}(0)+\hat{V}_{\text{eff}}(t)$, where 
$\ham_{\text{eff}}(0)$ is 
taken as the last superblock Hamiltonian approximating
$\ham(0)$. $\hat{V}_{\text{eff}}(t)$ as an approximation to 
$\hat{V}$ is built using 
the representations of operators in the final block bases:
\begin{equation}
    i\frac{\partial}{\partial t} \ket{\psi(t)} =
    [\ham_{\text{eff}}-E_{0}+\hat{V}_{\text{eff}}(t)] \ket{\psi(t)} .
    \label{eq:Schroedinger}
\end{equation}
The initial condition is obviously to take 
$\ket{\psi(0)}=\ket{\psi_{0}}$ obtained by the preliminary DMRG run.
Forward integration can be carried out by step-size adaptive methods 
such as Runge-Kutta integration based on the infinitesimal 
time evolution operator
\begin{equation}
    \ket{\psi(t+\Delta t)} = (1- \imag 
    \ham(t) \Delta t) \ket{\psi(t)}.     
    \label{eq:nonunitaryevolution}
\end{equation}

As an application, Cazalilla and Marston have considered a quantum dot 
weakly coupled to noninteracting leads of spinless fermions,
where time-dependency is introduced through a time-dependent chemical 
potential coupling to the number of particles left and right of the 
dot, 
\begin{equation}
    \hat{V}(t) = - \delta\mu_{R}(t) \hat{N}_{R}- \delta\mu_{L}(t) 
    \hat{N}_{L} ,
    \label{eq:bias}
\end{equation}
which is switched on smoothly at $t=0$. Setting 
$\delta\mu_{L}=-\delta\mu_{R}$, one may gauge-transform this chemical 
potential away into a time-dependent complex hopping from and to the 
dot,
\begin{equation}
    t_{q}(t) = t_{q} \exp \left[ \imag \int_{-\infty}^t dt' \delta\mu_{L}(t') 
    \right] .
    \label{eq:timehopping}
\end{equation}
The current is then given by evaluating the imaginary part of the local hopping 
expectation value. Obviously, in a finite system currents will not 
stay at some steady state value, but go to zero on a time scale of 
the inverse system size, when lead depletion has occurred. 

In this approach the hope is that an 
effective Hamiltonian 
obtained by targeting the ground state of the $t=0$ Hamiltonian is 
capable to catch the states that will be visited by the 
time-dependent Hamiltonian during time evolution. Cazalilla and 
Marston argue that on short time scales the perturbation can only access few excited 
states which are well represented in the 
effective Hamiltonian. With one exception \cite{Luo03} this seems 
borne out in their main application, the transport through a quantum 
dot; in many other applications, this approach is not sufficient.

As one way out of this dilemma, it has been demonstrated 
\cite{Luo03} that using 
a density matrix that is given by a superposition of states 
$\ket{\psi(t_{i})}$ at various times of the evolution,
$\dm = \sum_{i=0}^{N_{t}} \alpha_{i} 
\ket{\psi(t_{i})}\bra{\psi(t_{i})}$
with $\sum \alpha_{i}=1$
for the determination of the reduced Hilbert space is much more precise, whereas simply 
increasing $M$ is not suffcient. Of course, these states    
are not known initially; it was proposed to
start from a small DMRG system, evolve it in time, take these state 
vectors, use them in the density matrix to determine the reduced 
Hilbert space, and then to move on to the 
next larger DMRG system where the procedure is repeated, which is 
very time-consuming.

It is important to note that the simple Runge-Kutta approach is not even unitary, and 
should be improved by using e.g.\ the unitary Crank-Nicholson 
time-evolution \cite{Dale04}.

Instead of considering time evolution as a differential equation, one 
may also consider the time-evolution operator $\exp (-\imag\ham t)$, 
which avoids the numerical delicacies of the Schr\"{o}dinger 
equation. \citet{Schm04} computes the transport through a small 
interacting nanostructure using this approach. To this end, 
he splits the problem into two parts: By obtaining a relatively large number of 
low-lying eigenstates exactly (within time-independent DMRG 
precision), one can calculate their time evolution exactly. For the 
subspace orthogonal to these eigenstates, he uses an efficient implementation 
of the matrix exponential $\ket{\psi(t+\Delta t)}=\exp (-\imag\ham \Delta 
t)\ket{\psi(t)}$ using a Krylov subspace approximation; the reduced 
Hilbert space is determined by finite-system sweeps, concurrently targeting all
$\ket{\psi(t_{i})}$ at fixed times $t_{i}$ up to the time presently reached in 
the calculation. The 
limitation to this algorithm comes from the fact that to target more 
and more states as time evolves, the effective Hilbert space has to be 
chosen increasingly large.

{\em Adaptive time-dependent DMRG.} 
Time-dependent DMRG using {\em adaptive} Hilbert spaces has first been 
proposed in essentially identical form by \citet{Dale04} and 
\citet{Whit04}. Both approaches are efficient implementations of an 
algorithm for the classical simulation of the time evolution of 
weakly entangled quantum states invented by \citet{Vida03,Vida03a} 
[time-evolving block-decimation (TEBD) algorithm]. The TEBD algorithm 
was originally formulated in the matrix product state language. For 
simplicity, I will explain the algorithm in its DMRG context; a 
detailed discussion of the very strong connection between adaptive time-dependent 
DMRG and the original simulation algorithm is given by \citet{Dale04}: it 
turns out that DMRG naturally attaches good quantum numbers to state 
spaces used by the TEBD algorithm, allowing for the usual drastic 
increases in performance due to the use of good quantum numbers.   

Time evolution in the adaptive time-dependent DMRG is generated using 
a Trotter-Suzuki decomposition as discussed for quantum TMRG (Sec.\ 
\ref{subsec:quanttransfer}). Assuming nearest-neighbour interactions 
only, the decomposition reads $\ham=\hat{H}_{1}+\hat{H}_{2}$.
$\hat{H}_{1}=\sum_{i=1}^{N/2} \hat{h}_{2i-1}$ and 
$\hat{H}_{2}=\sum_{i=1}^{N/2} \hat{h}_{2i}$; $\hat{h}_{i}$ is the 
local Hamiltonian 
linking sites $i$ and $i+1$; neighboring local Hamiltonians will in general not 
commute, hence $[\hat{H}_{1},\hat{H}_{2}]\neq 0$. However, all terms 
in $\hat{H}_{1}$ or $\hat{H}_{2}$ commute. The first order Trotter 
decomposition of the infinitesimal time-evolution operator then reads
\begin{equation}
    \exp (- \imag \ham \Delta t) = \exp (-\imag \ham_{1} \Delta t)
    \exp (-\imag \ham_{2} \Delta t) + O(\Delta t^2) .
    \label{eq:firstorderTrotter}
\end{equation}    
Expanding $\ham_{1}$ and $\ham_{2}$ into the local Hamiltonians, one 
infinitesimal time step $t\rightarrow t +\Delta t$ may be carried out 
by performing the local time-evolution on (say) all even bonds first 
and then all odd bonds. In DMRG, a bond can be time-evolved exactly 
if it is formed by the two explicit sites typically occurring in DMRG 
states. We may therefore carry out an exact time-evolution by 
performing one finite-system sweep forward and backward 
through an entire one-dimensional chain, with time-evolutions on all 
even bonds on the forward sweep and all odd bonds on the backward 
sweep, at the price of the Trotter error of $O(\Delta t^2)$. This 
procedure necessitates that $\ket{\psi(t)}$ is available in the right 
block bases, which is ensured by carrying out the reduced basis 
transformations on $\ket{\psi(t)}$ that in standard DMRG form the 
basis of White's prediction method \cite{Whit96b}. The decisive idea of 
\citet{Vida03,Vida03a} was now to carry out a new Schmidt decomposition 
and make a new choice of the most relevant block basis states for $\ket{\psi(t)}$ 
after each local bond update. Therefore, as the quantum state changes 
in time, so do the block bases such that an optimal representation of 
the time-evolving state is ensured. Choosing new block bases changes 
the effective Hilbert space, hence the name {\em adaptive} 
time-dependent DMRG. 

An appealing feature of this algorithm is that it can be very easily 
implemented in existing finite-system DMRG. One uses standard 
finite-system DMRG to generate a high-precision initial state 
$\ket{\psi(0)}$ and continues to run finite-system sweeps, one for 
each infinitesimal time step, merely replacing the large 
sparse-matrix diagonalization at each step of the sweep by local bond 
updates for the odd and even bonds, respectively. 

Errors are reduced by using the second order Trotter decomposition
\begin{equation}
    e^{- \imag \ham \Delta t} = e^{-\imag \ham_{1} \Delta t/2} 
     e^{-\imag \ham_{2} \Delta t}
     e^{-\imag \ham_{1} \Delta t/2} + O(\Delta t^3) .
    \label{eq:secondorderTrotter}
\end{equation}    
As $\Delta t^{-1}$ steps are needed to reach a fixed time $t$, the 
overall error is then of order $O(\Delta t^2)$. 
One efficient way of implementing is to group the half-time steps of 
two subsequent infinitesimal evolution steps together, i.e.\ carry out 
the standard first-order procedure, but calculate expectation values 
as averages of expectations values measured before and after a
$\exp (-\imag \ham_{2} \Delta t)$ step.

Further errors are produced by the reduced basis 
transformations and the sequential update of the representation of block states in the 
Schmidt decomposition after each bond update during an infinitesimal 
time step. 
Applications of adaptive time-dependent DMRG so far for the 
time-dependent Bose-Hubbard model \cite{Dale04} and spin-1 Heisenberg 
chains \cite{Whit04} have demonstrated that it allows to access 
large time scales very reliably. The errors mentioned can be well 
controlled by increasing $M$ and show significant accumulation only 
after relatively long times.  

{\em Time-dependent correlations.} 
Time-dependent DMRG also yields an alternative approach to 
time-dependent correlations which have been evaluated in frequency 
space in Sec.\ \ref{sec:dynamics}. Alternatively, one may calculate
expressions as
\begin{equation}
    \bra{\psi} c^\dagger_{i}(t) c_{j}(0) \ket{\psi} = \bra{\psi} e^{+\imag\ham 
    t} c_{i}^\dagger e^{-\imag \ham t} c_{j} \ket{\psi}
\end{equation}    
constructing both $\ket{\psi}$ and $\ket{\phi}=c_{j}\ket{\psi}$ using 
standard DMRG (targeting both), and calculating both $\ket{\psi(t)}$ and
$\ket{\phi(t)}$ using time-dependent DMRG. The desired correlator is 
then simply given as
\begin{equation}
    \bra{\psi(t)} c_{i}^\dagger \ket{\phi(t)}
\end{equation}
and can be calculated for all $i$ and $t$ as time 
proceeds. Frequency-momentum space is reached by Fourier 
transformation. Finite system-sizes and edge effects impose physical 
constraints on the largest times and distances $|i-j|$ or minimal 
frequencies and wave vectors accessible,  
but unlike dynamical DMRG, one run of the time-dependent algorithm is 
sufficient to cover the entire accessible frequency-momentum range. 
\citet{Whit04} report a very successful calculation of 
the one-magnon dispersion relation in a $S=1$ Heisenberg 
antiferromagnetic chain using adaptive time-dependent DMRG, but 
similar calculations are in principle feasible in all time-dependent 
methods.

\subsection{Time-evolution and 
dissipation at $T>0$}
While our discussion so far has been restricted to time evolution at 
$T=0$, it is possible to generalize to the (possibly dissipative) time 
evolution at $T>0$ \cite{Vers04a,Zwol04}. Both proposals are in the 
spirit of adaptive Hilbert space methods. One prepares a 
$T=\infty$ completely mixed state $\hat{\rho}_{\infty}$, 
from which a finite-temperature mixed state $\exp (-\beta\ham)$ at 
$\beta^{-1}=T>0$ is created by imaginary-time evolution, i.e.\ 
\begin{equation}
    \exp (-\beta\ham) = (e^{-\tau\ham})^N \hat{\rho}_{\infty} 
    (e^{-\tau\ham})^N,
\end{equation}    
where $\beta = 2N\tau$ and $\tau\rightarrow 0$. The infinitesimal-time evolution operator 
$e^{-\tau\ham}$ is Trotter decomposed into locally acting bond 
evolution operators. The finite-temperature state is then evolved in 
real time using Hamiltonian or Lindbladian dynamics. The differences 
reside essentially in the representation of mixed states and the truncation 
procedure.

\citet{Vers04a} consider the 
$T=0$ matrix product state of Eq.\ (\ref{eq:mpsperiodic}), where the 
$A_{i}[\sigma_{i}]$ are interpreted (see Sec.\ \ref{subsec:mpa}) as maps from 
the tensor product of two auxiliary states (dimension $M^2$) to a 
physical state space of dimension $N_{\text{site}}$. This can be 
generalized to finite-temperature matrix product {\em density 
operators}
\begin{equation}
    \hat{\rho} = \sum_{\{\fat{\sigma}\}\{\fat{\sigma}'\}} \text{Tr}
    \left[\prod_{i=1}^L \hat{M}_{i} 
        [\sigma_{i}\sigma_{i}']
        \right] \ket{\fat{\sigma}}\bra{\fat{\sigma}'} .
        \label{eq:mpdoperiodic}
\end{equation}    
Here, the $ \hat{M} [\sigma_{i}\sigma_{i}']$ now are maps from the 
tensor product of four auxiliary state spaces (two left, two right of 
the physical site) of dimension $M^4$ to the 
$N_{\text{site}}^2$-dimensional local density-operator state space. 
The most 
general case allowed by quantum mechanics is for $\hat{M}$ to be a 
completely positive map. The dimension of $\hat{M}$ seems to be 
prohibitive, but it can be decomposed into a sum of $d$ tensor 
products of $A$-maps as
\begin{equation}
    \hat{M}_{i} [\sigma_{i}\sigma_{i}'] = \sum_{a_{i}=1}^{d}
    A_{i} [\sigma_{i}a_{i}] \otimes A_{i}^*  [\sigma_{i}'a_{i}].
    \label{eq:completemapdecomp}
\end{equation}
In general, $d\leq N_{\text{site}} M^2$, but it will be important to 
realise that for thermal states $d= N_{\text{site}}$ only. At $T=0$, 
one recovers the standard matrix product state, i.e.\ $d=1$. In order 
to actually simulate $\hat{\rho}$, \citet{Vers04a} consider the purification
\begin{equation}
    \ket{\psi_{MPDO}} = \sum_{\{\fat{\sigma}\}\{\bf{a}\}}
    \text{Tr}
        \left[\prod_{i=1}^L \hat{A}_{i} [\sigma_{i}a_{i}]
            \right] \ket{\fat{\sigma}\bf{a}} ,
    \label{eq:purification}
\end{equation}
such that $\hat{\rho}=\text{Tr}_{\text{\bf{a}}} \ket{\psi_{MPDO}}\bra{\psi_{MPDO}}$.
Here, ancilla state spaces $\{ \ket{a_{i}} \}$ of dimension 
$d$ have been introduced.

In this form, a completely mixed state is 
obtained from matrices $A_{i}[\sigma_{i}a_{i}] \propto \id \cdot 
\delta_{\sigma_{i},a_{i}}$, where $M$ may be 1 and normalization has 
been ignored. This shows $d= N_{\text{site}}$. This state is now 
subjected to infinitesimal 
evolution in imaginary time, $e^{-\tau\ham}$. As it acts on $\sigma$ only, 
the dimension of the ancilla state 
spaces need not be increased. Of course, for $T=0$ the state may be 
efficiently prepared using standard methods.

The imaginary-time evolution is carried out after a Trotter 
decomposition into infinitesimal time steps on bonds. The local bond 
evolution operator $\hat{U}_{i,i+1}$ is conveniently decomposed into a sum of 
$d_{U}$ tensor products of on-site evolution operators,
\begin{equation}
    \hat{U}_{i,i+1} = \sum_{k=1}^{d_{U}} \hat{u}_{i}^k \otimes \hat{u}_{i+1}^k .
    \label{eq:onsitedecomp}
\end{equation}
$d_{U}$ is typically small, say 2 to 4. Applying now 
the time evolution at, say, all odd bonds exactly, the auxiliary 
state spaces are enlarged from dimension $M$ to $Md_{U}$. One now 
has to find the optimal approximation $\ket{\tilde{\psi}(t+\Delta t)}$ 
to $\ket{\psi(t+\Delta t)}$ using auxiliary state spaces of dimension 
$M$ only. Hence, the state spaces of dimension $Md_{U}$ 
must be truncated optimally to minimize $\| \ket{\tilde{\psi}(t+\Delta 
t)} - \ket{\psi(t+\Delta t)}\|$. If one uses the matrices composing 
the state at $t$ as 
initial guess and keeps all $A$-matrices but one fixed, one obtains a 
linear equation system for this $A$; sweeping through all $A$-matrices 
several times is sufficient to reach the fixed 
point which is the variational optimum. As temperature is lowered, 
$M$ will be increased to maintain a desired precision.
Once the thermal state is constructed, real-time evolutions governed 
by Hamiltonians can be 
calculated similarly. In the case of Lindbladian time evolutions, they 
are carried out directly on states of the form of Eq.\ (\ref{eq:mpdoperiodic}),
which is formally equivalent to a matrix product state on a local 
$N_{\text{site}}^2$ dimensional state space.

The approach of \citet{Zwol04} is based on the observation that local 
(density) operators $\sum \rho_{\sigma\sigma'} \ket{\sigma}\bra{\sigma'}$ 
can be represented as $N_{\text{site}}\times N_{\text{site}}$
hermitian matrices. They now reinterpret the matrix coefficients as 
the coefficients of a local ``superket'' defined on a 
$N_{\text{site}}^2$-dimensional local state space. Any globally 
acting (density) operator is now a global superket expanded as a linear combination 
of tensor products of local superkets,
\begin{equation}
    \hat{\rho} = \sum_{\overline{\sigma}_{1}=1}^{N_{\text{site}}^2} \ldots 
    \sum_{\overline{\sigma}_{L}=1}^{N_{\text{site}}^2} 
    c_{\overline{\sigma}_{1}\ldots\overline{\sigma}_{L}} 
    \ket{\overline{\sigma}_{1}\ldots\overline{\sigma}_{L}} .
    \label{eq:superket}
\end{equation}
Here, $\ket{\overline{\sigma}_{i}} = \ket{\sigma_{i}}\ket{\sigma_{i}'}$, the state 
space of the local superket. We have now reached complete formal 
equivalence between the description of a pure state and a mixed state,
such that the TEBD algorithm \cite{Vida03,Vida03a} or its DMRG 
incarnation \cite{Dale04,Whit04} can be applied to the time-evolution 
of this state. 

Choosing the local totally mixed state as one basis state of the local 
superket state space, say $\overline{\sigma}_{i}=1$, the global totally 
mixed state has only one 
non-zero expansion coeffcient,  $c_{1\ldots1}=1$ upon suitable 
normalization. This shows that the $T=\infty$ state, as in the 
previous algorithm, takes a 
particularly simple form in this expansion; it can also be brought 
very easily to some matrix product form corresponding to a DMRG 
system consisting of two blocks and two sites. The 
time-evolution operators, whether of Hamiltonian or Lindbladian origin, 
now map from operators to operators and are hence referred to as 
``superoperators''. If one represents operators as superkets, these 
superoperators are formally equivalent to standard operators. As in 
the $T=0$ case, therefore, they  
are now Trotter decomposed, applied in infinitesimal time steps to all 
bonds sequentially, with Schmidt decompositions being carried out to 
determine block (super)state spaces. The number of states kept is 
determined from the acceptable truncation error. Imaginary time evolutions 
starting from the totally mixed state generate thermal states of a given
temperature, that may then be subjected to some evolution in real 
time. It turns out again that the number of states to be kept grows with 
inverse temperature.

For the simulation of dissipative systems both approaches are 
effectively identical but for the optimal approximation of 
\citet{Vers04a} replacing the somewhat less precise truncation of 
the approach of \citet{Zwol04} (see the discussion of the precision 
achievable for various DMRG setups in Sec.\ \ref{subsec:mpa}).
\section{OUTLOOK}
\label{sec:outlook}
What lies ahead for DMRG? On the one hand, DMRG is quickly becoming a 
standard tool routinely applied to study all low-energy properties of
strongly correlated one-dimensional quantum systems which I presume 
will be done using black box DMRG codes. On the other hand, there are 
several axes of DMRG research which I expect to be quite fruitful in 
the future: DMRG might emerge as a very powerful tool in quantum 
chemistry in the near future; its potential for time-dependent 
problems is far from being understood. Last but not least I feel that 
two-dimensional model Hamiltonians will increasingly be within reach 
of DMRG methods, at least for moderate system sizes that are still 
beyond the possibilities of exact diagonalization and quantum Monte 
Carlo. 

The strong link of DMRG to quantum information theory has only 
recently been started to be 
exploited -- DMRG variants to carry out complex calculations of 
interest in quantum information theory might emerge, while DMRG 
itself could be applied in a more focused manner using the new 
insights about its intrinsic rationale. In that sense, DMRG would be 
at the forefront of a growing entanglement between condensed matter 
physics and quantum computation.

\section*{Acknowledgments}

I would like to thank all the people with whom I have been 
discussing DMRG matters over the years -- they are too many to be 
named. This work has been supported by 
The Young Academy at the Berlin-Brandenburg Academy of Sciences 
and the Leopoldina.

\end{document}